\documentclass[a4paper,12pt]{article}
\usepackage{epsfig,graphicx,xcolor,amsbsy,amssymb,latexsym,amsfonts,amsmath,tcolorbox,setspace}
\usepackage{pstricks}
\usepackage{color}
\usepackage{soul}
\usepackage{placeins}
\usepackage{wrapfig,framed,caption}
\usepackage[makeroom]{cancel}

\usepackage[style=numeric,sorting=none,style=numeric-comp,maxnames=10]{biblatex}

\definecolor{shadecolor}{gray}{0.925}

\usepackage{eurosym}
\usepackage[a4paper]{geometry}
\usepackage{hyperref}
\usepackage{graphicx}
\usepackage{tikz}
\usetikzlibrary{decorations.pathreplacing,decorations.markings,snakes}
\usetikzlibrary{decorations.pathmorphing}
\usepackage{xcolor}
\usepackage{amsmath,amssymb,amsfonts,pstricks,setspace}
\usepackage[enableskew]{youngtab}
\usepackage{ytableau}

\numberwithin{equation}{section}

\newcommand{\bea}{\begin{eqnarray}\displaystyle}
\newcommand{\eea}{\end{eqnarray}}

\newcommand{\buildH}[2]{H^{{(#1),\{0\}}}_{(#2)}}
\newcommand{\buildW}[2]{W^{{(#1)}}_{(#2)}}

\newcommand{\Qra}{Q_{\widehat{a}_1}}

\newcommand{\Qa}{Q_{\widehat{a}}}
\newcommand{\Qs}{Q_{S}}

\newcommand{\Qt}{Q_{\tau}}
\newcommand{\Qr}{Q_{\rho}}

\newcommand{\partfe}[2]{\mathfrak{z}_{#1}^{#2}}
\newcommand{\partfen}[2]{\mathfrak{p}_{#1}^{#2}}
\newcommand{\partfenz}[2]{\mathfrak{q}_{#1}^{#2}}
\newcommand{\partfend}[2]{\mathfrak{h}_{#1}^{#2}}

\newcommand{\sth}[1]{H_{#1}}
\newcommand{\stw}[1]{W_{#1}}

\newcommand{\Qna}[1]{Q_{\widehat{a}_{#1}}}

\newcommand{\fa}{\phi_{0}}
\newcommand{\fb}{\phi_{-2}}

\usepackage{float}
\newtcolorbox{summary}[2][]{colbacktitle=blue!10!white, colback=yellow!10!white,coltitle=blue!70!black, title={#2},fonttitle=\bfseries,#1}

\title{
\begin{flushright}{\vspace{-2.5cm}\small LYCEN 2022-03\\}\end{flushright}
\vspace{2.3cm}
{\bf Little String Instanton Partition Functions and Scalar Propagators}\\[40pt]}

\author{\large \textsc{Baptiste~Filoche\footnote{\tt b.filoche@ip2i.in2p3.fr}}~,~\,and\,~\textsc{Stefan~Hohenegger\footnote{\tt s.hohenegger@ipnl.in2p3.fr}}}

\begin{document}

\maketitle
\thispagestyle{empty}
\begin{center}
\renewcommand{\thefootnote}{\fnsymbol{footnote}}\vspace{-0.5cm}
${}^{\footnotemark[1]\,\footnotemark[2]}$ Univ Lyon, Univ Claude Bernard Lyon 1, CNRS/IN2P3, IP2I Lyon, UMR 5822, F-69622, Villeurbanne, France\\[2.5cm]
\end{center}

\begin{abstract}
We discuss a class of Little String Theories (LSTs) whose low energy descriptions are supersymmetric gauge theories on the $\Omega$-background with gauge group $U(N)$ and matter in the adjoint representation. We show that the instanton partition function of these theories can be written in terms of Kronecker-Eisenstein series, which in a particular limit of the deformation parameters of the $\Omega$-background organise themselves into Greens functions of free scalar fields on a torus. We provide a concrete identification between (differences of) such propagators and Nekrasov subfunctions. The latter are also characterised by counting specific holomorphic curves in a Calabi-Yau threefold $X_{N,1}$ which engineers the LST. Furthermore, using the formulation of the partition function in terms of the Kronecker-Eisenstein series, we argue for new recursive structures which relate higher instanton contributions to products of lower ones.
\end{abstract}

\newpage

\tableofcontents

\vskip1cm

\section{Introduction}
Dualities in string theory are important tools to uncover and study symmetries and (algebraic) structures in gauge theories in various dimensions, in particular in the non-perturbative regime. Even dualities that are not of a strong-weak type oftentimes provide new viewpoints and alternative descriptions that are more adapted to tackle certain problems. A particularly rich class of examples of this type are so-called \emph{Little String Theories} (LSTs) \cite{Witten:1995zh,Aspinwall:1997ye,Seiberg:1997zk,Intriligator:1997dh,Hanany:1997gh,Brunner:1997gf} (reviews can be found in \cite{Aharony:1999ks,Kutasov:2001uf}). These are quantum theories which at low energies resemble quantum field theories with point-like degrees of freedom, however, whose UV-completion requires string-like degrees of freedom. In the context of string theory, such theories can be obtained through a decoupling limit that removes the gravitational sector, but retains the string scale. 

Concretely, an interesting class of such LSTs \cite{Hohenegger:2015btj,Hohenegger:2016eqy,Bastian:2017ing,Bastian:2018dfu} can be constructed in M-theory through $N$ parallel M5-branes compactified on a circle $\mathbb{S}^1_\tau$ (of radius $\tau$) and spread out on a circle $\mathbb{S}^1_\rho$ (of radius $\rho$), which probe a transverse $\text{ALE}_{A_{M-1}}$ space. Here $\text{ALE}_{A_{M-1}}$ has the structure of a $\mathbb{Z}_M$ orbifold and $M=1$ corresponds to the flat space $\mathbb{R}^4$. The low-energy world-volume theory on the M-branes is a quiver gauge theory on $\mathbb{R}^4\times \mathbb{T}^2$ with gauge group $U(N)^M$ and matter in the bifundamental representation (or the adjoint in the case of $M=1$). This M-theory setting is dual \cite{Haghighat:2013gba,Haghighat:2013tka,Hohenegger:2013ala} to a web of intersecting D5- and NS5-branes in type II string theory \cite{Aharony:1997bh} and can further be dualised to F-theory compactified on a class of non-compact toric Calabi-Yau threefolds called $X_{N,M}$ \cite{Haghighat:2013tka,Hohenegger:2013ala,Hohenegger:2016eqy}. These manifolds have the structure of a double elliptic fibration and more details on the mathematical construction can be found in \cite{Kanazawa:2016tnt}. These various dual descriptions provide numerous technical tools to perform explicit computations for the low energy gauge theory and in particular allow to compute explicitly the full non-perturbative partition function $\mathcal{Z}_{N,M}$ of the theory in various different fashions \cite{Hohenegger:2013ala} (see also \cite{Bastian:2018fba} for other limits that engineer five-dimensional gauge theories from $X_{N,M}$). For example, $\mathcal{Z}_{N,M}$ is captured by the topological string partition function on $X_{N,M}$, which can be computed explicitly \cite{Haghighat:2013gba,Haghighat:2013tka,Hohenegger:2013ala,Bastian:2017ing} using the (refined\footnote{'Refined' in this context refers to the \emph{refined topological string} \cite{Gopakumar:1998ii,Gopakumar:1998jq,Hollowood:2003cv} and specifically the refined topological vertex depends on two deformation parameters (called $\epsilon_1,\epsilon_2\in\mathbb{R}$), which are required to render the topological string partition function well defined. From the perspective of the gauge theory, these parametrise the Nekrasov $\Omega$-background which was used in \cite{Nekrasov:2002qd,Moore:1997dj,Lossev:1997bz} to compute the instanton partition function. See also \cite{Antoniadis:2010iq,Antoniadis:2013bja,Antoniadis:2013mna,Antoniadis:2015spa,Samsonyan:2017xdi,Angelantonj:2017qeh,Angelantonj:2019qfw,Angelantonj:2022dsx} for a physical interpretation of the refinement parameters from a world-sheet perspective in string theory.}) \emph{topological vertex formalism} \cite{Aganagic:2003db,Iqbal:2007ii}. This has allowed to study the spectrum of the underlying theories and as a consequence numerous surprising symmetries and algebraic structures have been discovered, for instance:
\begin{itemize}
\item[\emph{(i)}] By studying flop transformations of certain curves of $X_{N,M}$ it was shown in \cite{Hohenegger:2016yuv} that the extended K\"ahler moduli space of this Calabi-Yau threefold contains regions, which can be identified with $X_{N',M'}$ for any integers $(N',M')$ such that $NM=N'M'$ and $\text{gcd}(N,M)=\text{gcd}(N',M')$. It was furthermore argued in \cite{Bastian:2017ing} (see also \cite{Haghighat:2018gqf}) that the topological string partition functions of these two geometries are identical. Physically, this implies that the two LSTs whose low energy descriptions are given by quiver gauge theories with gauge groups $U(N)^M$ and $U(N')^{M'}$ (with $(N,M)$ and $(N',M')$ related as previously stated) are dual to one another, thus leading to webs of dual gauge theories \cite{Bastian:2017ary,Bastian:2018dfu}. 
\item[\emph{(ii)}] It was shown in \cite{Ahmed:2017hfr} (see also \cite{Hohenegger:2020gio}) that upon compactification on a further circle, the LSTs described above realise monopole strings in five dimensions. It was furthermore argued that the BPS counting function of a particular sector of these monopole strings can be expressed as the partition function of a symmetric orbifold conformal field theory whose target space is the symmetric product of moduli spaces of monopoles with fixed charges. This interpretation requires specific relations among different contributions to the topological string partititon function $\mathcal{Z}_{N,M}$ (or its associated free energy).
\item[\emph{(iii)}] Exploiting symmetries and geometric transformations of $X_{N,1}$ (\emph{i.e.} for $M=1$) that leave the partition function $\mathcal{Z}_{N,1}$ invariant, it was argued in \cite{Bastian:2018jlf} that the LSTs engineered by this manifold possess a non-trivial dihedral symmetry. This symmetry acts as linear transformations on the K\"ahler parameters of $X_{N,1}$ but is intrinsically non-perturbative from the perspective of the gauge theory.
\item[\emph{(iv)}] In \cite{Bastian:2019hpx,Bastian:2019wpx} (Fourier) expansions of subsectors of the (single particle) non-perturbative free energy of the theory for $M=1$ were proposed, which are compatible with the non-perturbative symmetries found in  \cite{Bastian:2018jlf}. Furthermore, it was shown in \cite{Hohenegger:2019tii,Hohenegger:2020slq} that this free energy can be decomposed in a way which resembles a Feynman-diagrammatic expansion: in the so-called unrefined limit (\emph{i.e.} where the two deformation parameters mentioned above satisfy $\epsilon_1=-\epsilon_2=\epsilon$) the leading instanton contribution (from the perspective of the underlying $U(N)$ theory), was written as products of propagators of a chiral free scalar field on the torus. The arguments of these two-point functions were differences of the gauge parameters of the $U(N)$ group. Furthermore, each propagator was multiplied by 'external states' consisting of contributions to the non-pertrubative free energy for $N=1$. This form was verified up to $N=5$ and a closed form for generic $N$ was conjectured. It was furthermore noticed that higher instanton contributions follow a similar pattern concerning the separation of gauge parameters and 'external states'. However, new 'decorations' of the propagator factors appeared, either in the form of derivatives or multiplicative factors, which were identified with dihedral \emph{graph functions} (MGF) with bivalent vertices \cite{DHoker:2015wxz,DHoker:2016mwo}. These were interpreted as contributions of integrated vertices in the Feynman diagrammatic decomposition. However, no (unique) building pattern for these contributions could be found in \cite{Hohenegger:2020slq}.
\end{itemize}
In this paper we go beyond the observation of \cite{Hohenegger:2019tii,Hohenegger:2020slq} explained in \emph{(iv)} and develop it into a concrete correspondance between specific contributions to the unrefined topological string partition function of $X_{N,1}$ and (differences of) two-point functions of chiral scalar fields on the torus. These contributions $\mathcal{T}_{\alpha_j\alpha_i}$ are labelled by two integer partitions $(\alpha_j,\alpha_i)$ and we can characterise them in two different fashions: 
\begin{itemize}
\item {\bf geometric characterisation:} We can cut the web diagram of $X_{N,1}$ open by placing pairs of branes on (certain) Lagrangian cycles of the manifold \cite{Aganagic:2000gs}. Following the discussion of \cite{Iqbal:2003ix} for a similar geometry, the $\mathcal{T}_{\alpha_j\alpha_i}$ then count holomorphic curves in the presence of these Lagrangian branes. In the current case, however, the $\mathcal{T}_{\alpha_j\alpha_i}$ count infinitely many curves due to the double elliptic fibration structure of $X_{N,1}$ (see \cite{Bastian:2017ing}).
\item {\bf gauge theory characterisation:} From the perspective of the low-energy $U(N)$ gauge theory, the $\mathcal{T}_{\alpha_j\alpha_i}$ calculate Nekrasov subfunctions of vector- and hypermultiplet contributions in the instanton partition function. Indeed, in the dual type II picture, these are BPS contributions of open strings stretched between individual D5-branes.  
\end{itemize}
Concretely, the unrefined $\mathcal{T}_{\alpha_j\alpha_i}$ can be written as sums of Kronecker-Eisenstein series (see Appendix~\ref{App:ModularGraphForms} for the definition) which depend on the gauge parameters of the low energy theory as well as a Coulomb branch parameter (called $S$). The former appear shifted by the remaining deformation parameter $\epsilon$, with integer coefficients. We find a form for these coefficients that is dictated by the instanton partition function of the Chern-Simons theory associated with the open topological string on $\mathbb{F}_0$ \cite{Iqbal:2003ix,Iqbal:2003zz}. The specific combinations of the Kronecker-Eisenstein series appearing in $\mathcal{T}_{\alpha_j\alpha_i}$ in turn can be formulated as propagators of chiral scalar fields on a torus. These results make the observation \emph{(iv)} much more concrete and we explain the precise relation to previous work in detail. Moreover, the current work pin-points the origin of the scalar field propagators observed in \cite{Hohenegger:2019tii,Hohenegger:2020slq} both geometrically as well as from the perspective of the low energy gauge theory. From the latter point of view in fact our results suggest a correspondence between elements of the partition function of a $U(N)$ gauge theory on $\mathbb{R}^4\times \mathbb{T}^2$ and combinations of simple Greens function of a free chiral scalar field on the torus, which is reminiscent of AGT-like correspondences found in \cite{Aganagic:2015cta} (and subsequent works). Furthermore, the Kronecker-Eisenstein series are building blocks of \emph{elliptic modular graph functions} and \emph{forms} (eMGF) \cite{DHoker:2020hlp,DHoker:2022dxx}, which generalise the concept of modular graph functions and forms \cite{DHoker:2015wxz,DHoker:2016mwo}. Our current work therefore further elaborates on the conjectured relation \cite{Hohenegger:2019tii,Hohenegger:2020slq} between the instanton sector of supersymmetric gauge theories and MGF.

Moving away from the unrefined limit (\emph{i.e.} keeping $\epsilon_{1,2}$ generic) we can still re-write the non-perturbative partition function in terms of Kronecker-Eisenstein series (which, however, in general do not arrange themselves into propagators of chiral scalars). Using this re-writing, we uncover new structures of the partition function. We focus on the leading singularity in $\epsilon_2$ of the level $r\in\mathbb{N}$ instanton contribution to the  instanton partition function of the gauge theory\footnote{We shall call this contribution the Nekrasov-Shatashvili (NS)-limit \cite{Nekrasov:2009rc,Mironov:2009uv}.} for $N=1$. We show that this contribution is up to a factor $r!$ the $r$th power of the leading instanton contribution. We provide evidence that this recursive structure also generalises to the NS-limit for all $N>1$. This structure therefore allows to recursively calculate the leading $\epsilon_2$ singularities of the non-perturbative partition function. 

This paper is organised as follows: In Section~\ref{Sect:BraneWebs}, we give a brief review of the topological string partition function on the Calabi-Yau threefold $X_{N,1}$, discussing in particular in detail the case $N=1$. In Section~\ref{Sect:KroneckerEisensteinSeries}, we present the decomposition of the topological string partition function of $X_{N,1}$ in terms of Kronecker-Eisenstein series and present its interpretation in the unrefined limit. In Section~\ref{Sect:RecursiveStructureNS}, we study the NS-limit of the partition function and provide evidence for a recursive structure among different instanton levels. In Section~\ref{Sect:DecmopositionFreeEnergy}, we develop further a diagrammatic decomposition of the partition function and compare to previous results in the literature. We focus in particular on the cases $N=2$ and $N=3$. Finally, Section~\ref{Sect:Conclusion} contains our conclusions and an outlook for future directions. Mathematical definitions and details on some computations have been relegated to three appendices.
\section{Review: Topological String Partition Functions}\label{Sect:BraneWebs}
In this Section we briefly review the topological string partition function on a class of non-compact toric Calabi-Yau threefolds, which capture the non-perturbative partition function of certain Little String Theories.
\subsection{Topological String Partition Function for $X_{N,1}$}
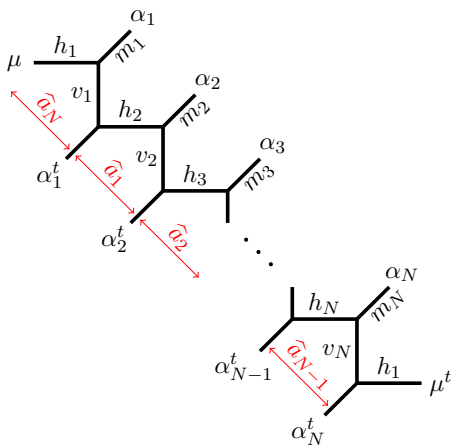
\begin{wrapfigure}{l}{0.4\textwidth}
${}$\\[-1cm]
\begin{center}
\scalebox{0.85}{\parbox{7.2cm}{\begin{tikzpicture}
\draw[ultra thick] (-1,0) -- (0,0) -- (0,-1) -- (1,-1) -- (1,-2) -- (2,-2) -- (2,-2.5);
\node[rotate=-45] at (2.5,-3) {\Large $\cdots$};
\draw[ultra thick] (3,-3.5) -- (3,-4) -- (4,-4) -- (4,-5) -- (5,-5);
\node at (-1.3,0) {\small $\mu$};
\node at (5.3,-5) {\small $\mu^t$};
\draw[ultra thick] (0,0) -- (0.5,0.5); 
\draw[ultra thick] (1,-1) -- (1.5,-0.5); 
\draw[ultra thick] (2,-2) -- (2.5,-1.5); 
\draw[ultra thick] (4,-4) -- (4.5,-3.5); 
\draw[ultra thick] (0,-1) -- (-0.5,-1.5); 
\draw[ultra thick] (1,-2) -- (0.5,-2.5); 
\draw[ultra thick] (3,-4) -- (2.5,-4.5); 
\draw[ultra thick] (4,-5) -- (3.5,-5.5); 
\node at (0.7,0.7) {\small $\alpha_1$};
\node at (1.7,-0.3) {\small $\alpha_2$};
\node at (2.7,-1.3) {\small $\alpha_3$};
\node at (4.7,-3.3) {\small $\alpha_N$};
\node at (-0.75,-1.75) {\small $\alpha_1^t$};
\node at (0.25,-2.75) {\small $\alpha_2^t$};
\node at (2.25,-4.75) {\small $\alpha_{N-1}^t$};
\node at (3.25,-5.75) {\small $\alpha_{N}^t$};
\node at (-0.5,0.25) {\small $h_1$};
\node at (0.5,-0.75) {\small $h_2$};
\node at (1.5,-1.75) {\small $h_3$};
\node at (3.5,-3.75) {\small $h_{N}$};
\node at (4.5,-4.75) {\small $h_{1}$};
\node at (-0.25,-0.5) {\small $v_1$};
\node at (0.75,-1.5) {\small $v_2$};
\node at (3.7,-4.5) {\small $v_N$};
\node[rotate=45] at (0.5,0.2) {\small $m_1$};
\node[rotate=45] at (1.5,-0.8) {\small $m_2$};
\node[rotate=45] at (2.5,-1.8) {\small $m_3$};
\node[rotate=45] at (4.5,-3.8) {\small $m_N$};
\draw[red,<->] (-1.35,-0.45) -- (-0.45,-1.35);
\node[red,rotate=-45] at (-0.725,-0.725) {\small $\widehat{a}_N$};
\draw[red,<->] (-0.35,-1.45) -- (0.55,-2.35);
\node[red,rotate=-45] at (0.275,-1.725) {\small $\widehat{a}_1$};
\draw[red,<->] (0.65,-2.45) -- (1.55,-3.35);
\node[red,rotate=-45] at (1.275,-2.725) {\small $\widehat{a}_2$};
\draw[red,<->] (2.65,-4.45) -- (3.55,-5.35);
\node[red,rotate=-45] at (3.275,-4.725) {\small $\widehat{a}_{N-1}$};
\end{tikzpicture}}}
\end{center}
\caption{\sl Web diagram of the Calabi-Yau manifold $X_{N,1}$ with a labelling of the various K\"ahler parameters. The diagram is glued to itself along the horizontal line (labelled with $\mu$) and the diagonal lines (labelled $\alpha_{1,\ldots,N}$).}
\label{Fig:Toric}
${}$\\[-0.5cm]
\end{wrapfigure}

We start by discussing the Calabi-Yau threefolds $X_{N,1}$ which are characterised by a web diagram as shown in Figure~\ref{Fig:Toric}. The diagram is double periodic since the curves labelled with $\mu$ and $\mu^t$ as well as $(\alpha_1,\ldots,\alpha_N)$ and $(\alpha_1^t,\ldots,\alpha_N^t)$ are respectively identified. In Figure~\ref{Fig:Toric} we have also indicated a labelling of the areas of the individual curves $(h_1,\ldots,h_N,v_1,\ldots,v_N,m_1,\ldots,m_N)$ (as the suitably normalised integrals over the K\"ahler form of the Calabi-Yau manifold). 

\noindent
${}$\\[-1cm]
However, due to the double periodicity of the diagram, not all of these areas are independent
\begin{align}
&v_1=v_2=\ldots=v_N\,,&&m_1=m_2=\ldots=m_N\,,
\end{align}
such that the diagram only has $N+2$ independent K\"ahler parameters. A convenient basis of the latter has been introduced in \cite{Bastian:2017ary} (see also \cite{Bastian:2017ing,Bastian:2018dfu}), labelled by $(\tau,S,\rho,\widehat{a}_1,\ldots,\widehat{a}_{N-1})$ with
\begin{align}
&\tau=v_1+m_1\,,\hspace{1cm}S=v_1\,,\hspace{1cm}\rho=\sum_{i=1}^N \widehat{a}_i\,,\nonumber\\
&\widehat{a}_i=h_{i+1}+v_i\,,\hspace{0.5cm}\forall i=1,\ldots,N\,.\label{BasisStrip}
\end{align}
Here we have used the identification $h_{N+1}=h_1$. In \cite{Bastian:2017ing} (building on earlier work in \cite{Haghighat:2013gba,Hohenegger:2013ala,Haghighat:2013tka}), a basic building block was computed which allows to explicitly write the refined topological string partition function of $X_{N,1}$\footnote{In fact, the building block computed in \cite{Bastian:2017ing} allows to compute the partition function \cite{Bastian:2018dfu} of more general Calabi-Yau manifolds (see \cite{Hohenegger:2016eqy}) that capture M-brane configurations probing transverse orbifolds.} as sums over the integer partitions $(\alpha_1,\ldots,\alpha_N)$ (which govern the gluing of the strip in Figure~\ref{Fig:Toric} along the diagonal lines), concretely 
\begin{align}
&\mathcal{Z}_{N,1}(\rho,S,\widehat{a}_{1,\ldots,N-1},\tau;\epsilon_{1,2})\nonumber\\
&=W_N(\emptyset)\,\widehat{Z}\sum_{\alpha_1,\ldots,\alpha_N}\left(\prod_{k=1}^N(-Q_{m_k})^{|\alpha_k|}\right)\,\prod_{i,j=1}^N\frac{\mathcal{J}_{\alpha_i \alpha_j}(\widehat{Q}_{i,i-j};q,t)\mathcal{J}_{\alpha_j \alpha_i}((\widehat{Q}_{i,i-j})^{-1} Q_\rho;q,t)}{\mathcal{J}_{\alpha_i\alpha_j}(\overline{Q}_{i,i-j}\sqrt{q/t};q,t)\mathcal{J}_{\alpha_j\alpha_i}(\dot{Q}_{i,j-i}\sqrt{t/q};q,t)}\,.\label{DefPartFunctionGeneral}
\end{align}
The notation and conventions (following \cite{Bastian:2017ing}) are further explained in Appendix~\ref{App:BuildingPartitionFunction}. As we shall discuss in more detail in Section~\ref{Sect:DecompositionKES}, using the particular choice of basis (\ref{BasisStrip}), the partition function can be expanded in powers of $\Qt=e^{2\pi i\tau}$
\begin{align}
\mathcal{Z}_{N,1}(\rho,S,\widehat{a}_{1,\ldots,N-1},\tau;\epsilon_{1,2})&=W_N(\emptyset)\sum_{\alpha_1,\ldots,\alpha_N}Q_\tau^{|\alpha_1|+\ldots+|\alpha_N|}\,\mathcal{P}_{\alpha_1,\ldots,\alpha_N}(\rho,S,\widehat{a}_{1,\ldots,N-1};\epsilon_{1,2})\nonumber\\
&=W_N(\emptyset)\,\sum_{r=0}^\infty Q_\tau^{r}\,K^{N,(r)}(\rho,S,\widehat{a}_{1,\ldots,N-1};\epsilon_{1,2})\,.\label{PartFctN1}
\end{align}
From the perspective of the $U(N)$ gauge theory (with hypermultiplet matter in the adjoint representation) that is engineered by $X_{N,1}$, $\Qt$ is identified with the instanton parameter and the sum over $r$ in (\ref{PartFctN1}) is the instanton expansion of the partition function. The parameters $\epsilon_{1,2}$ are deformations which are required to render the topological string partition well defined. They enter through the \emph{refined} topological vertex \cite{Aganagic:2003db,Iqbal:2007ii}, which was used in \cite{Haghighat:2013gba,Hohenegger:2013ala,Haghighat:2013tka,Bastian:2017ing} to compute $\mathcal{Z}_{N,1}$. From the perspective of the gauge theory, they are identified with the parameters of the Nekrasov $\Omega$-background \cite{Nekrasov:2002qd,Moore:1997dj,Lossev:1997bz}, which allows to compute the instanton contributions \cite{Hohenegger:2013ala}. While $\mathcal{Z}_{N,1}$ in (\ref{DefPartFunctionGeneral}) is computed for generic values of $\epsilon_{1,2}$ (with $q=e^{2\pi i\epsilon_1}$ and $t=e^{-2\pi i\epsilon_2}$ and $\epsilon_{1,2}\in\mathbb{R}$), there are two limits which play an important role in this paper
\begin{itemize}
\item \emph{unrefined limit:} In this limit the two deformation parameters are identified as
\begin{align}
\epsilon_1=-\epsilon_2=\epsilon\,,\label{UnrefinedLimit}
\end{align} 
which entails $q=t$. 
\item \emph{Nekrasov-Shatashvili (NS) \cite{Nekrasov:2009rc,Mironov:2009uv} limit} $\epsilon_2\to 0$: As we shall discuss, in this case the various instanton contributions $K^{N,(r)}$ to the partition function $\mathcal{Z}_{N,1}$ develop poles. In the following, we shall therefore take the NS-limit to mean the leading singularity in a (formal) Laurent series expansion around $\epsilon_2=0$. Concretely, starting from
\begin{align}
K^{N,(r)}(\rho,S,\widehat{a}_{1,\ldots,N-1};\epsilon_{1,2})=\sum_{k=0}^\infty\,\epsilon_2^{2k-r}\,K^{N,(r)}_{k}(\rho,S,\widehat{a}_{1,\ldots,N-1};\epsilon_1)\,,\label{E2expansion}
\end{align}
and following a similar definition in \cite{Hohenegger:2015cba,Hohenegger:2015btj,Hohenegger:2016eqy} for the free energy, we shall call $K^{N,(r)}_{0}$ the Nekrasov-Shatashvili (NS) limit of the partition function. For later convenience, we remark that the $K_k^{N,(r)}$ can further be expanded in a (formal) Laurent series around $\epsilon_1=0$ 
\begin{align}
K^{N,(r)}_{k}(\rho,S,\widehat{a}_{1,\ldots,N-1};\epsilon_1)=\sum_{p=0}^\infty\epsilon_1^{2p-r+k} K_{p,k}^{N,(r)}(\rho,S,\widehat{a}_{1,\ldots,N-1})\,.\label{ZnFormEpsi1Laurent}
\end{align}
\end{itemize}
\subsection{Instanton Partition Function for $N=1$}
As an explicit example and as an important building block for the later computations, we discuss in some more detail the topological string partition function for $X_{1,1}$, \emph{i.e.} the case $N=1$. The web diagram along with the (independent) K\"ahler parameters is shown in
\begin{wrapfigure}{l}{0.2\textwidth}
${}$\\[-1cm]
\begin{center}
\scalebox{0.85}{\parbox{3cm}{\begin{tikzpicture}
\draw[ultra thick] (0,0) -- (1,0) -- (1,-1) -- (2,-1) ;
\draw[ultra thick] (1,0) -- (1.5,0.5);
\draw[ultra thick] (1,-1) -- (0.5,-1.5);
\node at (-0.2,0) {{\small \bf $\mu$}};
\node at (2.25,-1) {{\small \bf $\mu^t$}};
\node at (1.65,0.65) {{\small \bf $\alpha_1$}};
\node at (0.2,-1.7) {{\small \bf $\alpha_1^t$}};
\node at (0.5,0.25) {{\small $h$}};
\node at (1.5,-0.75) {{\small $h$}};
\node at (0.8,-0.5) {{\small $v$}};
\node[rotate=45] at (1.5,0.2) {\small $m$};
\end{tikzpicture}}}
\end{center}
\caption{\sl Web diagram of $X_{1,1}$.}
\label{Fig:ToricN1}
${}$\\[-1cm]
\end{wrapfigure} 

\noindent
${}$\\[-1cm]
 Figure~\ref{Fig:ToricN1}. The basis (\ref{BasisStrip}) is obtained by
\begin{align}
&\rho=h+v\,,&&S=v\,,&&\tau=m+v\,.
\end{align}
The partition function (\ref{DefPartFunctionGeneral}) can be written in the form
\begin{align}
\mathcal{Z}_{1,1}(\rho,S,\tau;\epsilon_{1,2})=&\sum_{\alpha_1}\left(-Q_\tau/Q_S\right)^{|\alpha_1|}\,W_1(\emptyset)\cdot\hat{Z}\nonumber\\
&\times \frac{\mathcal{J}_{\alpha_1\alpha_1}(Q_\rho/Q_S;q,t)\,\mathcal{J}_{\alpha_1\alpha_1}(Q_S;q,t)}{\mathcal{J}_{\alpha_1\alpha_1}(Q_\rho\,\sqrt{q/t};q,t)\mathcal{J}_{\alpha_1\alpha_1}(Q_\rho\,\sqrt{t/q};q,t)}\,.
\end{align}
which can be simplified using the identities (\ref{IdCombineJ}) in Appendix~\ref{App:BuildingPartitionFunction} to yield
\begin{align}
\mathcal{Z}_{1,1}(\rho,S,\tau;\epsilon_{1,2})=W_1(\emptyset)\,\sum_{\alpha_1}\,Q_\tau^{|\alpha_1|}\,\frac{\vartheta_{\alpha_1\alpha_1}(Q_S;\rho)}{\vartheta_{\alpha_1\alpha_1}(\sqrt{q/t},\rho)}=W_1(\emptyset)\sum_{\alpha_1}Q_\tau^{|\alpha_1|}\,K^{1,(|\alpha_1|)}(\rho,S;\epsilon_{1,2})\,.\label{PartFct11}
\end{align}
Comparing with (\ref{PartFctN1}), we can extract the contribution for a fixed integer partition $\alpha_1$
\begin{align}
&\mathcal{P}_{\alpha_1}(\rho,S;\epsilon_{1,2})=\frac{\vartheta_{\alpha_1\alpha_1}(Q_S;\rho)}{\vartheta_{\alpha_1\alpha_1}(\sqrt{q/t};\rho)}=\nonumber\\
&\prod_{(i,j)\in\alpha_1}\!\!\!\frac{\theta_1\!\!\left(S+\epsilon_1(\alpha_{1,j}^t-i+\frac{1}{2})-\epsilon_2(\alpha_{1,i}-j+\frac{1}{2});\rho\right)\theta_1\!\!\left(S-\epsilon_1(\alpha_{1,j}^t-i+\frac{1}{2})+\epsilon_2(\alpha_{1,i}-j+\frac{1}{2});\rho\right)}{\theta_1\left(\epsilon_1(\alpha_{1,j}^t-i+1)-\epsilon_2(\alpha_{1,i}-j);\rho\right)\theta_1\left(-\epsilon_1(\alpha_{1,j}^t-i)+\epsilon_2(\alpha_{1,i}-j+1);\rho\right)}\,,\label{DefZalphaEps}
\end{align}
where we have used (\ref{DefCurlyTheta}). We remark that $\mathcal{P}_{\alpha_1}$ can be written as a product over contributions, that can be associated with individual boxes of the Young diagram of the partition $\alpha_1$.

This observation becomes even more apparent in the unrefined limit (\ref{UnrefinedLimit}): indeed for $q=t$ eq.~(\ref{DefZalphaEps}) can be simplified to give
\begin{align}
\mathcal{P}_{\alpha_1}(\rho,S;\epsilon_{1}=-\epsilon_2=\epsilon)=\prod_{(i,j)\in\alpha_1}\frac{\theta_1\left(S+\epsilon \,h_{\alpha_1}(i,j);\rho\right)\theta_1\left(-S+\epsilon\, h_{\alpha_1}(i,j);\rho\right)}{\theta_1\left(\epsilon\, h_{\alpha_1}(i,j)\right)^2}\,,\label{DefZalphaEpsUnrefined}
\end{align}
where we have used $\theta_1(x;\rho)=-\theta_1(x^{-1};\rho)$ and $h_{\alpha_1}(i,j)$ denotes the \emph{hook-length} of the box with the coordinates $(i,j)$ in the Young diagram of $\alpha_1$ (see \emph{e.g.} \cite{Iqbal:2010awa,})
\begin{align}
h_{\alpha_1}(i,j)=\alpha_{1,i}+\alpha^t_{1,j}-i-j+1\,.
\end{align}

\section{Instanton Partition Function and Kronecker-Eisenstein Series}\label{Sect:KroneckerEisensteinSeries}
In \cite{Hohenegger:2019tii,Hohenegger:2020slq} it has been argued that the expansions of the free energy associated with the partition function (\ref{DefPartFunctionGeneral}) can be organised in a way that resembles a (Feynman) diagrammatic expansion. This argument is based on studying the leading instanton contributions of the free energy which can be written using propagators of a free scalar field on the torus (along with additional contributions resembling loops). In this section we shall show a more direct relation between the instanton partition function and (elliptic) Modular Graph Functions.
\subsection{Decomposition of the Instanton Partition Function}\label{Sect:DecompositionKES}
In this Section we study in more detail the partition function (\ref{DefPartFunctionGeneral}) with the goal to re-write it using as building blocks Kronecker-Eisenstein series (which are reviewed in Appendix~\ref{App:ModularGraphForms}). We start by following \cite{Bastian:2017ing} and rewrite (\ref{DefPartFunctionGeneral}) in terms of the functions $\vartheta_{\alpha_1\alpha_2}$ defined in (\ref{DefCurlyTheta}) using the identities (\ref{IdCombineJ}) (see \cite{Hohenegger:2013ala})
\begin{align}
\mathcal{Z}_{N,1}(\rho,S,\widehat{a}_{1,\ldots,N-1},\tau;&\epsilon_{1,2})=W_N(\emptyset)\sum_{\alpha_1,\ldots,\alpha_N}\left(\prod_{r=1}^N(Q_\tau)^{|\alpha_r|}\frac{\vartheta_{\alpha_r\alpha_r}(Q_S;\rho)}{\vartheta_{\alpha_r\alpha_r}(\sqrt{q/t};\rho)}\right)\nonumber\\
&\times \prod_{1\leq i<j\leq N}\frac{\vartheta_{\alpha_j\alpha_i}\left(\Qs \prod_{k=i}^{j-1}\Qna{k};\rho\right)\vartheta_{\alpha_j\alpha_i}\left(\Qs^{-1} \prod_{k=i}^{j-1}\Qna{k};\rho\right)}{\vartheta_{\alpha_j\alpha_i}\left(\sqrt{\frac{q}{t}} \prod_{k=i}^{j-1}\Qna{k};\rho\right)\vartheta_{\alpha_j\alpha_i}\left(\sqrt{\frac{t}{q}} \prod_{k=i}^{j-1}\Qna{k};\rho\right)}\,.\label{PartN1}
\end{align}
From this, we can read off $\mathcal{P}_{\alpha_1,\ldots,\alpha_N}(\rho,S,\widehat{a}_{1,\ldots,N-1};\epsilon_{1,2})$ in (\ref{PartFctN1})
\begin{align}
\mathcal{P}_{\alpha_1,\ldots,\alpha_N}=\left(\prod_{k=1}^N\,\mathcal{P}_{\alpha_k}(\rho,S;\epsilon_{1,2})\right)\prod_{1\leq i<j\leq N}\mathcal{T}_{\alpha_j\alpha_i}(\rho,S,\widehat{a}_{1,\ldots,N-1};\epsilon_{1,2})\,,\label{DecompositionPBlocks}
\end{align}
where $\mathcal{P}_{\alpha_k}$ is defined in (\ref{DefZalphaEps}) and we introduced the shorthand notation
\begin{align}
&\mathcal{T}_{\alpha_j\alpha_i}(\rho,S,\widehat{a}_{1,\ldots,N-1};\epsilon_{1,2}):=\frac{\vartheta_{\alpha_j\alpha_i}\left(\Qs \prod_{k=i}^{j-1}\Qna{k};\rho\right)\vartheta_{\alpha_j\alpha_i}\left(\Qs^{-1} \prod_{k=i}^{j-1}\Qna{k};\rho\right)}{\vartheta_{\alpha_j\alpha_i}\left(\sqrt{\frac{q}{t}} \prod_{k=i}^{j-1}\Qna{k};\rho\right)\vartheta_{\alpha_j\alpha_i}\left(\sqrt{\frac{t}{q}} \prod_{k=i}^{j-1}\Qna{k};\rho\right)}\,,&&\forall 1\leq i<j\leq N\,.\label{DefTNotation}
\end{align}
We remark the symmetry
\begin{align}
\mathcal{T}_{\alpha_j\alpha_i}(\rho,S,\widehat{a}_{1,\ldots,N-1};\epsilon_{1,2})=\mathcal{T}_{\alpha^t_i\alpha^t_j}(\rho,S,\widehat{a}_{1,\ldots,N-1};\epsilon_{1,2})\,,\label{SymExchangePartitions}
\end{align}
which follows from the invariance of the partition function $\mathcal{Z}_{N,1}$ under a rotation of the web diagram in Figure~\ref{Fig:Toric} by $180$ degrees. Furthermore, for fixed integer partitions $(\alpha_j,\alpha_i)$ we can rewrite $\mathcal{T}_{\alpha_j\alpha_i}$ in terms of Jacobi theta-functions (see eq.~(\ref{DefJacobiTheta1}) for the definition) by using eq.(\ref{DefCurlyTheta})
{\allowdisplaybreaks
\begin{align}
\mathcal{T}_{\alpha_j\alpha_i}&=\prod_{(r,s)\in\alpha_j}\bigg(\frac{\theta_1\left(\mathfrak{a}_{ij}+S+\epsilon_1\left(\alpha_{i,s}^t-r+\frac{1}{2}\right)-\epsilon_2(\alpha_{j,r}-s+\frac{1}{2});\rho\right)}{\theta_1\left(\mathfrak{a}_{ij}+\epsilon_1\left(\alpha_{i,s}^t-r+1\right)-\epsilon_2(\alpha_{j,r}-s);\rho\right)}\nonumber\\
&\hspace{2.2cm}\times\frac{\theta_1\left(\mathfrak{a}_{ij}-S+\epsilon_1\left(\alpha_{i,s}^t-r+\frac{1}{2}\right)-\epsilon_2(\alpha_{j,r}-s+\frac{1}{2});\rho\right)}{\theta_1\left(\mathfrak{a}_{ij}+\epsilon_1\left(\alpha_{i,s}^t-r\right)-\epsilon_2(\alpha_{j,r}-s+1);\rho\right)}\bigg)\nonumber\\
&\times \prod_{(r,s)\in\alpha_i}\bigg(\frac{\theta_1\left(\mathfrak{a}_{ij}+S-\epsilon_1\left(\alpha_{j,s}^t-r+\frac{1}{2}\right)+\epsilon_2(\alpha_{i,r}-s+\frac{1}{2});\rho\right)}{\theta_1\left(\mathfrak{a}_{ij}-\epsilon_1\left(\alpha_{j,s}^t-r\right)+\epsilon_2(\alpha_{i,r}-s+1);\rho\right)}\nonumber\\
&\hspace{2.2cm}\times\frac{\theta_1\left(\mathfrak{a}_{ij}-S-\epsilon_1\left(\alpha_{j,s}^t-r+\frac{1}{2}\right)+\epsilon_2(\alpha_{i,r}-s+\frac{1}{2});\rho\right)}{\theta_1\left(\mathfrak{a}_{ij}-\epsilon_1\left(\alpha_{j,s}^t-r+\frac{1}{2}+1\right)+\epsilon_2(\alpha_{i,r}-s);\rho\right)}\bigg)\,,\label{FormTProdTheta1}
\end{align}}
\noindent with the shorthand notation $\mathfrak{a}_{ij}=\sum_{k=i}^{j-1}\widehat{a}_k$. The quotients of Jacobi-theta functions in turn can be expressed using the Kronecker-Eisenstein series $\Omega(u,v;\rho)$, which are defined in (\ref{DefOmega})
\begin{align}
&\mathcal{T}_{\alpha_j\alpha_i}=\left(-\frac{\fb(S+\epsilon_+;\rho)}{4\pi^2}\right)^{|\alpha_i|+|\alpha_j|}\nonumber\\
&\hspace{1.5cm}\times\prod_{(r,s)\in\alpha_j}\left[\Omega\left(\mathfrak{a}_{ij}+\mathfrak{e}_{r,s}^{\alpha_j,\alpha_i}-\epsilon_2,S+\epsilon_+;\rho\right)\, \Omega\left(\mathfrak{a}_{ij}+\mathfrak{e}_{r,s}^{\alpha_j,\alpha_i}+\epsilon_1,-S-\epsilon_+;\rho\right)\right]\nonumber\\
&\hspace{1.5cm}\times  \prod_{(r,s)\in\alpha_i}\left[\Omega\left(\mathfrak{a}_{ij}- \mathfrak{e}_{r,s}^{\alpha_i,\alpha_j}+\epsilon_2,S+\epsilon_+;\rho\right) \Omega\left(\mathfrak{a}_{ij}- \mathfrak{e}_{r,s}^{\alpha_i,\alpha_j}-\epsilon_1,-S-\epsilon_+;\rho\right)\right]\,,\label{TOmegaRewrite}
\end{align}
where we have used $\epsilon_{1,2}\in\mathbb{R}$. Furthermore, $\fb$ is a standard Jacobi form of weight $-2$ and index $1$ (see (\ref{DefStandardJacobi}) for the definition) and 
\begin{align}
&\epsilon_{\pm}=\frac{\epsilon_1\pm \epsilon_2}{2}\,,&&\mathfrak{e}_{r,s}^{\alpha_j,\alpha_i}=\epsilon_1\left(\alpha_{i,s}^t-r\right)-\epsilon_2(\alpha_{j,r}-s)\hspace{1cm}\forall(r,s)\in\alpha_j\,. \label{DefEsFacts}
\end{align}
Apart from the overall factor in the first line of (\ref{TOmegaRewrite}), the contribution $\mathcal{T}_{\alpha_j\alpha_i}$ is decomposed in terms of Kronecker-Eisenstein series, where a pair of $\Omega$'s is associated to each box of the Young tableaux of the two partitions. We shall discuss in the following Subsection the unrefined limit $\epsilon_{2}=-\epsilon_1$ (and $\epsilon_+=0$), in which case (\ref{TOmegaRewrite}) not only simplifies but can furthermore be written as products of scalar field propagators on the torus. We shall furthermore attempt in Section~\ref{Sect:InterpretGeometricGauge} an interpretation of this result both from a geometric and a gauge theory perspective. 

\subsection{Unrefined Limit}\label{Sect:UnrefinedLimit}
The form of $\mathcal{T}_{\alpha_j\alpha_i}$ in (\ref{TOmegaRewrite}) can be further simplified in the unrefined limit (\ref{UnrefinedLimit}) (\emph{i.e.} $\epsilon_+=0$)
\begin{align}
&\mathcal{T}_{\alpha_j\alpha_i}(\rho,S,\widehat{a}_{1,\ldots,N-1};\epsilon)=\left(-\frac{\fb(S;\rho)}{4\pi^2}\right)^{|\alpha_i|+|\alpha_j|}\nonumber\\
&\hspace{0.5cm}\times\prod_{\kappa=\pm1}\left[\left(\prod_{(r,s)\in\alpha_j}\Omega\left(\mathfrak{a}_{ij}+\epsilon\, n^{\alpha_j,\alpha_i}_{r,s},\kappa\, S;\rho\right)\right)\, \left(\prod_{(r,s)\in\alpha_i}\Omega\left(\mathfrak{a}_{ij}-\epsilon\, n^{\alpha_i,\alpha_j}_{r,s},\kappa\, S;\rho\right)\right)\right]\,,\label{DefTIntegerShift}
\end{align}
with
\begin{align}
&n^{\alpha_j,\alpha_i}_{r,s}=\alpha_{i,s}^t+\alpha_{j,r}-r-s+1\,,&&\forall (r,s)\in\alpha_j\,.
\end{align}
For concreteness, we have collected in the following table the integer numbers
\begin{align}
\underline{n}^{\alpha_j\alpha_i}=\{n_{r,s}^{\alpha_j,\alpha_i}|(r,s)\in\alpha_j\}\cup\{-n_{r,s}^{\alpha_i,\alpha_j}|(r,s)\in\alpha_i\}=-\underline{n}^{\alpha_i\alpha_j}\,,\label{DefFormIntegers}
\end{align}
\emph{i.e.} all $\epsilon$-shifts appearing in (\ref{DefTIntegerShift}) for fixed $(\alpha_i,\alpha_j)$ for the simplest integer partitions:
\begin{center}
\begin{tabular}{|c|c|c|}\hline
&&\\[-12pt]
$\alpha_i$ & $\alpha_j$ & $\underline{n}^{\alpha_j\alpha_i}$\\[2pt]\hline\hline
&&\\[-12pt]
\scalebox{0.65}{\parbox{0.65cm}{$\ydiagram{1}$}} & $\emptyset$ & $(0)$\\[4pt]\hline 
&&\\[-10pt]
\scalebox{0.65}{\parbox{1.3cm}{$\ydiagram{2}$}} & $\emptyset$ & $(0,-1)$\\[6pt]
\scalebox{0.65}{\parbox{0.65cm}{$\ydiagram{1,1}$}} & $\emptyset$ & $(1,0)$\\[12pt]
\scalebox{0.65}{\parbox{0.65cm}{$\ydiagram{1}$}} & \scalebox{0.65}{\parbox{0.65cm}{$\ydiagram{1}$}}  & $(1,-1)$\\[8pt]\hline
&&\\[-10pt]
\scalebox{0.65}{\parbox{1.95cm}{$\ydiagram{3}$}} & $\emptyset$ & $(0,-1,-2)$\\[8pt]
\scalebox{0.65}{\parbox{1.3cm}{$\ydiagram{2,1}$}} & $\emptyset$ & $(1,0,-1)$\\[14pt]
 \scalebox{0.65}{\parbox{1.3cm}{$\ydiagram{1,1,1}$}} &  $\emptyset$ & $(2,1,0)$\\[18pt]\hline
\end{tabular}
\hspace{-0.15cm}
\begin{tabular}{|c|c|c|}\hline
&&\\[-12pt]
$\alpha_i$ & $\alpha_j$ & $\underline{n}^{\alpha_j\alpha_i}$\\[2pt]\hline\hline
 &&\\[-10pt]
  \scalebox{0.65}{\parbox{1.3cm}{$\ydiagram{2}$}} &  \scalebox{0.65}{\parbox{0.65cm}{$\ydiagram{1}$}} & $(0,1,-2)$\\[8pt]
 \scalebox{0.65}{\parbox{0.65cm}{$\ydiagram{1,1}$}} &  \scalebox{0.65}{\parbox{0.65cm}{$\ydiagram{1}$}} & $(0,-1,2)$\\[12pt]\hline
 &&\\[-10pt]
 \scalebox{0.65}{\parbox{2.6cm}{$\ydiagram{4}$}} &  $\emptyset$ & $(0,-1,-2,-3)$\\[12pt]
 \scalebox{0.65}{\parbox{1.95cm}{$\ydiagram{3,1}$}} &  $\emptyset$ & $(1,0,-1,-2)$\\[16pt]
 \scalebox{0.65}{\parbox{1.3cm}{$\ydiagram{2,2}$}} &  $\emptyset$ & $(1,0,0,-1)$\\[16pt]
 \scalebox{0.65}{\parbox{1.3cm}{$\ydiagram{2,1,1}$}} &  $\emptyset$ & $(2,1,0,-1)$\\[20pt]
\hline
\end{tabular}
\hspace{-0.15cm}
\begin{tabular}{|c|c|c|}\hline
&&\\[-12pt]
$\alpha_i$ & $\alpha_j$ & $\underline{n}^{\alpha_j\alpha_i}$\\[2pt]\hline\hline
&&\\[-10pt]
 \scalebox{0.65}{\parbox{0.65cm}{$\ydiagram{1,1,1,1}$}} &  $\emptyset$ & $(3,2,1,0)$\\[22pt]
 \scalebox{0.65}{\parbox{1.95cm}{$\ydiagram{3}$}} &  \scalebox{0.65}{\parbox{0.65cm}{$\ydiagram{1}$}} & $(1,0,-1,-3)$\\[9pt]
 \scalebox{0.65}{\parbox{1.3cm}{$\ydiagram{2,1}$}} &  \scalebox{0.65}{\parbox{0.65cm}{$\ydiagram{1}$}} & $(2,0,0,-2)$\\[14pt]
\scalebox{0.65}{\parbox{0.65cm}{$\ydiagram{1,1,1}$}} & \scalebox{0.65}{\parbox{0.65cm}{$\ydiagram{1}$}} & $(3,1,0,-1)$\\[16pt]
\scalebox{0.65}{\parbox{1.3cm}{$\ydiagram{2}$}} & \scalebox{0.65}{\parbox{1.3cm}{$\ydiagram{2}$}} & $(2,1,-1,-2)$\\[6pt]
\scalebox{0.65}{\parbox{0.65cm}{$\ydiagram{1,1}$}} & \scalebox{0.65}{\parbox{1.3cm}{$\ydiagram{2}$}} & $(3,0,0,-1)$\\[10pt]\hline
\end{tabular}
\end{center}
The coefficients $\underline{n}^{\alpha_j\alpha_i}$ also exhibit the symmetry $\underline{n}^{\alpha_j\alpha_i}=\underline{n}^{\alpha_i^t\alpha_j^t}$ (and thus with (\ref{DefFormIntegers}) $\underline{n}^{\alpha_j\alpha_i}=-\underline{n}^{\alpha_j^t\alpha_i^t}$), which is a consequence of the symmetry (\ref{SymExchangePartitions}) of the contributions $\mathcal{T}_{\alpha_j\alpha_i}$ to the partition function. Furthermore, we have checked up to order $|\alpha_i|+|\alpha_j|=6$ that the set of integers $\underline{n}^{\alpha_j\alpha_i}$ can also be characterised in the following fashion: let $x\in\mathbb{R}$ and
\begin{align}
f_{\alpha^t_i,\alpha_j}(x)=\sum_{(r,s)\in\alpha_j}x^{n_{r,s}^{\alpha_j,\alpha_i}}+\sum_{(r,s)\in\alpha_i}x^{-n_{r,s}^{\alpha_i,\alpha_j}}=\sum_{a\in \underline{n}^{\alpha_j\alpha_i}}x^{a}\,,\label{FgenDefCS}
\end{align} 
then this generating function can be written as
\begin{align}
&f_{\alpha^t_i,\alpha_j}(x)=(x-2+x^{-1})\,f_{\alpha^t_i}(x)\,f_{\alpha_j}(x)+f_{\alpha^t_i}(x)+f_{\alpha_j}(x)\,,&&\text{with} &&f_\mu(x)=\sum_{r=1}^{\ell(\mu)}\sum_{k=1}^{\mu_j}x^{k-r}\,,
\end{align}
where $\ell(\mu)$ denotes the length of the partition $\mu=(\mu_1,\ldots,\mu_{\ell(\mu)})$. From the perspective of the topological vertex, the generating function arises in the product of two Schur polynomials \cite{Iqbal:2004ne}
\begin{align}
&f_{\mu,\nu}(x)=s_{\Box}(x^{\mu+\mathfrak{k}})\,s_{\Box}(x^{\nu+\mathfrak{k}})-\frac{x}{(1-x)^2}\,,
\end{align}
where we have used the shorthand notation $x^{\mu+\mathfrak{k}}=(x^{\mu_1-1/2},x^{\mu_2-3/2},x^{\mu_3-5/2},\ldots)$ for an integer partition $\mu$. It furthermore satisfies the following properties \cite{Iqbal:2003ix}
\begin{align}
&f_{\mu^t,\nu^t}(x)=f_{\mu,\nu}(x^{-1})\,,&&f_{\mu,\nu}(x)=f_{\nu,\mu}(x)\,,
\end{align}
which are indeed compatible with the symmetries we have found for the coefficients $\underline{n}^{\alpha_j\alpha_i}$. From a geometric perspective \cite{Iqbal:2003ix,Iqbal:2003zz}, the generating function (\ref{FgenDefCS}) appears in the partition function of the Chern-Simons theory that is associated with the open topological string theory on $\mathbb{F}_0$ (see~\cite{Aganagic:2002qg}), whose web diagram (along with a labelling of the geometric parameters associated with the fiber and base) is also shown in the following:
\begin{align}
&Z_{\text{CS}}(Q_B,Q_F;q)=\sum_{\alpha_1,\alpha_2}Q_B^{|\alpha_1|+|\alpha_2|} R_{\alpha_1,\alpha_2}(Q_F)^2\,,&&\scalebox{0.8}{\parbox{2.7cm}{\begin{tikzpicture}[scale = 1.25]
\draw[ultra thick] (0,0) -- (1,0) -- (1,1) -- (0,1) -- (0,0);
\draw[ultra thick] (0,0) -- (-0.5,-0.5);
\draw[ultra thick] (1,0) -- (1.5,-0.5);
\draw[ultra thick] (1,1) -- (1.5,1.5);
\draw[ultra thick] (0,1) -- (-0.5,1.5);
\node at (0.5,-0.3) {$B$};
\node at (0.5,1.3) {$B$};
\node at (-0.3,0.5) {$F$};
\node at (1.3,0.5) {$F$};
\end{tikzpicture}}
}
\end{align}
Here the sum is over integer partitions $\alpha_{1,2}$, while $Q_B=e^{2\pi i T_B}$ and $Q_F=e^{2\pi iT_F}$ and the general building block associated with a curve of type $(-2,0)$ (see \cite{Iqbal:2003ix} and Section~\ref{Sect:HoloCurves} for more details) is given by
\begin{align}
\parbox{7cm}{\begin{align}R_{\alpha_1,\alpha_2}(Q)&=\sum_{\alpha} Q^{|\alpha|}\,\mathcal{W}_{\alpha_1\alpha}(q)\,\mathcal{W}_{\alpha\alpha_2}(q)\nonumber\\
&=R_{\emptyset\emptyset}(Q)\,\mathcal{W}_{\alpha_1}(q)\,\mathcal{W}_{\alpha_2}(q)\,\text{exp}\left(\sum_{n=1}^\infty \frac{f_{\alpha_1\alpha_2}(q^n)}{n}\,Q^n\right)\,.\nonumber
\end{align}}\hspace{1.5cm}\scalebox{0.8}{\parbox{2cm}{\begin{tikzpicture}[scale = 1.25]
\draw[ultra thick] (-0.5,1.5) -- (0,1) -- (0,0) -- (-0.5,-0.5);
\node at (-0.3,0.5) {$Q$};
\draw[ultra thick] (0,0) -- (0.5,0);
\node at (0.8,0) {$\alpha_2$};
\draw[ultra thick] (0,1) -- (0.5,1);
\node at (0.8,1) {$\alpha_2$};
\end{tikzpicture}}
}
\nonumber
\end{align}
Here $R_{\emptyset\emptyset}$ is the closed string partition function on $T^*(\mathbb{P}^1)\times \mathbb{C}$ and we refer the  reader to \cite{Iqbal:2003ix} for the definition of $\mathcal{W}_{\alpha_i}$ and further details.

Besides the simplification of the contributions $\mathcal{T}_{\alpha_j\alpha_i}$ in (\ref{DefTIntegerShift}), the unrefined limit (\ref{TOmegaRewrite}) affords another important rewriting. Indeed, using the relation (\ref{IdOmegaWeierstrass}), we can express pairs of Kronecker-Eisenstein series in (\ref{DefTIntegerShift}) as differences of Weierstrass elliptic functions (defined in (\ref{DefWeierstrassEllipticP})), which with (\ref{KESpropagators}) can further be express in terms of differences of Greens functions of free scalar fields on a torus with modular parameter $\rho$
\begin{align}
&\mathcal{T}_{\alpha_j\alpha_i}(\rho,S,\widehat{a}_{1,\ldots,N-1};\epsilon)=\left(-\frac{\fb(S;\rho)}{4\pi^2}\right)^{|\alpha_i|+|\alpha_j|}\nonumber\\
&\hspace{0.3cm}\times \left(\prod_{(r,s)\in\alpha_j}\left(\mathbb{G}''(\mathfrak{a}_{ij}+\epsilon\,n_{r,s}^{\alpha_j,\alpha_i};\rho)-\mathbb{G}''(S;\rho)\right)\right)\left(\prod_{(r,s)\in\alpha_i}\left(\mathbb{G}''(\mathfrak{a}_{ij}-\epsilon\,n_{r,s}^{\alpha_i,\alpha_j};\rho)-\mathbb{G}''(S;\rho)\right)\right)\,.\label{ContributionPartitionFunctionPropagators}
\end{align}
In this form $\mathcal{T}_{\alpha_j\alpha_i}$ associates a pair of propagator factors to each box of the Young tableaux of the two partitions $(\alpha_i,\alpha_j)$: one of these depends on $S$ and the other on $\mathfrak{a}_{ij}$ (which are gauge parameter from the perspective of the gauge theory engineered by $X_{N,1}$) shifted by $\epsilon$ in a way that is dictated by the integers $\underline{n}^{\alpha_j,\alpha_i}$. Before discussing further consequences of this result as well as the the relation to previous work (notably \cite{Hohenegger:2019tii,Hohenegger:2020slq}) in subsequent Sections, we shall comment on (\ref{ContributionPartitionFunctionPropagators}) from the perspective of the underlying Calabi-Yau geometry $X_{N,1}$ and the associated supersymmetric gauge theory in the following Subsection.
\subsection{Geometric- and Gauge Theory Interpretation}\label{Sect:InterpretGeometricGauge}
Eq.~(\ref{ContributionPartitionFunctionPropagators}) allows to express a contribution to the topological string partition function that is labelled by the integer partitions $(\alpha_j,\alpha_i)$ in terms of (differences of) two-point functions of a free scalar field on the torus. 
\subsubsection{Holomorphic Curves and Lagrangian Branes}\label{Sect:HoloCurves}
The local geometry associated with a web diagram of the form depicted in Figure~\ref{Fig:Toric}, but with de-
\begin{wrapfigure}{l}{0.58\textwidth}
${}$\\[-1cm]
\begin{center}
\scalebox{0.85}{\parbox{11cm}{\begin{tikzpicture}
\draw[ultra thick] (0,0) -- (6,0) -- (6,2) -- (0,2) -- (0,0);
\draw[ultra thick] (0,0) -- (2,2) -- (2,0) -- (4,2) -- (4,0) -- (6,2);
\node at (7,1) {\Large $\cdots$};
\draw[ultra thick] (8,0) -- (10,0) -- (10,2) -- (8,2) -- (8,0);
\draw[ultra thick] (8,0) -- (10,2);
\draw[ultra thick,green!50!black] (1.4,1.8) -- (1.4,2.5);
\node at (1.4,2.8) {$\alpha_1$};
\draw[ultra thick,green!50!black] (1.8,1.4) -- (1.8,-0.5);
\node at (1.8,-0.8) {$\alpha_1^t$};
\draw[ultra thick,green!50!black] (3,1.4) -- (3,2.5);
\node at (3,2.8) {$\alpha_2$};
\draw[ultra thick,green!50!black] (3.4,1) -- (3.4,-0.5);
\node at (3.4,-0.8) {$\alpha_2^t$};
\draw[ultra thick,green!50!black] (4.6,1) -- (4.6,2.5);
\node at (4.6,2.8) {$\alpha_3$};
\draw[ultra thick,green!50!black] (5,0.6) -- (5,-0.5);
\node at (5,-0.8) {$\alpha_3^t$};
\draw[ultra thick,green!50!black] (8.2,0.6) -- (8.2,2.5);
\node at (8.2,2.8) {$\alpha_N$};
\draw[ultra thick,green!50!black] (8.6,0.2) -- (8.6,-0.5);
\node at (8.6,-0.8) {$\alpha_N^t$};
\draw[ultra thick,red] (-0.5,1.8) -- (1.4,1.8) -- (1.8,1.4) -- (3,1.4) -- (3.4,1) -- (4.6,1) -- (5,0.6) -- (6.25,0.6) ;
\draw[ultra thick,red] (7.75,0.6) -- (8.2,0.6) -- (8.6,0.2) -- (10.5,0.2);
\end{tikzpicture}}}
\end{center}
${}$\\[-1cm]
\caption{\sl Web diagram associated with $X_{N,1}$. The external green lines are pairwise identified.}
\label{Fig:ToricStrip}
${}$\\[-0.5cm]
\end{wrapfigure} 

\noindent
${}$\\[-1cm]
compactified external legs has been discussed in \cite{Iqbal:2004ne}. Generalising this discussion to our case, the relevant web diagram is shown in Figure~\ref{Fig:ToricStrip}. As explained in \cite{Iqbal:2004ne}, coloured lines connecting any two vertices represent $\mathbb{P}^1$'s, while locally the two non-compact directions of $X_{N,1}$ correspond to the direct sum of two line-bundles over each such $\mathbb{P}^1$. Concretely, there are two possible cases, namely \emph{(i)} $\mathcal{O}(-1)\oplus \mathcal{O}(-1)\to \mathbb{P}^1$ and \emph{(ii)} $\mathcal{O}(-2)\oplus \mathcal{O}(0)\to \mathbb{P}^1$ and we shall therefore call the respective $\mathbb{P}^1$'s either of type $(-1,-1)$ or $(-2,0)$. In the way the diagram in Figure~\ref{Fig:ToricStrip} is structured, all individual $\mathbb{P}^1$'s shown in red are of type $(-1,-1)$. Furthermore, curves made from chains of $\mathbb{P}^1$'s also have normal bundles $\mathcal{O}(-1)\oplus\mathcal{O}(-1)$ or $\mathcal{O}(-2)\oplus\mathcal{O}(0)$ depending on whether they consist of an odd or an even number of $\mathbb{P}^1$'s of type $(-1,-1)$. Therefore, focusing on curves on the chain of $\mathbb{P}^1$ drawn in red in Figure~\ref{Fig:ToricStrip}, there are two types
\begin{itemize}
\item[\emph{(i)}] curves connecting any pair of green lines on the same side of the diagram have local geometry $\mathcal{O}(-2)\oplus\mathcal{O}(0)\to \mathbb{P}^1$ (and we shall call them type $(-2,0)$)
\item[\emph{(ii)}] curves connecting any pair of green lines on opposite sides of the diagram have local geometry $\mathcal{O}(-1)\oplus\mathcal{O}(-1)\to \mathbb{P}^1$ (and we shall call them type $(-1,-1)$)
\end{itemize}
As explained in \cite{Bastian:2017ing}, to compute the topological string partition function in (\ref{DefPartFunctionGeneral}), we count holomorphic curves after placing Lagrangian branes \cite{Aganagic:2000gs} on pairs of the external legs labelled by $(\alpha_i,\alpha_i^t)$ (with $i=1,\ldots,N$). The contribution to the open topological string amplitude from curves of type \emph{(i)} with branes on external legs labelled $(\alpha_i,\alpha_j)$ or $(\alpha_i^t,\alpha_j^t)$ (\emph{i.e.} on the same side of the diagram) or of type \emph{(ii)} with branes on external legs labelled $(\alpha_i,\alpha^t_j)$ (\emph{i.e.} on opposite sides of the diagram) have explicitly been given in \cite{Bastian:2017ing}. Due to the compactified nature of the strip geometry in Figure~\ref{Fig:ToricStrip}, this calculation takes into account that there are infinitely many holomorphic cuves between any two legs of the diagram, by going around the entire strip multiple times. For given $(\alpha_j,\alpha_i)$ (with $i< j$), the collective contributions of all such curves in fact gives the $\mathcal{T}_{\alpha_j\alpha_i}$ in (\ref{DefTNotation}). The re-writing of the latter in terms of scalar field propagators (\ref{ContributionPartitionFunctionPropagators}) therefore provides a new interpretation of this geometric contribution in the unrefined limit. 
\subsubsection{Nekrasov Subfunctions}
The contributions to $\mathcal{T}_{\alpha_j\alpha_i}$ described geometrically in the previous Subsubsection can also be interpreted from a gauge theory perspective. Indeed, as was explained in \cite{Haghighat:2013gba,Hohenegger:2013ala,Haghighat:2013tka,Bastian:2018dfu}, the Calabi-Yau manifold $X_{N,1}$ engineers (among others \cite{Bastian:2017ary,Bastian:2018dfu}) an $U(N)$ gauge theory with hypermultiplet matter in the adjoint representation. This theory can also be described by $N$ parallel D5-branes separated on a circle $\mathbb{S}^1_\rho$ (with radius $\rho$) intersecting an NS5-brane transversally (see \cite{Haghighat:2013gba,Hohenegger:2013ala} for more details on the geometry). In this picture, the $\mathcal{T}_{\alpha_j\alpha_i}$ encode the BPS contributions of strings ending on two distinct D5-branes and winding around the circle $\mathbb{S}^1_\rho$. Indeed, $\mathcal{T}_{\alpha_j\alpha_i}$ is precisely made up from Nekrasov subfunctions of vector- and hypermultiplet contributions to the instanton gauge theory partition function~\cite{Hohenegger:2013ala,Bastian:2018dfu}.\footnote{Nekrasov subfunctions (also called subfactors) encode the contributions of different multiplets to the non-perturbative partition function of gauge theories. They have been studied in four-dimensions \cite{Nekrasov:2002qd}, five-dimensions \cite{Nekrasov:2003rj} and six-dimensions \cite{Iqbal:2015fvd,Nieri:2015dts,Hayling:2017cva} and reviews can for example be found in \cite{Mironov:2009qt,Gaiotto:2009we,Bastian:2018dfu}.} Notice in this regard the following limits of $\mathcal{T}_{\alpha_j\alpha_i}$ written in (\ref{ContributionPartitionFunctionPropagators}) (\emph{i.e.} in the unrefined limit)
\begin{itemize}
\item $\mathcal{T}_{\alpha_j\alpha_i}$ has a second order pole for $\mathfrak{a}_{ij}\to -\epsilon n$ for any $n\in\underline{n}^{\alpha_j,\alpha_i}$ \emph{i.e.} the limit where the gauge parameter associated with the vector multiplet becomes a particular integer multiple of $\epsilon$.\footnote{It has been observed in \cite{Bastian:2017jje} that the free energy associated with $\mathcal{Z}_{N,1}$ simplifies and reveals additional algebraic structures if the gauge parameters are identified with integer multiples of $\epsilon$.} Indeed, in the context of the Nekrasov subfunctions, the parameters $\mathfrak{a}_{ij}$ correspond to the vacuum expectation values of the scalar fields in the five-dimensional vector multiplets. For $\epsilon=0$, this condition becomes $\mathfrak{a}_{ij}\to 0$ and geometrically corresponds to the vanishing of a curve of type $(-2,0)$ in the web diagram.
\item $\mathcal{T}_{\alpha_j\alpha_i}$ is zero for $\mathfrak{a}_{ij}+\epsilon n=S$, for any $n\in\underline{n}^{\alpha_j,\alpha_i}$. For $\epsilon=0$, this condition geometrically corresponds to the vanishing of a curve of type $(-1,-1)$ in the web diagram, connecting  the legs $\alpha_i^t$ with $\alpha_{i+1}$ in Figure~\ref{Fig:ToricStrip}.
\item $\mathcal{T}_{\alpha_j\alpha_i}$ takes a finite value for $S\to 0$, corresponding to the vanishing of the mass of the adjoint multiplet associated with strings stretching between branes labelled $\alpha_i$ and $\alpha_j$. For $\epsilon=0$, this condition geometrically also corresponds to the vanishing of a curve of type $(-1,-1)$ in the web diagram, connecting  the legs $\alpha_i$ with $\alpha_i^t$ in Figure~\ref{Fig:ToricStrip}.
\end{itemize}
The re-writing of the contribution $\mathcal{T}_{\alpha_j\alpha_i}$ in eq.~(\ref{ContributionPartitionFunctionPropagators}), therefore suggests a relation between the Nekrasov subfunctions in a gauge theory on $\mathbb{R}^4\times \mathbb{T}^2$ and simple combinations of propagators in a two-dimensional theory of a free scalar field on a torus. This relation is very reminiscent of a relation between the $\mathcal{N}=(2,0)$ LST compactified on a Riemann surface and the 3-point conformal blocks of a $q$-deformed ADE-Toda (or Liouville) CFT, which was found in \cite{Aganagic:2015cta}: in \cite{Aganagic:2013tta} a triality relation between $q$-deformed Liouville theory (introduced in \cite{Shiraishi:1995rp,Awata:1996xt}) on a sphere, three-dimensional $\mathcal{N}=2$ gauge theories with $U(N)$ gauge group and $M$ flavours on the $\Omega$-background, and a five-dimensional $\mathcal{N}=1$ $U(M)$ gauge theory with $2M$ fundamental hyermultiplets was established. Notably the five-dimensional instanton sum is captured by the conformal blocks of the two-dimensional CFT. As anticipated by the AGT-correspondence \cite{Alday:2009aq}\footnote{The Alday-Gaiotto-Tachikawa correspondence \cite{Alday:2009aq,Wyllard:2009hg} has been originally established for $\mathcal{N}=2$ supersymmetric gauge theories in four dimensions and extended to five-dimensional gauge theories in \cite{Awata:2009ur,Schiappa:2009cc,Awata:2010yy} and to five-dimensional quiver gauge theories in \cite{Kimura:2015rgi,Kimura:2017hez} (see also the excellent review \cite{Kimura:2020jxl}).}, this relation was extended in \cite{Aganagic:2014oia} to a triality involving conformal blocks of $A_n$ Toda-CFTs on a sphere and families of four-dimensional $\mathcal{N}=2$ CFTs as well as two-dimensional theories living on the vortex configurations of the latter. In \cite{Aganagic:2015cta} (see also \cite{Haouzi:2017vec} for subsequent work for LSTs on more general defects) it was shown that the partition function of the $\mathcal{N}=(2,0)$ ADE LST on a sphere is captured by the conformal blocks of a $q$-deformed ADE Toda theory (see also \cite{Kimura:2016dys} for relations to $q$-conformal blocks on a torus). This result suggests that the combinations of free scalar-field propagators appearing in $\mathcal{T}_{\alpha_j\alpha_i}$ (\ref{ContributionPartitionFunctionPropagators}) have an interpretation in terms of the latter two-dimensional theory. We leave investigation of this relation for future work.

\section{Recursive Structure in the NS Limit}\label{Sect:RecursiveStructureNS}
We can use the form of the partition function~(\ref{PartN1}) described in the previous Section to find further symmetries and structures: in this Section, we argue for a recursive structure in the leading $\epsilon_2$ pole of the partition (which we call the NS-limit) which receives deformations at higher $\epsilon_2$-orders.
\subsection{Recursion for $N=1$}\label{Sect:RecursionN1}
\subsubsection{NS Limit}
We start by considering the partition function $\mathcal{Z}_{1,1}$, more concretely the order $r$-instanton contribution $K^{1,(r)}$ as defined in (\ref{PartFct11}) and which can be written in terms of $\mathcal{P}_{\alpha_1}$ in (\ref{DefZalphaEps})
\begin{align}
&K^{1,(r)}(\rho,S;\epsilon_{1,2})=\sum_{\alpha_1\text{ with }|\alpha_1|=r}\mathcal{P}_{\alpha_1}(\rho,S;\epsilon_{1,2})\,,&&\forall r\in\mathbb{N}\,.\label{DefSumKr}
\end{align}
Since $\theta_1(z;\rho)=2\pi \eta^3(\rho)\, z+\mathcal{O}(z^3)$, for a fixed integer partition $\alpha_1$, $\mathcal{P}_{\alpha_1}$ has poles in $\epsilon_2$, which stem from the boxes in the Young tableaux with either $\alpha_{1,j}^t=i-1$ or $\alpha_{1,j}^t=i$. There are no boxes for which the first condition is satisfied, but the second condition holds for all boxes, which have no boxes below them, \emph{e.g.}
\begin{align}
&\alpha_{1,j}^t=i:&&\text{\emph{e.g.}}\hspace{1cm}\scalebox{0.6}{\parbox{3.8cm}{\ydiagram{6, 4, 3,3,2} *[*(black!40!white)]{5+1,0,0,0,0} *[*(black!40!white)]{0,3+2,0,0,0} *[*(black!40!white)]{0,0,0,2+1,0} *[*(black!40!white)]{0,0,0,0,0+2} }}&&\text{with} &&(i,j)=\{(1,6),(2,4),(2,5),(4,3),(5,1),(5,2)\}\,.\nonumber
\end{align}
Consequently, the highest pole in $\epsilon_2$ in (\ref{DefSumKr}) (and (\ref{DefZalphaEps})) is of order $r=|\alpha_1|$ and stems from the partition, whose Young tableaux consists only of such boxes, \emph{i.e.}
\begin{align}
&\scalebox{0.6}{\parbox{3.8cm}{\ydiagram{6}  }}&&\text{with}&& \begin{array}{l}\alpha_{1,i}-j+1\in\{|\alpha_1|,|\alpha_1|-1\,\ldots,1\}\,, \\[2pt] \alpha_{1,j}^t-i+1=1\end{array}&&\text{ for }&&\begin{array}{l}i=1\,,\\[2pt]j\in\{1,\ldots,|\alpha_1|\}\,.\end{array}\label{LeadingYoungDiagram}
\end{align}
We can therefore write for $K_0^{1,(r)}$ (with $r=|\alpha_1|$)
\begin{align}
K_0^{1,(r)}(\rho,S;\epsilon_1)=\frac{1}{r!}\left(\frac{\theta_1(S+\epsilon_1/2;\rho)\theta_1(S-\epsilon_1/2;\rho)}{\theta_1(\epsilon_1;\rho)\theta'(0;\rho)}\right)^r\,,\label{RecursionRelN1}
\end{align}
which in fact allows to recursively express the NS-limit of the $r$-th instanton contribution as powers of the leading instanton level
\begin{align}
K_0^{1,(r)}(\rho,S;\epsilon_1)=\frac{1}{r!}\left(K_0^{1,(1)}(\rho,S;\epsilon_1)\right)^r\,.\label{NSFusionN1}
\end{align}
\subsubsection{Subleading Contributions}
The relation (\ref{NSFusionN1}) allows for $N=1$ to compute iteratively all instanton contributions starting from the first one in the NS-limit, \emph{i.e.} to leading order in $\epsilon_2$. For completeness, we consider whether a similar structure also exists to higher orders in $\epsilon_2$. In the following we shall provide evidence to this effect, by considering the next-to-leading order, \emph{i.e.} the contributions $K_{1}^{1,(r)}$, as defined in (\ref{E2expansion}) (which in turn can be expanded in a Laurent series around $\epsilon_1=0$, with coefficients $K_{p,k}^{1,(r)}(\rho,S)$ as defined in (\ref{ZnFormEpsi1Laurent})). These terms also receive contributions in (\ref{DefSumKr}) from partitions with Young tableaux other than (\ref{LeadingYoungDiagram}), featuring boxes with other boxes below them. Since keeping track of these various contributions becomes more and more tedious, in the following we shall provide evidence for further recursion relations by studying expansions of the $K_{p,k}^{1,(r)}(\rho,S)$ in terms of Jacobi forms for small values of $r$. Indeed, the $K_{p,k}^{1,(r)}(\rho,S)$ can again be expanded in polynomials of $(\fa,\fb)$ (see appendix~\ref{App:Modular} for the definitions) with coefficients given by homogeneous polynomials in the Eisenstein series $(E_2,E_4,E_6)$. By replacing $E_2$ by $\widehat{E}_2$, the $K_{p,k}^{1,(r)}(\rho,S)$ can be promoted to quasi-Jacobi forms of index $n$ and weight $2(p+k-r)$. Specifically, for $k=1$, we find the following expressions for $K^{1,(r)}_{p,1}$ to leading orders in $p$
\begin{center}
\begin{tabular}{|c||c|c|c|c|}\hline
&&&\\[-12pt]
$r$ & $p=0$ & $p=1$ & $p=2$ \\[2pt]\hline
&&&\\[-12pt]
$1$ & $\frac{\fa+2E_2 \fb}{48}$ & $\frac{(E_2^2-E_4)\fb}{1152}$ & $\frac{5(E_4-E_2^2)\fa+2(5E_2^2+3E_2 E_4-8E_6)\fb}{1105920}$\\[4pt]\hline
&&&\\[-12pt]
$2$ & $\frac{-\fb\left(2\fa+E_2 \fb\right)}{24}$ & $\frac{\fa^2-3E_2 \fa\fb+2(8E_4-E_2^2)\fb^2}{1152}$ & $\frac{35E_2\fa^2-8(5E_2^2+44E_4)\fa\fb-4(10E_2^2-147E_2 E_4-304E_6)\fb^2}{1105920}$\\[4pt]\hline
\end{tabular}
\end{center}
These expressions are compatible with the following product relations (which we have checked further up to $r=4$)
\begin{align}
K_{p,1}^{1,(r)}&=\frac{r_1+r_2}{r_1-r_2}\frac{r_1!r_2!}{(r_1+r_2)!}\,\sum_{q=0}^p\left(K_{q,0}^{1,(r_1)}K_{p-q,1}^{1,(r_2)}-K_{q,1}^{1,(r_1)}K_{p-q,0}^{1,(r_2)}\right)\,,\hspace{1cm}\forall r_{1,2} \text{ with }\begin{array}{l}r_1+r_2=r\,,\\r_1\neq r_2\,,\end{array}\nonumber\\
K_{p,1}^{1,(2r)}&=\frac{r^2}{3r-2}\sum_{q=0}^pK_{q,0}^{1,(2r)}\frac{(-1)^{q}(4^q-1)B_{2q}}{(2q-1)!!(2q)!!}\,E_{2q}+\frac{4(3r-1)(r!)^2}{(3r-2)(2r)!}\,\sum_{q=0}^p K_{q,0}^{1,(r)}K_{p-q,1}^{1,(r)}\,,\hspace{0.5cm}\forall r\geq 1\,,\nonumber
\end{align}
where $B_{2q}$ are the Bernoulli numbers. Resumming the Laurent series in $\epsilon_1$, these relations are equivalent to
\begin{align}
K_{1}^{1,(r)}&=\frac{r_1+r_2}{r_1-r_2}\frac{r_1!r_2!}{(r_1+r_2)!}\left(K_{0}^{1,(r_1)}K_{1}^{1,(r_2)}-K_{1}^{1,(r_1)}K_{0}^{1,(r_2)}\right)\,,\hspace{1.5cm}\forall \begin{array}{l}r_1+r_2=r\,,\\r_1\neq r_2\,,\end{array}\nonumber\\
K_{p,1}^{1,(2r)}&=\frac{r^2}{3r-2}K_{0}^{1,(2r)}\,\mathcal{E}(\rho,\epsilon_1)+\frac{4(3r-1)(r!)^2}{(3r-2)(2r)!}\,K_{0}^{1,(r)}K_{1}^{1,(r)}\,,\hspace{1.4cm}\forall r\geq 1\,,\label{KRelsNSSub}
\end{align} 
where we defined
\begin{align}
\mathcal{E}(\rho,\epsilon_1)=\sum_{q=0}^\infty\frac{(-1)^{q}(4^q-1)B_{2q}}{(2q-1)!!(2q)!!}\,E_{2q}(\rho)\,\epsilon_1^{2q-1}\,,
\end{align}
which depends on $\rho$, but is independent of $S$. To summarise, while the relations (\ref{KRelsNSSub}) still allow to compute instanton contribution from products of lower order ones, the corresponding coefficients depend on Eisenstein series as functions of $\rho$. The latter in fact, appear naturally upon expanding the $\theta_1$ in (\ref{DefZalphaEps}) in power series of $\epsilon_{2}$.
\subsection{Recursion for $N\geq 1$}
\subsubsection{NS Limit}
The analysis of the NS-limit for the case $N=1$ in Section~\ref{Sect:RecursionN1} can be generalised to $N\geq 1$ using (\ref{PartN1}). Indeed, with (\ref{PartFctN1}) we have
\begin{align}
K^{N,(r)}(\rho,S,\widehat{a}_{1,\ldots,N-1};\epsilon_{1,2})=\sum_{{\alpha_1,\ldots,\alpha_N}\atop{|\alpha_1|+\ldots+|\alpha_N|=r}}\mathcal{P}_{\alpha_1,\ldots,\alpha_N}(\rho,S,\widehat{a}_{1,\ldots,N-1};\epsilon_{1,2})\,.
\end{align}
Furthermore, from (\ref{FormTProdTheta1}) we find for generic values of $\widehat{a}_{1,\ldots,N-1}$
\begin{align}
&\lim_{\epsilon_2\to 0}\mathcal{T}_{\alpha_j\alpha_i}(\rho,S,\widehat{a}_{1,\ldots,N-1};\epsilon_{1,2})\longrightarrow \text{finite}&&\forall 1\leq i<j\leq N\,,
\end{align}
such that the only singularities in $K^{N,(r)}$ stem from the factors $\mathcal{P}_{\alpha_i}$ in (\ref{DecompositionPBlocks}). These, however, are exactly the contributions analysed in Section~\ref{Sect:RecursionN1}, such that we find immediately for the leading singularity in $\epsilon_2$
\begin{align}
K^{N,(r)}_0(\rho,S,\widehat{a}_{1,\ldots,N-1};\epsilon_1)=\sum_{{r_1,\ldots,r_N\geq 1}\atop{r_1+\ldots+r_N=r}}&\left(\prod_{k=1}^N\frac{1}{r_k!}\left(K_0^{1,(1)}(\rho,S;\epsilon_1)\right)^{r_k}\right)\nonumber\\
&\times \prod_{1\leq i<j\leq N}\mathcal{T}_{\beta_{r_j}\beta_{r_i}}(\rho,S,\widehat{a}_{1,\ldots,N-1};\epsilon_1,\epsilon_2=0)\,,\label{DefFormKNS}
\end{align}
where $\beta_{r_i}$ are partitions of $r_i$ with Young tableaux of the form (\ref{LeadingYoungDiagram}) (with $r_i$ boxes). Concretely, with the definitions (\ref{DefEsFacts}) 
\begin{align}
\mathfrak{e}_{r,s}^{\beta_j,\beta_i}\longrightarrow\left\{\begin{array}{lcl}\left\{\begin{array}{lcl} 0 & \text{for} & s\in\{1,\ldots,r_i\} \\ -\epsilon_1 &\text{for} & s\in\{r_i+1,\ldots,r_j\}\end{array}\right.&\text{if} & r_i< r_j \\[16pt] 0 & \text{if} & r_i\geq r_j\end{array}\right.
\end{align}
we find for the $\mathcal{T}_{\beta_{r_j}\beta_{r_i}}$
\begin{align}
&\mathcal{T}_{\beta_{r_j}\beta_{r_i}}(\epsilon_2=0)=\left(-\frac{\fb(S_+;\rho)}{4\pi^2}\right)^{r_i+r_j}\left\{\begin{array}{lcl}\omega_{+,0}^{2r_j}\,\omega_{+,1}^{r_i-r_j}\,\omega_{-,-1}^{r_j}\,\omega_{-,0}^{r_i-r_j}\,\omega_{-,1}^{r_j}&\text{for} & r_i\geq r_j\,, \\[8pt] \omega_{+,0}^{2r_i}\,\omega_{+,-1}^{r_j-r_i}\,\omega_{-,-1}^{r_i}\,\omega_{-,0}^{r_j-r_i}\,\omega_{-,1}^{r_i}&\text{for} & r_i< r_j\,,\end{array}\right.\nonumber
\end{align}
where we have introduced the shorthand notation $\omega_{\pm,n}:=\Omega(\mathfrak{a}_{ij}+n\epsilon_1,\pm S_+;\rho)$ and $S_+=S+\epsilon_1/2$ for better readability. For later use, we also note
\begin{align}
&\mathcal{T}_{\beta_{r_j}\beta_{r_i}}(\epsilon_{1,2}=0)=\left(-\frac{\fb(S;\rho)}{4\pi^2}\,\Omega(\mathfrak{a}_{ij},S;\rho)\,\Omega(\mathfrak{a}_{ij},-S;\rho)\right)^{r_i+r_j}\,.
\end{align}
\subsubsection{Leading Instanton Contribution}
\begin{figure}[h]
\scalebox{0.96}{\parbox{16.3cm}{\begin{tikzpicture}[scale = 1.50]
\draw[fill=black] (-1,0) circle (0.05);
\draw[fill=black] (1,0) circle (0.05);
\node[rotate=-38] at (0.6,0.8) {\Large $\cdots$};
\draw[fill=black] (-0.5,0.866) circle (0.05);
\draw[fill=black] (0.5,-0.866) circle (0.05);
\draw[fill=black] (-0.5,-0.866) circle (0.05);
\node at (-0.65,-1) {\footnotesize $\widehat{b}_1$};
\node at (0.7,-1) {\footnotesize $\widehat{b}_2$};
\node at (1.2,0) {\footnotesize $\widehat{b}_3$};
\node at (-0.7,1.1) {\footnotesize $\widehat{b}_{N-1}$};
\node at (-1.25,0) {\footnotesize $\widehat{b}_{N}$};
\begin{scope}[ultra thick,decoration={markings,mark=at position 0.55 with {\arrow{>}}}] 
\draw[ultra thick,postaction={decorate}] (-0.5,-0.82) -- (-0.5,0.82);
\draw[ultra thick,postaction={decorate}] (-0.455,-0.866) -- (0.455,-0.866);
\draw[ultra thick,postaction={decorate}] (-0.465,-0.84) -- (0.96,-0.02);
\draw[ultra thick,postaction={decorate}] (-0.535,-0.83) -- (-0.98,-0.04);
\end{scope}
\node at (1.75,0) {$+$};
\begin{scope}[xshift=3.7cm]
\draw[fill=black] (-1,0) circle (0.05);
\draw[fill=black] (1,0) circle (0.05);
\node[rotate=-38] at (0.6,0.8) {\Large $\cdots$};
\draw[fill=black] (-0.5,0.866) circle (0.05);
\draw[fill=black] (0.5,-0.866) circle (0.05);
\draw[fill=black] (-0.5,-0.866) circle (0.05);
\node at (-0.65,-1) {\footnotesize $\widehat{b}_1$};
\node at (0.7,-1) {\footnotesize $\widehat{b}_2$};
\node at (1.2,0) {\footnotesize $\widehat{b}_3$};
\node at (-0.7,1.1) {\footnotesize $\widehat{b}_{N-1}$};
\node at (-1.25,0) {\footnotesize $\widehat{b}_{N}$};
\begin{scope}[ultra thick,decoration={markings,mark=at position 0.55 with {\arrow{>}}}] 
\draw[ultra thick,postaction={decorate}] (0.455,-0.866) -- (-0.455,-0.866);
\draw[ultra thick,postaction={decorate}] (0.52,-0.825) -- (0.98,-0.036);
\draw[ultra thick,postaction={decorate}] (0.48,-0.825) -- (-0.475,0.83);
\draw[ultra thick,postaction={decorate}] (0.465,-0.84) -- (-0.96,-0.025);
\end{scope}
\end{scope}
\node at (5.8,0) {$+\ldots+$};
\begin{scope}[xshift=8.05cm]
\draw[fill=black] (-1,0) circle (0.05);
\draw[fill=black] (1,0) circle (0.05);
\node[rotate=-38] at (0.6,0.8) {\Large $\cdots$};
\draw[fill=black] (-0.5,0.866) circle (0.05);
\draw[fill=black] (0.5,-0.866) circle (0.05);
\draw[fill=black] (-0.5,-0.866) circle (0.05);
\node at (-0.65,-1) {\footnotesize $\widehat{b}_1$};
\node at (0.7,-1) {\footnotesize $\widehat{b}_2$};
\node at (1.2,0) {\footnotesize $\widehat{b}_3$};
\node at (-0.7,1.1) {\footnotesize $\widehat{b}_{N-1}$};
\node at (-1.25,0) {\footnotesize $\widehat{b}_{N}$};
\begin{scope}[ultra thick,decoration={markings,mark=at position 0.55 with {\arrow{>}}}] 
\draw[ultra thick,postaction={decorate}] (-0.98,-0.04) -- (-0.535,-0.83);
\draw[ultra thick,postaction={decorate}] (-0.96,-0.025) -- (0.465,-0.84);
\draw[ultra thick,postaction={decorate}] (-0.955,0) -- (0.955,0);
\draw[ultra thick,postaction={decorate}] (-0.98,0.04) -- (-0.535,0.83);
\end{scope}
\end{scope}
\end{tikzpicture}
}}
\caption{\sl Schematic graphical representation of the leading (\emph{i.e.} $r=1$) instanton contribution $K_0^{N,(1)}/\left(-K_0^{1,(1)}(\rho,S;\epsilon_1)\frac{\fb(S_+;\rho)}{4\pi^2}\right)$ in eq.~(\ref{FormK01toGraphical}). The oriented lines are defined in eq.~(\ref{DefLinesPropNS}).}
\label{Fig:GraphRepProps}
\end{figure}
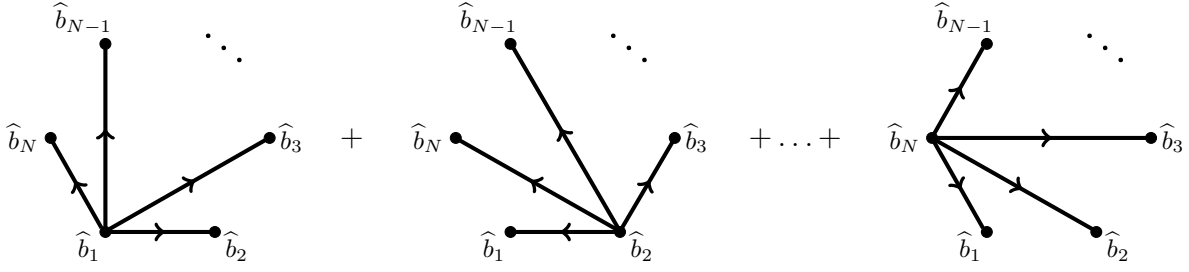

Using the results of the previous Subsubsection, we first calculate the leading instanton contribution $r=1$. Indeed, this corresponds to all contributions where only one of the $r_k=1$, while the remaining ones are vanishing
\begin{align}
&K_0^{N,(1)}(\rho,S,\widehat{a}_{1,\ldots,N-1};\epsilon_{1})=-\frac{\fb(S_+;\rho)}{4\pi^2}\,K_0^{1,(1)}(\rho,S;\epsilon_1)\nonumber\\
&\times \sum_{u=1}^N\left[\left(\prod_{i=1}^{u-1}\Omega\left(\mathfrak{a}_{iu}-\epsilon_1,S_+;\rho\right)\Omega\left(\mathfrak{a}_{iu},-S_+;\rho\right)\right) \left(\prod_{i=u+1}^N\Omega(\mathfrak{a}_{ui}+\epsilon_1,S_+;\rho)\,\Omega(\mathfrak{a}_{ui},-S_+;\rho)\right)\right]\,.
\end{align}
Using the definitions of the Kronecker-Eisenstein series, we can also write this expression as a single product
\begin{align}
&K_0^{N,(1)}(\rho,S,\widehat{a}_{1,\ldots,N-1};\epsilon_{1})\nonumber\\
&=-K_0^{1,(1)}(\rho,S;\epsilon_1)\frac{\fb(S_+;\rho)}{4\pi^2}\sum_{u=1}^{N}\prod_{{i=1}\atop {i\neq u}}^{N} \Omega(\widehat{b}_i-\widehat{b}_u+\theta_{iu}\,\epsilon_1,S_+;\rho)\,\Omega(\widehat{b}_i-\widehat{b}_u+\theta_{ui}\,\epsilon_1,-S_+;\rho)\,,\label{FormK01toGraphical}
\end{align}
with the notation 
\begin{align}
&\widehat{b}_1=0\,,&&\widehat{b}_j=\sum_{n=1}^j\widehat{a}_n\hspace{1cm}\forall j=2,\ldots,N\,,&&\text{and} &&\theta_{ij}=\left\{\begin{array}{lcl}1 & \text{if} & i>j\,,\\ 0 & \text{if} & i<j\,.\end{array}\right.\label{DefExternalPoints}
\end{align}
This notation lends itself to a graphical representation. We have schematically shown $K_0^{N,(1)}$ (up to the factor $-K_0^{1,(1)}(\rho,S;\epsilon_1)\frac{\fb(S_+;\rho)}{4\pi^2}$) in Figure~\ref{Fig:GraphRepProps}: here (oriented) lines are defined as

\begin{align}
\scalebox{1}{\parbox{2.7cm}{\begin{tikzpicture}[scale = 1.50]
\draw[fill=black] (0.19,0) circle (0.05);
\node at (0.2,0.3) {$\widehat{b}_j$};
\draw[fill=black] (-1.19,0) circle (0.05);
\node at (-1.2,-0.3) {$\widehat{b}_i$};
\begin{scope}[ultra thick,decoration={markings,mark=at position 0.55 with {\arrow{>}}}] 
\draw[ultra thick,postaction={decorate}] (-1.15,0) -- (0.15,0);
\end{scope}
\end{tikzpicture}
}}=\left\{\begin{array}{lcl}\Omega(\widehat{b}_j-\widehat{b}_i+\epsilon_1,S_+;\rho)\,\Omega(\widehat{b}_j-\widehat{b}_i,-S_+;\rho) & \text{if} & i<j\,,\\[4pt] \Omega(\widehat{b}_j-\widehat{b}_i,S_+;\rho)\,\Omega(\widehat{b}_j-\widehat{b}_i+\epsilon_1,-S_+;\rho) & \text{if} & i>j\,. \end{array}\right.\label{DefLinesPropNS}
\end{align}
Here the orientation of the lines is purely to keep track of the remaining deformation parameter $\epsilon_1$ in the NS-limit. We shall see below and in Section~\ref{Sect:DecmopositionFreeEnergy} that this picture dramatically simplifies in the unrefined limit $\epsilon_1=-\epsilon_2$ in which it is not be necessary to keep track of this orientation. In fact, we shall discover that the picture resembles much more directly the decomposition proposed in \cite{Hohenegger:2020slq} for the free energy.
\subsubsection{Next-to-Leading Instanton Contribution}\label{Sect:NSLimitNLO}
We next consider the instanton contribution $r=2$, in which case there are two distinct contributions: 
\begin{itemize}
\item $r_i=2$ for a single $i\in 1,\ldots,N$ (and all remaining $r_j=0$). These configurations contribute
\begin{align}
\frac{1}{2}\left(K_0^{1,(1)}\right)^2\left(-\frac{\fb(S_+;\rho)}{4\pi^2}\right)^2&\sum_{u=1}^N\bigg[\left(\prod_{i=1}^{u-1}\Omega\left(\mathfrak{a}_{iu}-\epsilon_1,S_+;\rho\right)\Omega\left(\mathfrak{a}_{iu},-S_+;\rho\right)\right)\nonumber\\
&\times \left(\prod_{i=u+1}^N\Omega(\mathfrak{a}_{ui}+\epsilon_1,S_+;\rho)\,\Omega(\mathfrak{a}_{ui},-S_+;\rho)\right)\bigg]^2\,,
\end{align}
which can be re-arranged in the same way as in eq.(\ref{FormK01toGraphical})
\begin{align}
\left(K_0^{1,(1)}\right)^2\,\frac{\fb^2(S_+;\rho)}{32\pi^4}\,\sum_{u=1}^{N}\left[\prod_{{i=1}\atop {i\neq u}}^{N} \Omega(\widehat{b}_i-\widehat{b}_u+\theta_{iu}\,\epsilon_1,S_+;\rho)\,\Omega(\widehat{b}_i-\widehat{b}_u+\theta_{ui}\,\epsilon_1,-S_+;\rho)\right]^2\,.\label{DefOrd2con1}
\end{align}
\item $r_i=1=r_j$ for two distinct $i\neq j$ (and all remaining $r_k=0$). These configurations contribute
\begin{align}
&\left(K_0^{1,(1)}\right)^2\left(-\frac{\fb(S_+;\rho)}{4\pi^2}\right)^2\sum_{1\leq i<j\leq N}\bigg[\Omega(\mathfrak{a}_{ij},S_+;\rho)^2\,\Omega(\mathfrak{a}_{ij}-\epsilon_1,S_-;\rho)\,\Omega(\mathfrak{a}_{ij}+\epsilon_1,S_-;\rho)\nonumber\\
&\times \left(\prod_{r=1}^{i-1}\Omega\left(\mathfrak{a}_{ri}-\epsilon_1,S_+;\rho\right)\Omega\left(\mathfrak{a}_{ri},-S_+;\rho\right)\right) \left(\prod_{{r=i+1}\atop{r\neq j}}^N\Omega(\mathfrak{a}_{ir}+\epsilon_1,S_+;\rho)\,\Omega(\mathfrak{a}_{ir},-S_+;\rho)\right)\nonumber\\
&\times \left(\prod_{{r=1}\atop{r\neq i}}^{j-1}\Omega\left(\mathfrak{a}_{rj}-\epsilon_1,S_+;\rho\right)\Omega\left(\mathfrak{a}_{rj},-S_+;\rho\right)\right) \left(\prod_{r=j+1}^N\Omega(\mathfrak{a}_{jr}+\epsilon_1,S_+;\rho)\,\Omega(\mathfrak{a}_{ir},-S_+;\rho)\right)\bigg]\,.
\end{align}
Using the definition (\ref{DefExternalPoints}), as well as (\ref{ThetaShift}), we can re-arrange this expression as
{\allowdisplaybreaks
\begin{align}
\left(K_0^{1,(1)}\right)^2\,&\frac{\fb^2(S_+;\rho)}{16\pi^4}\sum_{1\leq i<j\leq N}\bigg[\Omega(\widehat{b}_j-\widehat{b}_i+\epsilon_1,S_+;\rho)\,\Omega(\widehat{b}_j-\widehat{b}_i,-S_+;\rho)\nonumber\\
&\hspace{3.6cm}\times \Omega(\widehat{b}_i-\widehat{b}_j,S_+;\rho)\,\Omega(\widehat{b}_i-\widehat{b}_j+\epsilon_1,-S_+;\rho)\nonumber\\
&\times\prod_{{r=1}\atop {r\neq i,j}}^{N} \Omega(\widehat{b}_r-\widehat{b}_i+\theta_{ri}\,\epsilon_1,S_+;\rho)\,\Omega(\widehat{b}_r-\widehat{b}_i+\theta_{ir}\,\epsilon_1,-S_+;\rho)\nonumber\\
&\times\prod_{{r=1}\atop {r\neq i,j}}^{N} \Omega(\widehat{b}_r-\widehat{b}_j+\theta_{rj}\,\epsilon_1,S_+;\rho)\,\Omega(\widehat{b}_r-\widehat{b}_j+\theta_{jr}\,\epsilon_1,-S_+;\rho)\bigg]\,.\label{DefOrd2con2}
\end{align}}
\end{itemize}
The combination of these contributions yields
\begin{align}
K^{N,(2)}_0(\rho,S,\widehat{a}_{1,\ldots,N-1};\epsilon_1)=\frac{1}{2}\,\left(K^{N,(1)}_0(\rho,S,\widehat{a}_{1,\ldots,N-1};\epsilon_1)\right)^2\,,\label{RelGenK2}
\end{align}
which generalises (\ref{RecursionRelN1}) to generic $N$ for $r=2$. While the above expressions are very complicated, the relation (\ref{RelGenK2}) can be represented simpler in a graphical fashion. Indeed, dividing by a factor $\left(K_0^{1,(1)}\right)^2\left(-\frac{\fb(S_+;\rho)}{4\pi^2}\right)^2$, eq.~(\ref{RelGenK2}) for $N=2$ can be visualised as follows

\begin{align}
\frac{1}{2}\left(\scalebox{0.9}{\parbox{2.65cm}{\begin{tikzpicture}[scale = 1.50]
\draw[fill=black] (0.19,0) circle (0.05);
\node at (0.2,0.3) {$\widehat{b}_2$};
\draw[fill=black] (-1.19,0) circle (0.05);
\node at (-1.2,-0.3) {$\widehat{b}_1$};
\begin{scope}[ultra thick,decoration={markings,mark=at position 0.55 with {\arrow{>}}}] 
\draw[fill=black,ultra thick,postaction={decorate}] (-1.15,0) -- (0.15,0);
\end{scope}
\end{tikzpicture}
}}\,+\,\scalebox{0.9}{\parbox{2.65cm}{\begin{tikzpicture}[scale = 1.50]
\draw[fill=black] (0.19,0) circle (0.05);
\node at (0.2,0.3) {$\widehat{b}_2$};
\draw[fill=black] (-1.19,0) circle (0.05);
\node at (-1.2,-0.3) {$\widehat{b}_1$};
\begin{scope}[ultra thick,decoration={markings,mark=at position 0.55 with {\arrow{>}}}] 
\draw[fill=black,ultra thick,postaction={decorate}]  (0.15,0) -- (-1.15,0);
\end{scope}
\end{tikzpicture}
}}\right)^2=\frac{1}{2}\left(\scalebox{0.9}{\parbox{2.65cm}{\begin{tikzpicture}[scale = 1.50]
\draw[fill=black] (0.19,0) circle (0.05);
\node at (0.2,0.3) {$\widehat{b}_2$};
\draw[fill=black] (-1.19,0) circle (0.05);
\node at (-1.2,-0.3) {$\widehat{b}_1$};
\begin{scope}[ultra thick,decoration={markings,mark=at position 0.55 with {\arrow{>}}}] 
\draw[ultra thick,postaction={decorate}] (-1.15,0.035) -- (0.15,0.035);
\draw[ultra thick,postaction={decorate}] (-1.15,-0.035) -- (0.15,-0.035);
\end{scope}
\end{tikzpicture}
}}\,+\,\scalebox{0.9}{\parbox{2.65cm}{\begin{tikzpicture}[scale = 1.50]
\draw[fill=black] (0.19,0) circle (0.05);
\node at (0.2,0.3) {$\widehat{b}_2$};
\draw[fill=black] (-1.19,0) circle (0.05);
\node at (-1.2,-0.3) {$\widehat{b}_1$};
\begin{scope}[ultra thick,decoration={markings,mark=at position 0.55 with {\arrow{>}}}] 
\draw[ultra thick,postaction={decorate}]  (0.15,0.035) -- (-1.15,0.035);
\draw[ultra thick,postaction={decorate}]  (0.15,-0.035) -- (-1.15,-0.035);
\end{scope}
\end{tikzpicture}
}}\right)\,+\,\scalebox{0.9}{\parbox{2.65cm}{\begin{tikzpicture}[scale = 1.50]
\draw[fill=black] (0.19,0) circle (0.05);
\node at (0.2,0.3) {$\widehat{b}_2$};
\draw[fill=black] (-1.19,0) circle (0.05);
\node at (-1.2,-0.3) {$\widehat{b}_1$};
\begin{scope}[ultra thick,decoration={markings,mark=at position 0.75 with {\arrow{>}}}] 
\draw[ultra thick,postaction={decorate}]  (0.15,0.035) -- (-1.15,0.035);
\draw[ultra thick,postaction={decorate}]  (-1.15,-0.035) -- (0.15,-0.035);
\end{scope}
\end{tikzpicture}
}}
\end{align}
Here the first two terms on the right hand side are identified with (\ref{DefOrd2con1}), while the last term corresponds to the first two lines of (\ref{DefOrd2con2}) (since the last two lines in (\ref{DefOrd2con2}) are absent for $N=2$). The more involved example of $N=3$ is schematically represented by\footnote{Here we have again divided by a factor $\left(K_0^{1,(1)}\right)^2\left(-\frac{\fb(S_+;\rho)}{4\pi^2}\right)^2$.} 
{\allowdisplaybreaks\begin{align}
&\frac{1}{2}\left(\scalebox{0.9}{\parbox{2.8cm}{\begin{tikzpicture}[scale = 1.50]
\draw[fill=black] (0.19,0) circle (0.05);
\draw[fill=black] (-1.19,0) circle (0.05);
\draw[fill=black] (-0.5,0.866) circle (0.05);
\node at (-1.2,-0.25) {\footnotesize $\widehat{b}_1$};
\node at (0.3,-0.25) {\footnotesize $\widehat{b}_2$};
\node at (-0.48,1.15) {\footnotesize $\widehat{b}_3$};
\begin{scope}[ultra thick,decoration={markings,mark=at position 0.55 with {\arrow{>}}}] 
\draw[ultra thick,postaction={decorate}] (-1.15,0) -- (0.15,0);
\draw[ultra thick,postaction={decorate}] (-1.165,0.04) -- (-0.525,0.825);
\end{scope}
\end{tikzpicture}
}}\hspace{0.25cm}+\hspace{0.25cm}\scalebox{0.9}{\parbox{2.8cm}{\begin{tikzpicture}[scale = 1.50]
\draw[fill=black] (0.19,0) circle (0.05);
\draw[fill=black] (-1.19,0) circle (0.05);
\draw[fill=black] (-0.5,0.866) circle (0.05);
\node at (-1.2,-0.25) {\footnotesize $\widehat{b}_1$};
\node at (0.3,-0.25) {\footnotesize $\widehat{b}_2$};
\node at (-0.48,1.15) {\footnotesize $\widehat{b}_3$};
\begin{scope}[ultra thick,decoration={markings,mark=at position 0.55 with {\arrow{>}}}] 
\draw[ultra thick,postaction={decorate}] (0.15,0) -- (-1.15,0);
\draw[ultra thick,postaction={decorate}] (0.165,0.04) -- (-0.475,0.825);
\end{scope}
\end{tikzpicture}
}}\hspace{0.25cm}+\hspace{0.25cm}\scalebox{0.9}{\parbox{2.8cm}{\begin{tikzpicture}[scale = 1.50]
\draw[fill=black] (0.19,0) circle (0.05);
\draw[fill=black] (-1.19,0) circle (0.05);
\draw[fill=black] (-0.5,0.866) circle (0.05);
\node at (-1.2,-0.25) {\footnotesize $\widehat{b}_1$};
\node at (0.3,-0.25) {\footnotesize $\widehat{b}_2$};
\node at (-0.48,1.15) {\footnotesize $\widehat{b}_3$};
\begin{scope}[ultra thick,decoration={markings,mark=at position 0.55 with {\arrow{>}}}] 
\draw[ultra thick,postaction={decorate}] (-0.525,0.825) -- (-1.165,0.04);
\draw[ultra thick,postaction={decorate}] (-0.475,0.825) -- (0.165,0.04);
\end{scope}
\end{tikzpicture}
}}
\right)^2\nonumber\\[8pt]
&\hspace{2cm}=\frac{1}{2}\left(\scalebox{0.9}{\parbox{2.8cm}{\begin{tikzpicture}[scale = 1.50]
\draw[fill=black] (0.19,0) circle (0.05);
\draw[fill=black] (-1.19,0) circle (0.05);
\draw[fill=black] (-0.5,0.866) circle (0.05);
\node at (-1.2,-0.25) {\footnotesize $\widehat{b}_1$};
\node at (0.3,-0.25) {\footnotesize $\widehat{b}_2$};
\node at (-0.48,1.15) {\footnotesize $\widehat{b}_3$};
\begin{scope}[ultra thick,decoration={markings,mark=at position 0.55 with {\arrow{>}}}] 
\draw[ultra thick,postaction={decorate}] (-1.15,0.035) -- (0.15,0.035);
\draw[ultra thick,postaction={decorate}] (-1.15,-0.035) -- (0.15,-0.035);
\draw[ultra thick,postaction={decorate}] (-1.2,0.065) -- (-0.555,0.855);
\draw[ultra thick,postaction={decorate}] (-1.14,0.015) -- (-0.49,0.805);
\end{scope}
\end{tikzpicture}
}}\hspace{0.25cm}+\hspace{0.25cm}\scalebox{0.9}{\parbox{2.25cm}{\begin{tikzpicture}[scale = 1.50]
\draw[fill=black] (0.19,0) circle (0.05);
\draw[fill=black] (-1.19,0) circle (0.05);
\draw[fill=black] (-0.5,0.866) circle (0.05);
\node at (-1.2,-0.25) {\footnotesize $\widehat{b}_1$};
\node at (0.3,-0.25) {\footnotesize $\widehat{b}_2$};
\node at (-0.48,1.15) {\footnotesize $\widehat{b}_3$};
\begin{scope}[ultra thick,decoration={markings,mark=at position 0.55 with {\arrow{>}}}] 
\draw[ultra thick,postaction={decorate}] (0.15,0.035) -- (-1.15,0.035);
\draw[ultra thick,postaction={decorate}] (0.15,-0.035) -- (-1.15,-0.035);
\draw[ultra thick,postaction={decorate}] (0.2,0.065) -- (-0.445,0.855);
\draw[ultra thick,postaction={decorate}] (0.14,0.015) -- (-0.51,0.805);
\end{scope}
\end{tikzpicture}
}}\hspace{0.25cm}+\hspace{0.25cm}\scalebox{0.9}{\parbox{2.8cm}{\begin{tikzpicture}[scale = 1.50]
\draw[fill=black] (0.19,0) circle (0.05);
\draw[fill=black] (-1.19,0) circle (0.05);
\draw[fill=black] (-0.5,0.866) circle (0.05);
\node at (-1.2,-0.25) {\footnotesize $\widehat{b}_1$};
\node at (0.3,-0.25) {\footnotesize $\widehat{b}_2$};
\node at (-0.48,1.15) {\footnotesize $\widehat{b}_3$};
\begin{scope}[ultra thick,decoration={markings,mark=at position 0.55 with {\arrow{>}}}] 
\draw[ultra thick,postaction={decorate}] (-0.555,0.855) -- (-1.2,0.065);
\draw[ultra thick,postaction={decorate}] (-0.49,0.805) -- (-1.14,0.015);
\draw[ultra thick,postaction={decorate}] (-0.445,0.855) -- (0.2,0.065);
\draw[ultra thick,postaction={decorate}] (-0.51,0.805) -- (0.14,0.015);
\end{scope}
\end{tikzpicture}
}}
\right)\nonumber\\[8pt]
&\hspace{2.75cm}+\scalebox{0.9}{\parbox{2.8cm}{\begin{tikzpicture}[scale = 1.50]
\draw[fill=black] (0.19,0) circle (0.05);
\draw[fill=black] (-1.19,0) circle (0.05);
\draw[fill=black] (-0.5,0.866) circle (0.05);
\node at (-1.2,-0.25) {\footnotesize $\widehat{b}_1$};
\node at (0.3,-0.25) {\footnotesize $\widehat{b}_2$};
\node at (-0.48,1.15) {\footnotesize $\widehat{b}_3$};
\begin{scope}[ultra thick,decoration={markings,mark=at position 0.55 with {\arrow{>}}}] 
\draw[ultra thick,postaction={decorate}] (-1.165,0.04) -- (-0.525,0.825);
\draw[ultra thick,postaction={decorate}] (0.165,0.04) -- (-0.475,0.825);
\end{scope}
\begin{scope}[ultra thick,decoration={markings,mark=at position 0.75 with {\arrow{>}}}] 
\draw[ultra thick,postaction={decorate}] (0.15,0.035) -- (-1.15,0.035);
\draw[ultra thick,postaction={decorate}] (-1.15,-0.035) -- (0.15,-0.035);
\end{scope}
\end{tikzpicture}
}}\hspace{0.25cm}+\hspace{0.25cm}\scalebox{0.9}{\parbox{2.8cm}{\begin{tikzpicture}[scale = 1.50]
\draw[fill=black] (0.19,0) circle (0.05);
\draw[fill=black] (-1.19,0) circle (0.05);
\draw[fill=black] (-0.5,0.866) circle (0.05);
\node at (-1.2,-0.25) {\footnotesize $\widehat{b}_1$};
\node at (0.3,-0.25) {\footnotesize $\widehat{b}_2$};
\node at (-0.48,1.15) {\footnotesize $\widehat{b}_3$};
\begin{scope}[ultra thick,decoration={markings,mark=at position 0.55 with {\arrow{>}}}] 
\draw[ultra thick,postaction={decorate}] (-1.15,0) -- (0.15,0);
\draw[ultra thick,postaction={decorate}] (-0.475,0.825) -- (0.165,0.04);
\end{scope}
\begin{scope}[ultra thick,decoration={markings,mark=at position 0.75 with {\arrow{>}}}] 
\draw[ultra thick,postaction={decorate}] (-0.555,0.855) -- (-1.2,0.065);
\draw[ultra thick,postaction={decorate}] (-1.14,0.015) -- (-0.49,0.805);
\end{scope}
\end{tikzpicture}
}}\hspace{0.25cm}+\hspace{0.25cm}\scalebox{0.9}{\parbox{2.8cm}{\begin{tikzpicture}[scale = 1.50]
\draw[fill=black] (0.19,0) circle (0.05);
\draw[fill=black] (-1.19,0) circle (0.05);
\draw[fill=black] (-0.5,0.866) circle (0.05);
\node at (-1.2,-0.25) {\footnotesize $\widehat{b}_1$};
\node at (0.3,-0.25) {\footnotesize $\widehat{b}_2$};
\node at (-0.48,1.15) {\footnotesize $\widehat{b}_3$};
\begin{scope}[ultra thick,decoration={markings,mark=at position 0.55 with {\arrow{>}}}] 
\draw[ultra thick,postaction={decorate}] (0.15,0) -- (-1.15,0);
\draw[ultra thick,postaction={decorate}] (-0.525,0.825) -- (-1.165,0.04);
\end{scope}
\begin{scope}[ultra thick,decoration={markings,mark=at position 0.75 with {\arrow{>}}}] 
\draw[ultra thick,postaction={decorate}] (0.2,0.065) -- (-0.445,0.855);
\draw[ultra thick,postaction={decorate}] (-0.51,0.805) -- (0.14,0.015);
\end{scope}
\end{tikzpicture}
}}\,.
\end{align}}

\subsubsection{Higher Order Instanton Contributions}
The recursive relation (\ref{RelGenK2}) can be generalised for higher $r>2$ and we conjecture the following structure 
\begin{align}
&K^{N,(r)}_0(\rho,S,\widehat{a}_{1,\ldots,N-1};\epsilon_1)=\frac{1}{r!}\,\left(K^{N,(1)}_0(\rho,S,\widehat{a}_{1,\ldots,N-1};\epsilon_1)\right)^r\,,&&\forall r\in \mathbb{N}\,.\label{RelGenKN}
\end{align}
We remark that this relation is complementary to the self-similarity found in \cite{Hohenegger:2016eqy}, which allows to related the (single-particle) free energy for $N>1$ to the one for $N=1$. For simplicity, we shall demonstrate (\ref{RelGenKN}) only to leading order in $\epsilon_1$, \emph{i.e.} for the $K^{N,(r)}_{0,0}$ defined in (\ref{ZnFormEpsi1Laurent}). Indeed, in this case the structure in (\ref{FormK01toGraphical}) simplifies drastically\footnote{In fact also the graphical representation (\ref{DefLinesPropNS}) becomes much more streamlined since the orientation of the lines (\emph{i.e.} the arrows) become irrelevant.} and we have 
\begin{align}
\left(K^{N,(1)}_{0,0}(\rho,S,\widehat{a}_{1,\ldots,N-1})\right)^r=\left(-K_{0,0}^{1,(1)}\,\frac{\fb(S;\rho)}{4\pi^2}\sum_{u=1}^{N}\prod_{{i=1}\atop {i\neq u}}^{N} \Omega(\widehat{b}_i-\widehat{b}_u,S;\rho)\,\Omega(\widehat{b}_i-\widehat{b}_u,-S;\rho)\right)^r\,.\nonumber
\end{align}
Upon introducing the shorthand notation 
\begin{align}
\mathcal{G}_{ij}:=-\frac{\fb(S;\rho)}{4\pi^2}\,\Omega(\widehat{b}_j-\widehat{b}_i,S;\rho)\,\Omega(\widehat{b}_j-\widehat{b}_i,-S;\rho)=\mathcal{G}_{ji}
\end{align}
we can develop the power of $r$ in the following form
{\allowdisplaybreaks
\begin{align}
\left(K^{N,(1)}_{0,0}\right)^r&=\left(K_{0,0}^{1,(1)}\right)^r\,\sum_{{r_1,\ldots,r_N=0}\atop{r_1+\ldots+r_N=r}}^r\frac{r!}{r_1!\ldots r_N!}\,\prod_{u=1}^N\left(\prod_{{i=1}\atop{i\neq 0}}^N \mathcal{G}_{iu} \right)^{r_u}\nonumber\\
&=\left(-K_{0,0}^{1,(1)}\right)^r\,\sum_{{r_1,\ldots,r_N=0}\atop{r_1+\ldots+r_N=r}}^r\frac{r!}{r_1!\ldots r_N!}\,\prod_{1\leq i<j\leq N}\mathcal{G}_{ij}^{r_i+r_j}\,.
\end{align}}
Notice that in total the last expression contains $(N-1)r$ factors $\mathcal{G}$ (with different end-points), which are organised graphically as $r_i+r_j$ (unoriented) lines stretching between the nodes $\widehat{b}_i$ and $\widehat{b}_j$. Using (\ref{NSFusionN1}), we have finally
\begin{align}
\left(K^{N,(1)}_{0,0}\right)^r&=r!\sum_{{r_1,\ldots,r_N=0}\atop{r_1+\ldots+r_N=r}}^r\left[
\left(\prod_{u=1}^NK_{0,0}^{1,(r_u)}\right)\,\prod_{1\leq i<j\leq N}\mathcal{G}_{ij}^{r_i+r_j}\right]=r!\,K^{N,(r)}_{0,0}\,,
\end{align}
which is indeed the leading term in the relation (\ref{RelGenKN}).

\section{Diagrammatic Decomposition in the Unrefined Limit}\label{Sect:DecmopositionFreeEnergy}
In Section~\ref{Sect:KroneckerEisensteinSeries} we have seen that the topological string partition function on $X_{N,1}$ can be re-formulated using the Kronecker-Eisenstein series as an example of an elliptic modular graph form (eMGF). Moreover, in the unrefined limit, the latter conspire in a fashion to allow a rewriting of specific contributions to the non-perturbative gauge theory partition function in terms of (differences of) two-point functions of a (chiral) free scalar field on the torus (see eq.~(\ref{ContributionPartitionFunctionPropagators})). This result resembles a similar proposal made in \cite{Hohenegger:2019tii,Hohenegger:2020slq} for the decomposition of the free energy of $X_{N,1}$. In this Section we shall therefore discuss in more detail the unrefined limit of the partition function and explain how our results of Section~\ref{Sect:KroneckerEisensteinSeries} are not only compatible with \cite{Hohenegger:2019tii,Hohenegger:2020slq}, but also fix certain ambiguities that have been pointed out there for the exact decomposition. We shall start out by reviewing the results of \cite{Hohenegger:2019tii,Hohenegger:2020slq} (adapted to our notation) and, for simplicity, consider the concrete examples $N=2,3$.
\subsection{Unrefined Limit and Diagrammatic Decomposition}\label{Sect:UnrefinedDiagrammatic}
We start by using the result (\ref{ContributionPartitionFunctionPropagators}) for the unrefined limit of the contribution $\mathcal{T}_{\alpha_j\alpha_i}$ to give a diagrammatic representation of the (entire) non-perturbative partition function $\mathcal{Z}_{N,1}$. Indeed,

\begin{wrapfigure}{l}{0.35\textwidth}
${}$\\[-0.8cm]
\begin{center}
\scalebox{0.96}{\parbox{6cm}{\begin{tikzpicture}[scale = 1.50]
\draw (-1,0) circle (0.05);
\draw (1,0) circle (0.05);
\node[rotate=-30] at (0.55,0.8) {\Large $\cdots$};
\draw (-0.5,0.866) circle (0.05);
\draw (0.5,-0.866) circle (0.05);
\draw (-0.5,-0.866) circle (0.05);
\node at (-0.75,-1.1) {\footnotesize $(\widehat{b}_1,\alpha_1)$};
\node at (0.75,-1.1) {\footnotesize $(\widehat{b}_2,\alpha_2)$};
\node at (1.45,0) {\footnotesize $(\widehat{b}_3,\alpha_3)$};
\node at (-0.9,1.1) {\footnotesize $(\widehat{b}_{N-1},\alpha_{N-1})$};
\node at (-1.55,0) {\footnotesize $(\widehat{b}_{N},\alpha_N)$};
\draw[ultra thick] (-0.5,-0.82) -- (-0.5,0.82);
\draw[ultra thick] (-0.455,-0.866) -- (0.455,-0.866);
\draw[ultra thick] (-0.465,-0.84) -- (0.96,-0.02);
\draw[ultra thick] (-0.535,-0.83) -- (-0.98,-0.04);
\draw[ultra thick] (0.52,-0.825) -- (0.98,-0.036);
\draw[ultra thick] (0.48,-0.825) -- (-0.475,0.83);
\draw[ultra thick] (0.465,-0.84) -- (-0.96,-0.025);
\draw[ultra thick] (0.96,0.02) -- (-0.460,0.85);
\draw[ultra thick] (-0.955,0) -- (0.955,0);
\draw[ultra thick] (-0.98,0.04) -- (-0.535,0.83);
\draw[ultra thick] (0.98,0.04) -- (0.8,0.5);
\draw[ultra thick] (0.1,0.975) -- (-0.46,0.88);
\draw[ultra thick] (-0.47,-0.82) -- (0.45,0.75);
\draw[ultra thick] (0.5,-0.82) -- (0.65,0.65);
\draw[ultra thick]  (-0.965,0.025) -- (0.25,0.85);
\end{tikzpicture}
}}
\end{center}
${}$\\[-16pt]
\caption{\sl Diagrammatic Representation of $\widehat{\mathcal{P}}_{\alpha_1,\ldots,\alpha_N}$ in the unrefined limit.}
\label{Fig:GraphRepPropsUnrefined}
${}$\\[-2cm]
\end{wrapfigure} 

\noindent
for $\epsilon_1=-\epsilon_2=\epsilon$, the contribution $\mathcal{T}_{\alpha_j\alpha_i}$ simplifies as in (\ref{DefTIntegerShift}) and allows for a much more streamlined (graphical) representation of $\mathcal{P}_{\alpha_1,\ldots,\alpha_N}$ as defined in (\ref{DecompositionPBlocks}). To make the discussion more concrete, we first divide out overall factors in $\mathcal{P}_{\alpha_1,\ldots,\alpha_N}$ that do not depend on $\widehat{a}_{1,\ldots,N-1}$ by defining
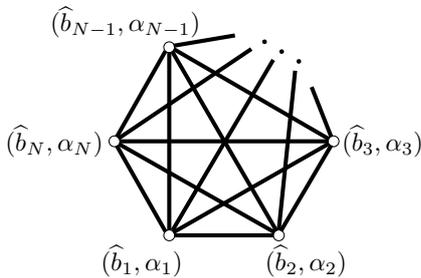
\begin{align}
\mathcal{P}_{\alpha_1,\ldots,\alpha_N}&=\left(\prod_{k=1}^{N}\mathcal{P}_{\alpha_k}(\rho,S;\epsilon)\right)\left(\frac{\fb(S;\rho)}{-4\pi^2}\right)^{(N-1)\sum_{k=1}^N |\alpha_k|}\nonumber\\
&\hspace{0.7cm}\times \widehat{\mathcal{P}}_{\alpha_1,\ldots,\alpha_N}(\rho,S,\widehat{a}_{1,\ldots,N-1};\epsilon)\,,\label{RefPaaGraph}
\end{align}
with 
\begin{align}
\widehat{\mathcal{P}}_{\alpha_1,\ldots,\alpha_N}=\prod_{1\leq i<j\leq N}\widehat{\mathcal{T}}_{\alpha_j\alpha_i}(\rho,S,\widehat{a}_{1,\ldots,N-1};\epsilon)\,. 
\end{align}
The reduced functions $\widehat{P}_{\alpha_1,\ldots,\alpha_N}$ are schematically represented in Figure~\ref{Fig:GraphRepPropsUnrefined}, which is similar to the graphical representation in Section~\ref{Sect:NSLimitNLO} in the NS-limit. However, in addition to the 'positions' $(\widehat{b}_1,\ldots,\widehat{b}_{N})$ we also associate the integer partitions $(\alpha_1,\ldots,\alpha_N)$ with each of the vertices, as shown in Figure~\ref{Fig:GraphRepPropsUnrefined}. To each line connecting two distinct such vertices $(\widehat{b}_i,\alpha_i)$ and $(\widehat{b}_j,\alpha_j)$ (with $i<j$) we then associate the following expression
\begin{align}
\scalebox{1}{\parbox{3.2cm}{\begin{tikzpicture}[scale = 1.50]
\draw (0.19,0) circle (0.05);
\node at (0.1,0.3) {\footnotesize $(\widehat{b}_{j},\alpha_j)$};
\draw (-1.18,0) circle (0.05);
\node at (-1.2,-0.3) {\footnotesize $(\widehat{b}_{i},\alpha_i)$};
\draw[ultra thick] (-1.15,0) -- (0.15,0);
\end{tikzpicture}
}}=\prod_{n\in\underline{n}^{\alpha_j\alpha_i}}\Omega\left(\widehat{b}_{j}-\widehat{b}_{i}+\epsilon\, n,S;\rho\right)\,\Omega\left(\widehat{b}_{j}-\widehat{b}_{i}+\epsilon\, n,-S;\rho\right)&&\text{for }i<j\,.\label{DefLinesPropUnrefined}
\end{align}
As already remarked in eq.~(\ref{ContributionPartitionFunctionPropagators}) the combination of Kronecker-Eisenstein series in (\ref{DefLinesPropUnrefined}) can be re-written as products of differences of propagators of free scalar fields on a torus
\begin{align}
&\scalebox{1}{\parbox{3.2cm}{\begin{tikzpicture}[scale = 1.50]
\draw (0.19,0) circle (0.05);
\node at (0.1,0.3) {\footnotesize $(\widehat{b}_{j},\alpha_j)$};
\draw (-1.18,0) circle (0.05);
\node at (-1.2,-0.3) {\footnotesize $(\widehat{b}_{i},\alpha_i)$};
\draw[ultra thick] (-1.15,0) -- (0.15,0);
\end{tikzpicture}
}}=\prod_{n\in\underline{n}^{\alpha_j\alpha_i}}\left(\mathbb{G}''(\widehat{b}_{j}-\widehat{b}_i+n\,\epsilon;\rho)-\mathbb{G}''(S;\rho)\right)\,,&&\text{for }i<j\,.\label{DefLinesPropagators}
\end{align}
We also remark that if $\alpha_i=\emptyset=\alpha_j$ (\ref{DefLinesPropagators}) is trivially $1$.

\subsection{Free Energy and Decomposition in Terms of Propagators}\label{Sect:FreeEnergy}
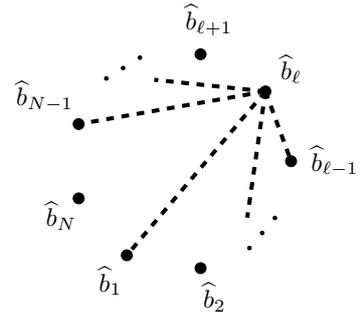
\begin{wrapfigure}{r}{0.32\textwidth}
${}$\\[-1.2cm]
\begin{center}
\scalebox{0.96}{\parbox{4.95cm}{\begin{tikzpicture}[scale = 1.50]
\draw[fill=black] (0.766044, 0.642788) circle (0.05); 
\node at (1,0.85) {\footnotesize $\widehat{b}_{\ell}$};
\draw[fill=black] (0.173648, 0.984808) circle (0.05); 
\node at (0.25,1.3) {\footnotesize $\widehat{b}_{\ell+1}$};
\node[rotate=32] at (-0.5, 0.866025) {\Large $\cdots$};
\draw[fill=black] (-0.939693,0.34202) circle (0.05); 
\node at (-1.25,0.6) {\footnotesize $\widehat{b}_{N-1}$};
\draw[fill=black] (-0.939693, -0.34202) circle (0.05); 
\node at (-1.1,-0.5) {\footnotesize $\widehat{b}_{N}$};
\draw[fill=black] (-0.5, -0.866025) circle (0.05); 
\node at (-0.65,-1.1) {\footnotesize $\widehat{b}_1$};
\draw[fill=black] (0.173648, -0.984808) circle (0.05); 
\node at (0.3,-1.25) {\footnotesize $\widehat{b}_2$};
\node[rotate=50] at (0.766044, -0.642788) {\Large$\cdots$};
\draw[fill=black] (1,0) circle (0.05); 
\node at (1.4,0) {\footnotesize $\widehat{b}_{\ell-1}$};
\draw[ultra thick,dashed ] (0.766044, 0.642788) -- (-0.5, -0.866025);
\draw[ultra thick,dashed ] (0.766044, 0.642788) -- (-0.939693,0.34202);
\draw[ultra thick,dashed ] (0.766044, 0.642788) -- (1,0);
\draw[ultra thick,dashed ] (0.766044, 0.642788) -- (0.6,-0.52);
\draw[ultra thick,dashed ] (0.766044, 0.642788) -- (-0.25,0.75);
\end{tikzpicture}
}}
\end{center}
${}$\\[-22pt]
\caption{\sl Diagrammatic Representation of a single term in the sum over $\mathcal{S}$ in (\ref{DefOoldPaper}) for fixed $\ell$. }
\label{Fig:GraphRepOldPaper}
${}$\\[-2.3cm]
\end{wrapfigure}

In \cite{Hohenegger:2020slq} the non-perturbative free energy
\begin{align}
\mathcal{F}_{N,1}(\rho,S,\widehat{a}_{1,\ldots,N-1},\tau;\epsilon_{1,2})=\ln\,\mathcal{Z}_{N,1}(\rho,S,\widehat{a}_{1,\ldots,N-1},\tau;\epsilon_{1,2})\,,
\end{align}
was discussed. Evidence was provided that the leading instanton contribution to $\mathcal{F}_{N,1}$ in the unrefined limit can be written in the form
\begin{align}
\sum_{s=0}^\infty \epsilon^{2s-2}H_{(s)}^{(1),\{0\}}(\rho,S)\sum_{k=0}^{N-1}\left(W_{(0)}^{(1)}(\rho,S)\right)^{N-1-k}\left(H_{(0)}^{(1),\{0\}}(\rho,S)\right)^k\,\mathcal{O}^{(N),k}(\widehat{a}_{1,\ldots,N-1},\rho)\,.\label{DecompOldPaper}
\end{align}
We refer the reader to \cite{Hohenegger:2020slq} for the precise notation and conventions (it is briefly reviewed in Appendix~\ref{App:PropBuildingBlocks}) and only mention here that $H_{(s)}^{(1),\{0\}}$ are functions appearing in the instanton expansion of the free energy $\mathcal{F}_{N=1,1}$, while $W_{(0)}^{(1)}$ are expansion coefficients of a quasi-Jacobi form that governs the BPS-counting of an M-brane configuration of single M2-branes ending on either side of an M5-brane~\cite{Hohenegger:2015btj,Ahmed:2017hfr}. Most importantly, the functions $\mathcal{O}^{(N),\alpha}$ which encode the entire gauge structure of the free energy can be written as\footnote{We have adapted the notation for $\widehat{b}_{1,\ldots,N}$ relative to \cite{Hohenegger:2020slq} to conform with the definition (\ref{DefExternalPoints}).}
\begin{align}
\mathcal{O}^{(N),k}(\widehat{a}_{1,\ldots,N-1},\rho)=\frac{1}{(2\pi)^{2k}}\sum_{\ell=0}\sum_{\mathcal{S}\in\{1,\ldots,N\}\setminus\{\ell\}\atop{|\mathcal{S}|=k}}\prod_{j\in\mathcal{S}}\left(\mathbb{G}''(\widehat{b}_\ell-\widehat{b}_j;\rho)+\frac{2\pi i}{\rho-\bar{\rho}}\right)\,,\label{DefOoldPaper}
\end{align}
where a single term in the sum over $\mathcal{S}$ (\emph{i.e.} for a fixed $\ell$) is schematically represented in Figure~\ref{Fig:GraphRepOldPaper}. 

\noindent
The dashed lines in this Figure represent a single propagator of a free scalar field on a torus, whose argument (for fixed $\ell$) is determined by its end points
\begin{align}
\scalebox{1}{\parbox{2.5cm}{\begin{tikzpicture}[scale = 1.50]
\draw[fill=black] (0.19,0) circle (0.05);
\node at (0.2,0.3) {\footnotesize $\widehat{b}_{j}$};
\draw[fill=black] (-1.18,0) circle (0.05);
\node at (-1.15,-0.3) {\footnotesize $\widehat{b}_\ell$};
\draw[ultra thick,dashed] (-1.15,0) -- (0.15,0);
\end{tikzpicture}
}}=\frac{1}{(2\pi i)^2}\left(\mathbb{G}''(\widehat{b}_\ell-\widehat{b}_j;\rho)+\frac{2\pi i}{\rho-\bar{\rho}}\right)\,.\label{DefLinesPropoldPaper}
\end{align}
The right hand side of this object is identical to $\mathcal{I}_0$ defined in (\ref{DefI0E2}): as can be seen from this expression, however, it is a quasi-Jacobi form due to the appearance of $E_2$. With the representation given in Figure~\ref{Fig:GraphRepOldPaper}, the leading instanton contribution (\ref{DecompOldPaper}) therefore takes the form of a (tree-level) $N$-point function of  a free scalar field on the torus, while $H_{(s)}^{(1),\{0\}}$ and $W_{(0)}^{(1)}$ encode additional information for the 'external states' of these correlators. Higher instanton contributions were shown in \cite{Hohenegger:2019tii} to allow for a similar (albeit not unique) decomposition as (\ref{DecompOldPaper}), in particular separating the gauge structure and the parameter $S$. However, in addition\footnote{ A more systematic overview over these new elements is given in appendix~\ref{App:2PointProps}.} derivatives of propagator factors appear as well as multiplicative factors of (polynomials of) Eisenstein series, which were identified with (holomorphic) dihedral graph functions with bivalent vertices~\cite{DHoker:2015wxz,DHoker:2016mwo,DHoker:2017pvk,Zerbini:2017usf,Zerbini:2018hgs,Gerken:2018jrq,Gerken:2019cxz,Gerken:2020yii,Gerken:2020xte}. These contributions have been interpreted in \cite{Hohenegger:2020slq} as contributions stemming from integrated points in scalar field correlation functions.

\subsection{Further Expansion of the Partition Function}\label{Sect:ExpandPartitionFunctionExamples}
In this Subsection we discuss the relation between the diagrammatical decomposition of the partition function (\ref{RefPaaGraph}) (with the graphical representation shown in Figure~\ref{Fig:GraphRepPropsUnrefined}) to the decomposition (\ref{DefOoldPaper}) appearing in the free energy (as graphically represented in Figure~\ref{Fig:GraphRepOldPaper}) for the examples $N=2$ and $N=3$. Indeed, we shall show that (upon further decomposition) the instanton contributions to the partition function can be decomposed in a form resembling Figure~\ref{Fig:GraphRepOldPaper}. To this end, for convenience, we define a(n a priori formal) Laurent series expansion of $K^{N,(r)}$ in terms of the remaining deformation parameter $\epsilon$
\begin{align}
K^{N,(r)}(\rho,S,\widehat{a}_{1,\ldots,N-1};\epsilon_{1}=-\epsilon_2=\epsilon)=\sum_{k=0}^\infty \epsilon^{2k-2r}\,L_k^{N,(r)}(\rho,S,\widehat{a}_{1,\ldots,N})\,.\label{DefLdecomp}
\end{align} 
We shall discuss the functions $L_k^{N,(r)}(\rho,S,\widehat{a}_{1,\ldots,N})$ (predominantly for $k=0$) in more detail for $N=2$ and $N=3$ in Sections~\ref{Sect:Config21} and \ref{Sect:Config31} resepctively.

\subsubsection{Configuration $(2,1)$}\label{Sect:Config21}
Studying explicit examples\footnote{We have studied examples up to $r=6$ and $k=3$.} suggests that $L_k^{2,(r)}(\rho,S,\widehat{a}_1)$ in eq.~(\ref{DefLdecomp}) can be decomposed as
\begin{align}
&L^{2,(r)}_{k}(\rho,S,\widehat{a}_1)=\partfe{k}{(r)}(\rho,S)+\sum_{n=1}^\infty\left(\Qra^n+\frac{Q^n_\rho}{\Qra^n}\right)\frac{\partfen{k}{(r)}(\rho,S,n)}{1-Q_\rho^n}\,,\label{PartitionFunctionLeading}
\end{align}
where the $\partfe{k}{(r)}$ can be promoted to quasi-Jacobi forms of index $2r$ and weight $-2r+2k$ and the $\partfen{k}{(r)}$ are homogeneous polynomials in $(\fa,\fb)$ of order $2r$ with coefficients given by polynomials in $(E_2,E_4,E_6,n)$. It was remarked in \cite{Bastian:2019hpx,Bastian:2019wpx} that (in the case of the single-particle free energy) the form (\ref{PartitionFunctionLeading}) is compatible with a non-perturbative dihedral symmetry found in \cite{Bastian:2018jlf}. Concretely, for $r=1$ we find to leading orders in $k$
\begin{align}
&\partfe{0}{(1)}=-\frac{\fb}{12} (\fa+2E_2 \fb)\,,&&\partfe{1}{(1)}=\frac{\fa}{288} (\fa+2E_2 \fb)\,,&&\partfe{2}{(1)}=-\frac{E_4\,\fa}{2880} (\fa+2E_2\fb)\,,\nonumber\\
&\partfen{0}{(1)}=-2n\,\fb^2\,, &&\partfen{1}{(1)}=\frac{n}{12}\,\fa\,\fb\,,&&\partfen{2}{(1)}=\frac{n}{120}\,E_4\,\fb^2\,,\nonumber
\end{align}
and similarly for $r=2$
{\allowdisplaybreaks
\begin{align}
&\partfe{0}{(2)}=\frac{\fb^2}{288} (\fa^2+4\,E_2\,\fa\fb+4E_4\fb^2)\,,\nonumber\\
&\partfe{0}{(2)}=-\frac{\fb}{13824}\left(7\fa^3+28 E_2 \fa^2 \fb+28 E_4 \fa\fb^2+96(E_2 E_4-E_6)\fb^3\right)\,,\nonumber\\
&\partfen{2}{(1)}=\frac{n}{6}\,\fb^3\,(\fa+2n^2\fb)\,,\hspace{1cm}\partfen{1}{(2)}=-\frac{n\fb^2}{288}\left(7\fa^2+32 n^2 \fa\fb +12(n^4+2 E_4) \fb^2\right)\,.\nonumber
\end{align}}
Higher orders can be computed in the same fashion. Moreover, following the discussion of Section~\ref{Sect:FreeEnergy} and using the notation in Appendices~\ref{App:ExternalStates} and \ref{App:2PointProps}, we can re-organise (\ref{PartitionFunctionLeading}) 
{\allowdisplaybreaks\begin{align}
L^{2,(1)}_{0}(\rho,S,\widehat{a}_1)&=-2\,\sth{1}^2\,\,\scalebox{1}{\parbox{2.55cm}{\begin{tikzpicture}[scale = 1.50]
\draw[fill=black] (0.19,0) circle (0.05);
\draw[fill=black] (-1.19,0) circle (0.05);
\node at (-1.15,-0.25) {\footnotesize $\widehat{b}_1$};
\node at (0.2,0.25) {\footnotesize $\widehat{b}_2$};
\draw[ultra thick,dashed] (-1.15,0) -- (0.15,0);
\end{tikzpicture}
}}\,+\,2\,\sth{1}\,\stw{1}\,,\nonumber\\
L^{2,(2)}_{0}(\rho,S,\widehat{a}_1)&=\frac{1}{3}\,\sth{1}^4\,\,\scalebox{1}{\parbox{2.55cm}{\begin{tikzpicture}[scale = 1.50]
\draw[fill=black] (-1.19,0) circle (0.05);
\draw[fill=black] (0.19,0) circle (0.05);
\node at (-1.15,-0.25) {\footnotesize $\widehat{b}_1$};
\node at (0.2,0.25) {\footnotesize $\widehat{b}_2$};
\draw[ultra thick,dashed] (-1.15,0) -- (0.15,0);
\draw[ultra thick] (-0.9,0.1) -- (-0.7,-0.1);
\draw[ultra thick] (-0.9,-0.1) -- (-0.7,0.1);
\draw[ultra thick] (-0.1,0.1) -- (-0.3,-0.1);
\draw[ultra thick] (-0.1,-0.1) -- (-0.3,0.1);
\end{tikzpicture}}}\,-\,
\frac{8}{3}\,\sth{1}^2\,\sth{2}\,\scalebox{1}{\parbox{2.55cm}{\begin{tikzpicture}[scale = 1.50]
\draw[fill=black] (0.19,0) circle (0.05);
\draw[fill=black] (-1.19,0) circle (0.05);
\node at (-1.15,-0.25) {\footnotesize $\widehat{b}_1$};
\node at (0.2,0.25) {\footnotesize $\widehat{b}_2$};
\draw[ultra thick,dashed] (-1.15,0) -- (0.15,0);
\end{tikzpicture}
}}\,+\,\frac{4}{3}\,\sth{1}^2\,\stw{2}\,,\nonumber\\
L^{2,(3)}_{0}(\rho,S,\widehat{a}_1)&=-\frac{\sth{1}^6}{90}\,\scalebox{1}{\parbox{2.55cm}{\begin{tikzpicture}[scale = 1.50]
\draw[fill=black] (-1.19,0) circle (0.05);
\draw[fill=black] (0.19,0) circle (0.05);
\node at (-1.15,-0.25) {\footnotesize $\widehat{b}_1$};
\node at (0.2,0.25) {\footnotesize $\widehat{b}_2$};
\draw[ultra thick,dashed] (-1.15,0) -- (0.15,0);
\draw[ultra thick] (-1.1,0.1) -- (-0.9,-0.1);
\draw[ultra thick] (-1.1,-0.1) -- (-0.9,0.1);
\draw[ultra thick] (-0.8,0.1) -- (-0.6,-0.1);
\draw[ultra thick] (-0.8,-0.1) -- (-0.6,0.1);
\draw[ultra thick] (-0.45,0.1) -- (-0.25,-0.1);
\draw[ultra thick] (-0.45,-0.1) -- (-0.25,0.1);
\draw[ultra thick] (-0.1,0.1) -- (0.1,-0.1);
\draw[ultra thick] (-0.1,-0.1) -- (0.1,0.1);
\end{tikzpicture}
}}\,+\,
\frac{4}{9}\,\sth{1}^4\,\sth{2}\,\scalebox{1}{\parbox{2.55cm}{\begin{tikzpicture}[scale = 1.50]
\draw[fill=black] (-1.19,0) circle (0.05);
\draw[fill=black] (0.19,0) circle (0.05);
\node at (-1.15,-0.25) {\footnotesize $\widehat{b}_1$};
\node at (0.2,0.25) {\footnotesize $\widehat{b}_2$};
\draw[ultra thick,dashed] (-1.15,0) -- (0.15,0);
\draw[ultra thick] (-0.9,0.1) -- (-0.7,-0.1);
\draw[ultra thick] (-0.9,-0.1) -- (-0.7,0.1);
\draw[ultra thick] (-0.1,0.1) -- (-0.3,-0.1);
\draw[ultra thick] (-0.1,-0.1) -- (-0.3,0.1);
\end{tikzpicture}}}\,-\,
3\sth{1}^3\,\sth{3}\,\scalebox{1}{\parbox{2.55cm}{\begin{tikzpicture}[scale = 1.50]
\draw[fill=black] (0.19,0) circle (0.05);
\draw[fill=black] (-1.19,0) circle (0.05);
\node at (-1.15,-0.25) {\footnotesize $\widehat{b}_1$};
\node at (0.2,0.25) {\footnotesize $\widehat{b}_2$};
\draw[ultra thick,dashed] (-1.15,0) -- (0.15,0);
\end{tikzpicture}
}}\nonumber\\
&
\hspace{0.45cm}\,+\,\frac{\sth{1}^6}{15}\,E_4\,\scalebox{1}{\parbox{2.55cm}{\begin{tikzpicture}[scale = 1.50]
\draw[fill=black] (0.19,0) circle (0.05);
\draw[fill=black] (-1.19,0) circle (0.05);
\node at (-1.15,-0.25) {\footnotesize $\widehat{b}_1$};
\node at (0.2,0.25) {\footnotesize $\widehat{b}_2$};
\draw[ultra thick,dashed] (-1.15,0) -- (0.15,0);
\end{tikzpicture}
}}\,+\,\frac{\sth{1}^6}{60}\,\mathfrak{d}E_4\,+\,\sth{1}^3\stw{3}\,.\label{Decomp2ptN2}
\end{align}}
In the context of the free energy, the appearance of the Eisenstein series as explicit pre-factors (such as in the decomposition of $L_{0}^{2,(3)}$) has been interpreted as contributions from integrated insertion points and a connection to modular graph functions has been proposed. While several observations have been presented that support this proposal, no precise pattern for the decomposition of the free energy (notably for the numerical prefactors) has been found in \cite{Hohenegger:2020slq} and also (\ref{Decomp2ptN2}) a priori seems difficult to generalise to higher orders. However, the form presented in (\ref{RefPaaGraph}) (along with (\ref{PartFctN1})) for the partition function implicitly fixes all higher instanton terms. For $k=0$ this relationship\footnote{These relations can be generalised to higher $k$, but lead to more complicated relations than (\ref{FormRelDecN2}), which we shall not exhibit in detail in this work.} can be cast into a very compact form, using a few further observations regarding the partition function: 
\begin{itemize}
\item The various factors of Eisenstein series in (\ref{Decomp2ptN2}) appear in combinations that are manifestly Jacobi forms. For example, as explained in Appendix~\ref{App:Modular} (see eq.~(\ref{HolomorphicCombinations})), the contribution proportional to $\sth{1}^6$ in $L_0^{2,(3)}$ is $\frac{1}{60}\,\left(4 E_4\,\scalebox{1}{\parbox{1.05cm}{\begin{tikzpicture}[scale = 1.50]
\draw[fill=black] (0.19,0) circle (0.05);
\draw[fill=black] (-0.39,0) circle (0.05);
\draw[ultra thick,dashed] (-0.35,0) -- (0.15,0);
\end{tikzpicture}
}}+\mathfrak{d}E_4\right)$, which is a meromorphic Jacobi form of weight $6$ and index $0$: in an expansion in powers of $\widehat{a}_1$, all contributions of $E_2$ precisely cancel out. We have also verified that to higher orders (\emph{i.e.} $r>3$) only combinations of the form (\ref{HolomorphicCombinations}) (for $n>1$) appear.
\item Similarly, in all the combinations $\left(-r\,\sth{r}\,\scalebox{1}{\parbox{1.05cm}{\begin{tikzpicture}[scale = 1.50]
\draw[fill=black] (0.19,0) circle (0.05);
\draw[fill=black] (-0.39,0) circle (0.05);
\draw[ultra thick,dashed] (-0.35,0) -- (0.15,0);
\end{tikzpicture}
}}+\stw{r}\right)$ that appear in $L^{2,(r)}_{0}$ all Eisenstein series $E_2$ cancel out, rendering them meromorphic Jacobi forms. This condition in fact uniquely dictates the contribution of any of the states $\stw{r}$ to the partititon function at this order of $\epsilon$. Since the derivatives of the propagator (\ref{PropDer}) are meromorphic Jacobi forms, this suggests that the expressions in (\ref{Decomp2ptN2}) are also meromorphic Jacobi forms.
\end{itemize}
This last property can in fact be made manifest: we first use the explicit form of $\stw{1}$ in (\ref{FormW1}) to write for the leading instanton contribution 
\begin{align}
L_{0}^{2,(1)}&=-2\,\sth{1}^2\left[\scalebox{1}{\parbox{1.8cm}{\begin{tikzpicture}[scale = 1.50]
\draw[fill=black] (0.19,0) circle (0.05);
\draw[fill=black] (-0.69,0) circle (0.05);
\node at (-0.65,-0.25) {\footnotesize $\widehat{b}_1$};
\node at (0.2,0.25) {\footnotesize $\widehat{b}_2$};
\draw[ultra thick,dashed] (-0.65,0) -- (0.15,0);
\end{tikzpicture}
}}+\frac{1}{24}\left(\frac{\fa}{\fb}+2\,E_2\right)\right]=-2\,\sth{1}^2\left[\scalebox{1}{\parbox{1.8cm}{\begin{tikzpicture}[scale = 1.50]
\draw[fill=black] (0.19,0) circle (0.05);
\draw[fill=black] (-0.69,0) circle (0.05);
\node at (-0.65,-0.25) {\footnotesize $\widehat{b}_1$};
\node at (0.2,0.25) {\footnotesize $\widehat{b}_2$};
\draw[ultra thick,dashed] (-0.65,0) -- (0.15,0);
\end{tikzpicture}
}}+\frac{E_2}{12}+\frac{\wp(S;\rho)}{24\zeta(2)}\right]\nonumber\\
&=\frac{\sth{1}^2}{(2\pi)^2}\,\left(\scalebox{1}{\parbox{2.4cm}{\begin{tikzpicture}[scale = 1.50]
\draw (0.19,0) circle (0.05);
\draw (-0.69,0) circle (0.05);
\node at (-0.7,-0.2) {\footnotesize $(\widehat{b}_1,\emptyset)$};
\node at (0.15,0.25) {\footnotesize $(\widehat{b}_2,\Box)$};
\draw[ultra thick] (-0.65,0) -- (0.15,0);
\end{tikzpicture}}
}\,+\,\scalebox{1}{\parbox{2.4cm}{\begin{tikzpicture}[scale = 1.50]
\draw (0.19,0) circle (0.05);
\draw (-0.69,0) circle (0.05);
\node at (-0.7,-0.2) {\footnotesize $(\widehat{b}_1,\Box)$};
\node at (0.15,0.25) {\footnotesize $(\widehat{b}_2,\emptyset)$};
\draw[ultra thick] (-0.65,0) -- (0.15,0);
\end{tikzpicture}}
}\right)\bigg|_{\epsilon=0}\,.
\end{align}
with the generalised propagator $\scalebox{1}{\parbox{1.05cm}{\begin{tikzpicture}[scale = 1.50]
\draw (0.19,0) circle (0.05);
\draw (-0.39,0) circle (0.05);
\draw[ultra thick] (-0.35,0) -- (0.15,0);
\end{tikzpicture}
}}$ defined in (\ref{DefLinesPropUnrefined}). The recursive structure (\ref{RelGenKN}) then leads to the pattern
\begin{align}
L_{0}^{2,(r)}(\rho,S,\widehat{a}_1)=\frac{1}{r!}\,\left[\frac{\sth{1}}{2\pi}\,\scalebox{1}{\parbox{2.4cm}{\begin{tikzpicture}[scale = 1.50]
\draw (0.19,0) circle (0.05);
\draw (-0.69,0) circle (0.05);
\node at (-0.7,-0.2) {\footnotesize $(\widehat{b}_1,\emptyset)$};
\node at (0.15,0.25) {\footnotesize $(\widehat{b}_2,\Box)$};
\draw[ultra thick] (-0.65,0) -- (0.15,0);
\end{tikzpicture}}
}\,\frac{\sth{1}}{2\pi}\hspace{0.3cm}+\hspace{0.3cm}\frac{\sth{1}}{2\pi}\,\scalebox{1}{\parbox{2.4cm}{\begin{tikzpicture}[scale = 1.50]
\draw (0.19,0) circle (0.05);
\draw (-0.69,0) circle (0.05);
\node at (-0.7,-0.2) {\footnotesize $(\widehat{b}_1,\Box)$};
\node at (0.15,0.25) {\footnotesize $(\widehat{b}_2,\emptyset)$};
\draw[ultra thick] (-0.65,0) -- (0.15,0);
\end{tikzpicture}}
}\,\frac{\sth{1}}{2\pi}\right]^r\bigg|_{\epsilon=0}\,.\label{FormRelDecN2}
\end{align}


\subsubsection{Configuration $(3,1)$}\label{Sect:Config31}
For simplicity, we shall restrict ourselves to the coefficients $L_{0}^{3,(r)}$ in (\ref{PartitionFunctionLeading}). Studying low values of $r$ suggest that it can be decomposed in the following form
\begin{align}
L_{0}^{3,(r)}(&\rho,S,\widehat{a}_1,\widehat{a}_2)=\partfe{0}{(r)}(\rho,S)\nonumber\\
&+\sum_{n=1}^\infty\left(\Qna{1}^n+\Qna{2}^n+\left(\Qna{1}\Qna{2}\right)^n+\frac{Q^n_\rho}{\Qna{1}^n}+\frac{Q^n_\rho}{\Qna{2}^n}+\frac{Q^n_\rho}{(\Qna{1}\Qna{2})^n}\right)\frac{\partfen{0}{(r)}(\rho,S,n)}{1-Q_\rho^n}\nonumber\\
&+\sum_{n=1}^\infty\left[Q_\rho^n\left(\Qna{1}^n+\Qna{2}^n+\frac{Q^n_\rho}{(\Qna{1}\Qna{2})^n}\right)+\left(\Qna{1}\Qna{2}\right)^n+\frac{Q^n_\rho}{\Qna{1}^n}+\frac{Q^n_\rho}{\Qna{2}^n}\right]\frac{\partfenz{0}{(r)}(\rho,S,n)}{(1-Q_\rho^n)^2}\nonumber\\
&+\sum_{n_1,n_2=1}^\infty\left(\Qna{1}^{n_1+n_2}\Qna{2}^{n_1}+\Qna{2}^{n_1+n_2}\Qna{1}^{n_1}+\frac{Q_\rho^{n_1+n_2}}{\Qna{1}^{n_1+n_2}\Qna{2}^{n_1}}+\frac{Q_\rho^{n_1+n_2}}{\Qna{2}^{n_1+n_2}\Qna{1}^{n_1}}+\frac{Q_\rho^{n_1}\Qna{1}^{n_2}}{\Qna{2}^{n_1}}+\frac{Q_\rho^{n_1}\Qna{2}^{n_2}}{\Qna{1}^{n_1}}\right)\nonumber\\
&\hspace{1.5cm}\times\left(\frac{\partfend{0}{(r)}(\rho,S,n_1,n_2) n_2(2n_1+n_2)}{(1-Q_\rho^{n_1})(1-Q_\rho^{n_1+n_2})}+\frac{\partfend{0}{(r)}(\rho,S,n_1,-n_1-n_2)\,(n_1^2-n_2^2)}{(1-Q_\rho^{n_1})(1-Q_\rho^{n_2})}\right)\,.\label{Part31}
\end{align}
Concretely, we find for $r=1$
\begin{align}
&\partfe{0}{(1)}=-\frac{\fb}{192}\left(\fa+2E_2\fb\right)^2\,,&&\partfen{0}{(1)}=-\frac{n\fb^2}{12}\left(\fa+2E_2\fb\right)\,,
&&\partfenz{0}{(1)}=-\fb^3\,n^2\,,&&\partfend{0}{(1)}=-\fb^3\,,\nonumber
\end{align}
and for $r=2$
{\allowdisplaybreaks
\begin{align}
\partfe{0}{(2)}&=\fb^2\bigg[\frac{5E_4^2-8E_2E_4+6E_4E_2^2}{13824}\fb^4-\frac{2E_6-5E_2E_4}{6912}\fa\fb^2+\frac{2E_4+7E_2^2}{27648}\fa^2\fb^2\nonumber\\
&\hspace{1.6cm}+\frac{E_2}{9216}\fa^3\fb+\frac{\fa^4}{73728}\bigg]\,,\nonumber\\
\partfen{0}{(1)}&=n\,\fb^3\bigg[\frac{2E_4 n^2-2E_6+3 E_2 E_4}{864}\,\fb^3+\frac{5E_4+4 n^2 E_2}{1728}\,\fa\fb^2+\frac{7E_2+2n^2}{3456}\,\fb\fa^2+\frac{\fa^3}{2304}\bigg]\,,\nonumber\\
\partfenz{0}{(2)}&=n\,\fb^4\left[\frac{n(3E_4+4n^4)}{144}\,\fb^2+\frac{7n^3}{144}\,\fb\fa+\frac{7n}{576}\,\fa^2\right]\,,\nonumber\\
\partfend{0}{(2)}&=\fb^4\left[\left(\frac{E_4}{48}+\frac{(n_1+n_2)^2(n_1^2+n_1n_2+n_2^2)}{36}\right)\fb^2+\frac{7n_1^2+12n_1n_2+7n_2^2}{144}\,\fa\fb+\frac{7\fa^2}{576}\right]\,.\nonumber
\end{align}}
Using the notation of Appendix~\ref{App:3PointProps}, we can write (\ref{Part31}) for $r=1$
\begin{align}
L_{0}^{3,(1)}&=\sth{1}^{3}\hspace{0.1cm}\left(\,\scalebox{1}{\parbox{2.2cm}{\begin{tikzpicture}[scale = 1.50]
\draw[fill=black] (-0.89,0) circle (0.05);
\draw[fill=black] (-0.01,0) circle (0.05);
\draw[fill=black] (-0.48,0.69) circle (0.05);
\node at (-1,-0.25) {\footnotesize $\widehat{b}_1$};
\node at (0.1,-0.25) {\footnotesize $\widehat{b}_2$};
\node at (-0.48,0.95) {\footnotesize $\widehat{b}_3$};
\draw[ultra thick,dashed] (-0.85,0) -- (-0.05,0);
\draw[ultra thick,dashed] (-0.89,0) -- (-0.48,0.69);
\end{tikzpicture}
}}\hspace{0.25cm}+\hspace{0.25cm}\scalebox{1}{\parbox{2.2cm}{\begin{tikzpicture}[scale = 1.50]
\draw[fill=black] (-0.89,0) circle (0.05);
\draw[fill=black] (-0.01,0) circle (0.05);
\draw[fill=black] (-0.48,0.69) circle (0.05);
\node at (-1,-0.25) {\footnotesize $\widehat{b}_1$};
\node at (0.1,-0.25) {\footnotesize $\widehat{b}_2$};
\node at (-0.48,0.95) {\footnotesize $\widehat{b}_3$};
\draw[ultra thick,dashed] (-0.01,0) -- (-0.48,0.69);
\draw[ultra thick,dashed] (-0.85,0) -- (-0.05,0);
\end{tikzpicture}
}}\hspace{0.25cm}+\hspace{0.25cm}\scalebox{1}{\parbox{2.2cm}{\begin{tikzpicture}[scale = 1.50]
\draw[fill=black] (-0.89,0) circle (0.05);
\draw[fill=black] (-0.01,0) circle (0.05);
\draw[fill=black] (-0.48,0.69) circle (0.05);
\node at (-1,-0.25) {\footnotesize $\widehat{b}_1$};
\node at (0.1,-0.25) {\footnotesize $\widehat{b}_2$};
\node at (-0.48,0.95) {\footnotesize $\widehat{b}_3$};
\draw[ultra thick,dashed] (-0.89,0) -- (-0.48,0.69);
\draw[ultra thick,dashed] (-0.01,0) -- (-0.48,0.69);
\end{tikzpicture}
}}\,\right)\nonumber\\
&\hspace{0.5cm}-2\sth{1}^2\,\stw{1}\hspace{0.1cm}\left(\,\scalebox{1}{\parbox{2.2cm}{\begin{tikzpicture}[scale = 1.50]
\draw[fill=black] (-0.89,0) circle (0.05);
\draw[fill=black] (-0.01,0) circle (0.05);
\draw[fill=black] (-0.48,0.69) circle (0.05);
\node at (-1,-0.25) {\footnotesize $\widehat{b}_1$};
\node at (0.1,-0.25) {\footnotesize $\widehat{b}_2$};
\node at (-0.48,0.95) {\footnotesize $\widehat{b}_3$};
\draw[ultra thick,dashed] (-0.85,0) -- (-0.05,0);
\end{tikzpicture}
}}\hspace{0.25cm}+\hspace{0.25cm}\scalebox{1}{\parbox{2.2cm}{\begin{tikzpicture}[scale = 1.50]
\draw[fill=black] (-0.89,0) circle (0.05);
\draw[fill=black] (-0.01,0) circle (0.05);
\draw[fill=black] (-0.48,0.69) circle (0.05);
\node at (-1,-0.25) {\footnotesize $\widehat{b}_1$};
\node at (0.1,-0.25) {\footnotesize $\widehat{b}_2$};
\node at (-0.48,0.95) {\footnotesize $\widehat{b}_3$};
\draw[ultra thick,dashed] (-0.01,0) -- (-0.48,0.69);
\end{tikzpicture}
}}\hspace{0.25cm}+\hspace{0.25cm}\scalebox{1}{\parbox{2.2cm}{\begin{tikzpicture}[scale = 1.50]
\draw[fill=black] (-0.89,0) circle (0.05);
\draw[fill=black] (-0.01,0) circle (0.05);
\draw[fill=black] (-0.48,0.69) circle (0.05);
\node at (-1,-0.25) {\footnotesize $\widehat{b}_1$};
\node at (0.1,-0.25) {\footnotesize $\widehat{b}_2$};
\node at (-0.48,0.95) {\footnotesize $\widehat{b}_3$};
\draw[ultra thick,dashed] (-0.89,0) -- (-0.48,0.69);
\end{tikzpicture}
}}\,\right)+3\sth{1}\,\stw{1}^2\,,\label{N3LeadingInstanton}
\end{align}
with the various elements defined in (\ref{Tier1Correlators2Prop}). Moreover, using the definition (\ref{DefLinesPropUnrefined}), as well as the definitions in Appendix~\ref{App:PropBuildingBlocks} we can equally represent this partition function as 
\begin{align}
L_{0}^{3,(1)}=\frac{\sth{1}^{3}}{(2\pi)^3}\hspace{0.1cm}\left(\,\scalebox{1}{\parbox{2.8cm}{\begin{tikzpicture}[scale = 1.50]
\draw (-0.89,0) circle (0.05);
\draw (-0.01,0) circle (0.05);
\draw (-0.48,0.69) circle (0.05);
\node at (-1.0,-0.25) {\footnotesize $(\widehat{b}_1,\Box)$};
\node at (0.1,-0.25) {\footnotesize $(\widehat{b}_2,\emptyset)$};
\node at (-0.48,0.95) {\footnotesize $(\widehat{b}_3,\emptyset)$};
\draw[ultra thick] (-0.87,0.04) -- (-0.5,0.65);
\draw[ultra thick] (-0.04,0.04) -- (-0.46,0.65);
\draw[ultra thick] (-0.845,0.) -- (-0.055,0);
\end{tikzpicture}
}}\hspace{0.25cm}+\hspace{0.25cm}
\scalebox{1}{\parbox{2.8cm}{\begin{tikzpicture}[scale = 1.50]
\draw (-0.89,0) circle (0.05);
\draw (-0.01,0) circle (0.05);
\draw (-0.48,0.69) circle (0.05);
\node at (-1.0,-0.25) {\footnotesize $(\widehat{b}_1,\emptyset)$};
\node at (0.1,-0.25) {\footnotesize $(\widehat{b}_2,\Box)$};
\node at (-0.48,0.95) {\footnotesize $(\widehat{b}_3,\emptyset)$};
\draw[ultra thick] (-0.87,0.04) -- (-0.5,0.65);
\draw[ultra thick] (-0.04,0.04) -- (-0.46,0.65);
\draw[ultra thick] (-0.845,0.) -- (-0.055,0);
\end{tikzpicture}
}}
\hspace{0.25cm}+\hspace{0.25cm}
\scalebox{1}{\parbox{2.8cm}{\begin{tikzpicture}[scale = 1.50]
\draw (-0.89,0) circle (0.05);
\draw (-0.01,0) circle (0.05);
\draw (-0.48,0.69) circle (0.05);
\node at (-1.0,-0.25) {\footnotesize $(\widehat{b}_1,\emptyset)$};
\node at (0.1,-0.25) {\footnotesize $(\widehat{b}_2,\emptyset)$};
\node at (-0.48,0.95) {\footnotesize $(\widehat{b}_3,\Box)$};
\draw[ultra thick] (-0.87,0.04) -- (-0.5,0.65);
\draw[ultra thick] (-0.04,0.04) -- (-0.46,0.65);
\draw[ultra thick] (-0.845,0.) -- (-0.055,0);
\end{tikzpicture}
}}
\,\right)_{\epsilon=0}\,.\label{N3LeadingInstantonSimple}
\end{align}
We note that the lines connecting points that both have the trivial partition $\emptyset$ associated with them, are trivial factors of $1$.

We briefly remark that similar decompositions of the partition function can also be written to higher instanton order. For example, using the shorthand notation explained in Appendix~\ref{App:3PointProps}, the two-instanton contribution can be written in the form 
\begin{align}
L_{0}^{3,(1)}&=\frac{\sth{1}^6}{72}\,\scalebox{1}{\parbox{2.8cm}{\begin{tikzpicture}[scale = 1.50]
\draw[fill=black] (-1.19,0) circle (0.05);
\draw[fill=black] (0.19,0) circle (0.05);
\draw[fill=black] (-0.48,1.19) circle (0.05);
\node at (-1.2,-0.25) {\footnotesize $\widehat{b}_1$};
\node at (0.3,-0.25) {\footnotesize $\widehat{b}_2$};
\node at (-0.48,1.45) {\footnotesize $\widehat{b}_3$};
\draw[thick,domain=0:-300,<-] plot ({0.15*cos(\x)-0.48},{0.15*sin(\x)+0.5});
\node at (-0.48,0.5) {\tiny $\sum$};
\draw[ultra thick,dashed] (-1.175,0.05) -- (-0.5,1.15);
\draw[ultra thick,dashed] (0.175,0.05) -- (-0.46,1.15);
\begin{scope}[rotate=60,xshift=1.1cm,yshift=1.01cm]
\draw[ultra thick] (-0.9,0.1) -- (-0.7,-0.1);
\draw[ultra thick] (-0.9,-0.1) -- (-0.7,0.1);
\end{scope}
\begin{scope}[rotate=60,xshift=0.7cm,yshift=1.01cm]
\draw[ultra thick] (-0.9,0.1) -- (-0.7,-0.1);
\draw[ultra thick] (-0.9,-0.1) -- (-0.7,0.1);
\end{scope}
\begin{scope}[rotate=-60,xshift=0.cm,yshift=0.175cm]
\draw[ultra thick] (-0.9,0.1) -- (-0.7,-0.1);
\draw[ultra thick] (-0.9,-0.1) -- (-0.7,0.1);
\end{scope}
\begin{scope}[rotate=-60,xshift=0.4cm,yshift=0.175cm]
\draw[ultra thick] (-0.9,0.1) -- (-0.7,-0.1);
\draw[ultra thick] (-0.9,-0.1) -- (-0.7,0.1);
\end{scope}
\end{tikzpicture}
}}\hspace{0.2cm}+\hspace{0.2cm}\frac{4}{3}\sth{1}^4\sth{2}\,\scalebox{1}{\parbox{2.8cm}{\begin{tikzpicture}[scale = 1.50]
\draw[fill=black] (-1.19,0) circle (0.05);
\draw[fill=black] (0.19,0) circle (0.05);
\draw[fill=black] (-0.48,1.19) circle (0.05);
\node at (-1.2,-0.25) {\footnotesize $\widehat{b}_1$};
\node at (0.3,-0.25) {\footnotesize $\widehat{b}_2$};
\node at (-0.48,1.45) {\footnotesize $\widehat{b}_3$};
\draw[thick,domain=0:-300,<-] plot ({0.15*cos(\x)-0.48},{0.15*sin(\x)+0.5});
\node at (-0.48,0.5) {\tiny $\sum$};
\draw[ultra thick,dashed] (-1.15,0) -- (0.15,0);
\draw[ultra thick,dashed] (-1.175,0.05) -- (-0.5,1.15);
\draw[ultra thick,dashed] (0.175,0.05) -- (-0.46,1.15);
\end{tikzpicture}
}}
\hspace{0.2cm}+\hspace{0.2cm}\frac{2}{9}\,\sth{1}^4\sth{2}\,\scalebox{1}{\parbox{2.8cm}{\begin{tikzpicture}[scale = 1.50]
\draw[fill=black] (-1.19,0) circle (0.05);
\draw[fill=black] (0.19,0) circle (0.05);
\draw[fill=black] (-0.48,1.19) circle (0.05);
\node at (-1.2,-0.25) {\footnotesize $\widehat{b}_1$};
\node at (0.3,-0.25) {\footnotesize $\widehat{b}_2$};
\node at (-0.48,1.45) {\footnotesize $\widehat{b}_3$};
\draw[thick,domain=0:-300,<-] plot ({0.15*cos(\x)-0.48},{0.15*sin(\x)+0.5});
\node at (-0.48,0.5) {\tiny $\sum$};
\draw[ultra thick,dashed] (-1.175,0.05) -- (-0.5,1.15);
\draw[ultra thick,dashed] (0.175,0.05) -- (-0.46,1.15);
\begin{scope}[rotate=60,xshift=1.1cm,yshift=1.01cm]
\draw[ultra thick] (-0.9,0.1) -- (-0.7,-0.1);
\draw[ultra thick] (-0.9,-0.1) -- (-0.7,0.1);
\end{scope}
\begin{scope}[rotate=60,xshift=0.7cm,yshift=1.01cm]
\draw[ultra thick] (-0.9,0.1) -- (-0.7,-0.1);
\draw[ultra thick] (-0.9,-0.1) -- (-0.7,0.1);
\end{scope}
\end{tikzpicture}
}}\nonumber\\
&\hspace{0.2cm}+\hspace{0.2cm}\frac{2}{9}\sth{1}^2\left(8\sth{2}^2+9\sth{1}^2\sth{2}\right)\,\scalebox{1}{\parbox{2.8cm}{\begin{tikzpicture}[scale = 1.50]
\draw[fill=black] (-1.19,0) circle (0.05);
\draw[fill=black] (0.19,0) circle (0.05);
\draw[fill=black] (-0.48,1.19) circle (0.05);
\node at (-1.2,-0.25) {\footnotesize $\widehat{b}_1$};
\node at (0.3,-0.25) {\footnotesize $\widehat{b}_2$};
\node at (-0.48,1.45) {\footnotesize $\widehat{b}_3$};
\draw[thick,domain=0:-300,<-] plot ({0.15*cos(\x)-0.48},{0.15*sin(\x)+0.5});
\node at (-0.48,0.5) {\tiny $\sum$};
\draw[ultra thick,dashed] (-1.175,0.05) -- (-0.5,1.15);
\draw[ultra thick,dashed] (0.175,0.05) -- (-0.46,1.15);
%
%
\end{tikzpicture}
}}\nonumber\\
&\hspace{0.2cm}+\hspace{0.2cm}\text{single 2-point functions}+\text{terms independent of $\widehat{a}_{1,2}$}
\end{align}
(All) these combinations can be equally represented in the form
\begin{align}
L_{0}^{3,(2)}&=\frac{\sth{1}^{6}}{(2\pi)^6}\hspace{0.1cm}\left(\,\scalebox{1}{\parbox{2.8cm}{\begin{tikzpicture}[scale = 1.50]
\draw (-0.89,0) circle (0.05);
\draw (-0.01,0) circle (0.05);
\draw (-0.48,0.69) circle (0.05);
\node at (-1.0,-0.25) {\footnotesize $(\widehat{b}_1,\Box)$};
\node at (0.1,-0.25) {\footnotesize $(\widehat{b}_2,\Box)$};
\node at (-0.48,0.95) {\footnotesize $(\widehat{b}_3,\emptyset)$};
\draw[ultra thick] (-0.87,0.04) -- (-0.5,0.65);
\draw[ultra thick] (-0.04,0.04) -- (-0.46,0.65);
\draw[ultra thick] (-0.845,0.) -- (-0.055,0);
\end{tikzpicture}
}}\hspace{0.25cm}+\hspace{0.25cm}
\scalebox{1}{\parbox{2.8cm}{\begin{tikzpicture}[scale = 1.50]
\draw (-0.89,0) circle (0.05);
\draw (-0.01,0) circle (0.05);
\draw (-0.48,0.69) circle (0.05);
\node at (-1.0,-0.25) {\footnotesize $(\widehat{b}_1,\emptyset)$};
\node at (0.1,-0.25) {\footnotesize $(\widehat{b}_2,\Box)$};
\node at (-0.48,0.95) {\footnotesize $(\widehat{b}_3,\Box)$};
\draw[ultra thick] (-0.87,0.04) -- (-0.5,0.65);
\draw[ultra thick] (-0.04,0.04) -- (-0.46,0.65);
\draw[ultra thick] (-0.845,0.) -- (-0.055,0);
\end{tikzpicture}
}}
\hspace{0.25cm}+\hspace{0.25cm}
\scalebox{1}{\parbox{2.8cm}{\begin{tikzpicture}[scale = 1.50]
\draw (-0.89,0) circle (0.05);
\draw (-0.01,0) circle (0.05);
\draw (-0.48,0.69) circle (0.05);
\node at (-1.0,-0.25) {\footnotesize $(\widehat{b}_1,\Box)$};
\node at (0.1,-0.25) {\footnotesize $(\widehat{b}_2,\emptyset)$};
\node at (-0.48,0.95) {\footnotesize $(\widehat{b}_3,\Box)$};
\draw[ultra thick] (-0.87,0.04) -- (-0.5,0.65);
\draw[ultra thick] (-0.04,0.04) -- (-0.46,0.65);
\draw[ultra thick] (-0.845,0.) -- (-0.055,0);
\end{tikzpicture}
}}
\,\right)_{\epsilon=0}\nonumber\\
&\hspace{-0.05cm}+\frac{\sth{1}^{6}}{2(2\pi)^6}\hspace{0.1cm}\left(\,\scalebox{1}{\parbox{3cm}{\begin{tikzpicture}[scale = 1.50]
\draw (-0.89,0) circle (0.05);
\draw (-0.01,0) circle (0.05);
\draw (-0.48,0.69) circle (0.05);
\node at (-1.0,-0.25) {\footnotesize $(\widehat{b}_1,\scalebox{0.4}{\parbox{1.3cm}{\vspace{-0.4cm}\ydiagram{2}}})$};
\node at (0.1,-0.25) {\footnotesize $(\widehat{b}_2,\emptyset)$};
\node at (-0.48,0.95) {\footnotesize $(\widehat{b}_3,\emptyset)$};
\draw[ultra thick] (-0.87,0.04) -- (-0.5,0.65);
\draw[ultra thick] (-0.04,0.04) -- (-0.46,0.65);
\draw[ultra thick] (-0.845,0.) -- (-0.055,0);
\end{tikzpicture}
}}\hspace{0.25cm}+\hspace{0.25cm}
\scalebox{1}{\parbox{3cm}{\begin{tikzpicture}[scale = 1.50]
\draw (-0.89,0) circle (0.05);
\draw (-0.01,0) circle (0.05);
\draw (-0.48,0.69) circle (0.05);
\node at (-1.0,-0.25) {\footnotesize $(\widehat{b}_1,\emptyset)$};
\node at (0.1,-0.25) {\footnotesize $(\widehat{b}_2,\scalebox{0.4}{\parbox{1.3cm}{\vspace{-0.4cm}\ydiagram{2}}})$};
\node at (-0.48,0.95) {\footnotesize $(\widehat{b}_3,\emptyset)$};
\draw[ultra thick] (-0.87,0.04) -- (-0.5,0.65);
\draw[ultra thick] (-0.04,0.04) -- (-0.46,0.65);
\draw[ultra thick] (-0.845,0.) -- (-0.055,0);
\end{tikzpicture}
}}
\hspace{0.25cm}+\hspace{0.25cm}
\scalebox{1}{\parbox{2.8cm}{\begin{tikzpicture}[scale = 1.50]
\draw (-0.89,0) circle (0.05);
\draw (-0.01,0) circle (0.05);
\draw (-0.48,0.69) circle (0.05);
\node at (-1.0,-0.25) {\footnotesize $(\widehat{b}_1,\emptyset)$};
\node at (0.1,-0.25) {\footnotesize $(\widehat{b}_2,\emptyset)$};
\node at (-0.48,0.95) {\footnotesize $(\widehat{b}_3,\scalebox{0.4}{\parbox{1.3cm}{\vspace{-0.4cm}\ydiagram{2}}})$};
\draw[ultra thick] (-0.87,0.04) -- (-0.5,0.65);
\draw[ultra thick] (-0.04,0.04) -- (-0.46,0.65);
\draw[ultra thick] (-0.845,0.) -- (-0.055,0);
\end{tikzpicture}
}}
\,\right)_{\epsilon=0}\,.\label{N3HigherInstantonSimple}
\end{align}
We have furthermore checked, that this result is compatible with the recursive structure (\ref{RelGenKN})
\begin{align}
L_{0}^{3,(r)}=\frac{\sth{1}^{3r}}{(2\pi)^{3r}r!}\hspace{0.1cm}\left(\,\scalebox{1}{\parbox{2.8cm}{\begin{tikzpicture}[scale = 1.50]
\draw (-0.89,0) circle (0.05);
\draw (-0.01,0) circle (0.05);
\draw (-0.48,0.69) circle (0.05);
\node at (-1.0,-0.25) {\footnotesize $(\widehat{b}_1,\Box)$};
\node at (0.1,-0.25) {\footnotesize $(\widehat{b}_2,\emptyset)$};
\node at (-0.48,0.95) {\footnotesize $(\widehat{b}_3,\emptyset)$};
\draw[ultra thick] (-0.87,0.04) -- (-0.5,0.65);
\draw[ultra thick] (-0.04,0.04) -- (-0.46,0.65);
\draw[ultra thick] (-0.845,0.) -- (-0.055,0);
\end{tikzpicture}
}}\hspace{0.25cm}+\hspace{0.25cm}
\scalebox{1}{\parbox{2.8cm}{\begin{tikzpicture}[scale = 1.50]
\draw (-0.89,0) circle (0.05);
\draw (-0.01,0) circle (0.05);
\draw (-0.48,0.69) circle (0.05);
\node at (-1.0,-0.25) {\footnotesize $(\widehat{b}_1,\emptyset)$};
\node at (0.1,-0.25) {\footnotesize $(\widehat{b}_2,\Box)$};
\node at (-0.48,0.95) {\footnotesize $(\widehat{b}_3,\emptyset)$};
\draw[ultra thick] (-0.87,0.04) -- (-0.5,0.65);
\draw[ultra thick] (-0.04,0.04) -- (-0.46,0.65);
\draw[ultra thick] (-0.845,0.) -- (-0.055,0);
\end{tikzpicture}
}}
\hspace{0.25cm}+\hspace{0.25cm}
\scalebox{1}{\parbox{2.8cm}{\begin{tikzpicture}[scale = 1.50]
\draw (-0.89,0) circle (0.05);
\draw (-0.01,0) circle (0.05);
\draw (-0.48,0.69) circle (0.05);
\node at (-1.0,-0.25) {\footnotesize $(\widehat{b}_1,\emptyset)$};
\node at (0.1,-0.25) {\footnotesize $(\widehat{b}_2,\emptyset)$};
\node at (-0.48,0.95) {\footnotesize $(\widehat{b}_3,\Box)$};
\draw[ultra thick] (-0.87,0.04) -- (-0.5,0.65);
\draw[ultra thick] (-0.04,0.04) -- (-0.46,0.65);
\draw[ultra thick] (-0.845,0.) -- (-0.055,0);
\end{tikzpicture}
}}
\,\right)^r_{\epsilon=0}\,.
\end{align}


\subsection{Recursive Structures and Product Relations}
In the previous Section we have shown how the (combinations of) two-point functions defined in eq.~(\ref{DefLinesPropUnrefined}) can be related to the propagators proposed in \cite{Hohenegger:2020slq} (see eq.~(\ref{DefLinesPropoldPaper})) by expanding leading contributions of the partition function in both ways for $N=2$ and $N=3$. To make contact to similar results at the level of the free energy $\mathcal{F}_{N,1}=\ln \mathcal{Z}_{N,1}$ requires to (re-)expand products of the propagators of the type in eq.~(\ref{DefLinesPropoldPaper}), with the same arguments. By comparing the series expansions in $\widehat{b}_j-\widehat{b}_i$ (assuming convergence of the series) we find for example
\begin{align}
\left(\scalebox{1}{\parbox{1.85cm}{\begin{tikzpicture}[scale = 1.50]
\draw[fill=black] (0,0) circle (0.05);
\node at (-0.01,0.25) {\footnotesize $\widehat{b}_{j}$};
\draw[fill=black] (-0.89,0) circle (0.05);
\node at (-0.89,-0.25) {\footnotesize $\widehat{b}_i$};
\draw[ultra thick,dashed] (-0.89,0) -- (-0.01,0);
\end{tikzpicture}
}}\right)^2\hspace{0.25cm}=\hspace{0.25cm}\frac{1}{6}\,\scalebox{1}{\parbox{1.85cm}{\begin{tikzpicture}[scale = 1.50]
\draw[fill=black] (0,0) circle (0.05);
\node at (-0.01,0.25) {\footnotesize $\widehat{b}_{j}$};
\draw[fill=black] (-0.89,0) circle (0.05);
\node at (-0.89,-0.25) {\footnotesize $\widehat{b}_i$};
\draw[ultra thick,dashed] (-0.89,0) -- (-0.01,0);
\draw[ultra thick] (-0.7,0.1) -- (-0.5,-0.1);
\draw[ultra thick] (-0.7,-0.1) -- (-0.5,0.1);
\draw[ultra thick] (-0.2,0.1) -- (-0.4,-0.1);
\draw[ultra thick] (-0.2,-0.1) -- (-0.4,0.1);
\end{tikzpicture}
}}\hspace{0.25cm}-\hspace{0.25cm}\frac{E_2}{6}
\scalebox{1}{\parbox{1.85cm}{\begin{tikzpicture}[scale = 1.50]
\draw[fill=black] (0,0) circle (0.05);
\node at (-0.01,0.25) {\footnotesize $\widehat{b}_{j}$};
\draw[fill=black] (-0.89,0) circle (0.05);
\node at (-0.89,-0.25) {\footnotesize $\widehat{b}_i$};
\draw[ultra thick,dashed] (-0.89,0) -- (-0.01,0);
\end{tikzpicture}
}}
\hspace{0.25cm}-\hspace{0.25cm}\frac{E_4-E_2^2}{144}\,.
\end{align}
Formally (\emph{i.e.} by comparing the leading expansion terms), we can generalise these relations to products of the infinite series $\mathcal{I}_{k_1}\,\mathcal{I}_{k_2}$ (for integers $k_{1,2}\geq 0$), as defined in eq.~(\ref{DefSeriesIk}), which can be re-expanded in a finite sum of $\mathcal{I}_k$ with coefficients that are functions of Eisenstein series. Based on the leading examples in $k_{1,2}$, we have found the following empiric relation
\begin{align}
&\mathcal{I}_{k_1}(u,\rho)\,\mathcal{I}_{k_2}(u,\rho)=-\frac{B_{2\bar k}}{2\bar k}\,\mathfrak{d}E_{2\bar k}(\rho)-\sum_{k=0}^{\bar k}\frac{B_{2\bar k-2k}E_{2\bar k - 2k}\mathcal{P}_{k}(k_1,k_2)}{(2k+1)!}\,\mathcal{I}_k(u,\rho)\,,&&\text{for} &&\bar k = k_1 + k_2 + 1\,,\label{FusionPropagators}
\end{align}
which we have checked up to $k_1=k_2=6$. Here the derivatives $\mathfrak{d}E_{2n}$ of the Eisenstein series have been defined in (\ref{DerEisenstein}) and $E_0=1$ is understood. Furthermore, $\mathcal{P}_k(k_1,k_2)$ are symmetric polynomials of order $2(k+1)$ with integer coefficients, for which more information is provided in Appendix~\ref{App:Polynomials}. In writing eq.~(\ref{FusionPropagators}) we have implicitly assumed that the sums $\mathcal{I}_k$ are convergent. We leave a more systematic (and rigorous) study of the relations (\ref{FusionPropagators}) to future work, but simply remark that up to the level we have checked (\emph{i.e.} up to $k_1+k_2=10$) they are sufficient to relate the results found here to the decompositions of the free energy provided in \cite{Hohenegger:2020slq} (and to fix any ambiguities pointed out there).

\section{Conclusions and Outlook}\label{Sect:Conclusion}
In this paper we have analysed the topological string partition function on a class of toric, non-compact Calabi-Yau threefolds $X_{N,1}$. Physically, these capture (among others) the instanton partition function of Little String Theories, whose low energy limit are gauge theories with gauge group $U(N)$ and matter in the adjoint representation. We have shown for generic deformation parameters $\epsilon_{1,2}$ that for fixed instanton level, the contribution to the partition function that depends on the gauge structure can be written in terms of simple elliptic modular graph functions: indeed, we have identified elementary building blocks $\mathcal{T}_{\alpha_j\alpha_i}$ (see eq.~(\ref{DefTNotation})) that are labelled by two integer partitions and which can be re-written in terms of Kronecker-Eisenstein series as in eq.~(\ref{TOmegaRewrite}). In the unrefined limit (\emph{i.e.} for $\epsilon_1=-\epsilon_2=\epsilon$), the latter re-arrange themselves in terms of differences of propagators of a free chiral boson on the torus. We have shown that this re-organisation of the partition function is responsible for the Feynman-diagrammatic decomposition of the non-perturbative free energy observed previously in the literature~\cite{Hohenegger:2019tii,Hohenegger:2020slq}. Moreover, we have provided an interpretation of the origin of the propagator factors both geometrically (in terms of counting holomorphic curves on the geometry $X_{N,1}$) and physically (in terms of Nekrasov subfunctions of the $U(N)$ gauge theory). 

In a different limit of the deformation parameters (\emph{i.e.} the leading singularity in $\epsilon_2$, which we have called the NS-limit), we have shown for $N=1$ a recursive structure, which allows to express higher instanton contributions as powers of the leading one. We have provided strong evidence that this structure also persists for higher $N$. Including corrections (in terms of $\epsilon_2$) to the NS limit, we have shown that a similar hierarchy structure exists in principle for the entire partition function, however, non-trivial coefficient functions appear. Implicitly, these relations govern the precise numerical factors that appear in Feynman diagrammatic expansion of the free energy.

The results of this work strengthen the idea that the non-perturbative partition function (and thus also the free energy) of a class of LSTs (whose low energy limit are supersymmetric gauge theories with gauge group $U(N)$) allows a decomposition in terms of propagators of free two-dimensional scalar fields. We have found a precise relation between Nekrasov-subfunctions of the low energy theory and combinations of Greens functions of free scalar fields on the torus. At the same time, AGT-like correspondences \cite{Alday:2009aq} for (compactifications of) LSTs have been studied \cite{Aganagic:2015cta}, establishing a connection to $q$-deformed Toda (and Liouville theories). In the future it will be important to understand the deeper physical connection to the results presented in this work. Moreover, it would be interesting to see if a possible connection to eMGF that we pointed out can cast further light on the computation of conformal blocks in the two-dimensional theory. Similarly, it would be interesting to understand how the results presented here are related to the reformulation of the non-perturbative partition function of a $\Gamma$-quiver gauge theory in terms of a correlator in an underlying $W(\Gamma)$-algebra that was found in \cite{Kimura:2016ebq}.

Furthermore, it will be interesting in the future to extend our analysis to more general models of LSTs (see \emph{e.g.} \cite{Bhardwaj:2015oru} for a classification of such theories using F-theory compactifications). Indeed, as a simple generalisation, models of M5-branes probing non-trivial $\mathbb{Z}_M$ orbifold geometries have been studied for which the partition function can be computed using the same technology that we have used here \cite{Bastian:2017ary,Hohenegger:2016eqy,Bastian:2018dfu,Bastian:2017ing}. It would be interesting to extend our results to this class of models. Moreover, it would be interesting to see if similar decompositions as the ones discussed here also exist in LSTs that are not of $A_N$-type.

\section*{Acknowledgements}
We would like to thank Taro Kimura and Oliver Schlotterer for valuable exchanges and discussions as well as very useful comments on a preliminary draft of this paper. SH would like to thank Jean-Emile Bourgine, Fabrizio Nieri and Pierre Vanhove for very useful and enlightening discussions. He would also like to thank the organisers of the workshop 'New Trends in Non-Perturbative Gauge/String Theory and Integrability' (Institut de Mathématiques de Bourgogne (IMB), Dijon, 27/06 -- 01/07/2022) for kind hospitality and for creating a very stimulating atmosphere, in which part of this work was being carried out.
\appendix
\section{Notation and Conventions}
\subsection{Non-Perturbative Topological String Partition Function}\label{App:BuildingPartitionFunction}
In this appendix we collect our conventions and notation used to write the partition function discussed in Section~\ref{Sect:BraneWebs}. Starting from the parameters $(h_1,\ldots,h_N,v_1,\ldots,v_N,m_1,\ldots,m_N)$ introduced in Figure~\ref{Fig:Toric}, we first define
\begin{align}
&Q_{h_k}=e^{2\pi ih_k}\,,&&Q_{v_k}=e^{2\pi i v_k}\,,&&Q_{m_k}=e^{2\pi im_k}\,,&&\widetilde{Q}_k=Q_{h_k}Q_{v_k}\,,&&\forall k=1,\ldots,N\,.
\end{align}
Throughout this appendix, we consider the indices $i,j,k=1,\ldots,N$ to be defined modulo $N$ (\emph{e.g.} $Q_{h_{N+1}}=Q_{h_1}$). With this convention, we furthermore define~\cite{Bastian:2017ing}
\begin{align}
&\widehat{Q}_{i,j}=Q_{h_i}\prod_{k=1}^{j-1}\widetilde{Q}_{i-k}\,,&&\dot{Q}_{i,j}=\widetilde{Q}_{i+1}\ldots\widetilde{Q}_{i+j}\,,&&\overline{Q}_{i,j}=\left\{\begin{array}{lcl} 1 & \text{if} & j=N\,, \\ Q_{h_i}Q_{v_{i-j}}\prod_{k=1}^{j-1}\widetilde{Q}_{i-k}&\text{if} &j\neq L\,.\end{array}\right.
\end{align}
For the basis of K\"ahler parameters $(\tau,S,\rho,\widehat{a}_1,\ldots,\widehat{a}_{N-1})$ defined in (\ref{BasisStrip}), we introduce
\begin{align}
&\Qt=e^{2\pi i\tau}\,,&&\Qs=e^{2\pi iS}\,,&&\Qr=e^{2\pi i\rho}\,,&&\Qna{i}=e^{2\pi i\widehat{a}_i}\,\hspace{1cm}\forall i=1,\ldots,N\,.
\end{align}
and for the deformation parameters $\epsilon_{1,2}$ of the refined topological vertex \cite{Aganagic:2003db,Iqbal:2007ii}
\begin{align}
&q=e^{2\pi i\epsilon_1}\,,&&\text{and} &&t=e^{-2\pi i\epsilon_2}\,.
\end{align}
In order to write the partition function (\ref{DefPartFunctionGeneral}) we next need to review a number of definitions~\cite{Hohenegger:2013ala,Bastian:2017ing}. For $\mu=(\mu_1,\ldots,\mu_{\ell(\mu)})$ an integer partition of length $\ell(\mu)$ and $\mu^t$ its transpose, let
\begin{align}
&|\mu|=\sum_{i=1}^{\ell(\mu)}\mu_i\,,&&||\mu||^2=\sum_{i=1}^{\ell(\mu)}\mu_i^2\,,&&||\mu^t||^2=\sum_{i=1}^{\ell(\mu^t)}(\mu_i^t)^2\,.
\end{align}
We then define
\begin{align}
\mathcal{J}_{\mu\nu}(x;q,t)&=\prod_{k=1}^\infty J_{\mu\nu}(\Qr^{k-1}x;q,t)\,,\nonumber\\
J_{\mu\nu}(x;q,t)&=\prod_{(i,j)\in\mu}\left(1-x q^{\nu_j^t-i+\frac{1}{2}}t^{\mu_i-j+\frac{1}{2}}\right)\prod_{(i,j)\in\nu}\left(1-x q^{-\mu_j^t+i-\frac{1}{2}}t^{-\nu_i+j-\frac{1}{2}}\right)\,,
\end{align}
where $(i,j)\in\mu$ denote the positions of the boxes in the Young diagram associated with the partition $\mu$. Products of $\mathcal{J}_{\mu\nu}$ can be simplified using the relations
\begin{align}
\mathcal{J}_{\mu\nu}(x;q,t)\,\mathcal{J}_{\nu\mu}(\Qr x^{-1};q,t)&=x^{\frac{|\mu|+|\nu|}{2}}\,q^{\frac{||\nu^t||^2-||\mu^t||^2}{4}}t^{\frac{||\mu||^2-||\nu||^2}{4}}\,\vartheta_{\mu\nu}(x;\rho)\,,\nonumber\\
\frac{(-1)^{|\mu|}t^{\frac{||\mu||^2}{2}}q^{\frac{||\mu^t||^2}{2}}\widetilde{Z}_\mu(q,t)\widetilde{Z}_{\mu^t}(t,q)}{\mathcal{J}_{\mu\mu}(\Qr \sqrt{t/q};q,t)\mathcal{J}_{\mu\mu}(\Qr \sqrt{q/t};q,t)}&=\frac{1}{\vartheta_{\mu\mu}(\sqrt{q/t};\rho)}=\frac{1}{\vartheta_{\mu\mu}(\sqrt{t/q};\rho)}\,,\label{IdCombineJ}
\end{align}
where we have defined the theta-functions (with $x=e^{2\pi iz}$)
\begin{align}
\vartheta_{\mu\nu}(x;\rho)&=\prod_{(i,j)\in\mu}\vartheta\left(x^{-1}q^{-\nu_j^t+i-\frac{1}{2}}t^{-\mu_i+j-\frac{1}{2}};\rho\right)\prod_{(i,j)\in\nu}\vartheta\left(x^{-1}q^{\mu_j^t-i+\frac{1}{2}}t^{\nu_i-j+\frac{1}{2}};\rho\right)\,,\nonumber\\
\vartheta(x;\rho)&=\left(x^{1/2}-x^{-1/2}\right)\prod_{k=1}^\infty(1-x \Qr^k)(1-x^{-1}\Qr^k)=\frac{i\Qr^{-1/8}\theta_1(z;\rho)}{\prod_{k=1}^\infty(1-\Qr^k)}\,,\label{DefCurlyTheta}
\end{align}
with $\theta_1$ the Jacobi-theta function. The factors $\widetilde{Z}_{\mu}$ in (\ref{IdCombineJ}) appear naturally in the partition function (\ref{DefPartFunctionGeneral}) as part of $\widehat{Z}$, namely
\begin{align}
&\widehat{Z}=\prod_{k=1}^N t^{\frac{||\alpha_k||^2}{2}}q^{\frac{||\alpha_k^t||^2}{2}}\,\widetilde{Z}_{\alpha_k}(q,t)\,\widetilde{Z}_{\alpha^t_k}(t,q)\,,&&\text{with}&&\widetilde{Z}_{\alpha}(q,t)=\prod_{(i,j)\in\alpha}\left(1-q^{\nu_j^t-i+1} t^{\nu_i-j}\right)^{-1}\,.
\end{align}
Finally, the normalisation factor $W_N(\emptyset)$ in (\ref{DefPartFunctionGeneral}) has an interpretation as a closed topological string amplitude and the precise definition (which shall not be needed in this paper) can be found in \cite{Bastian:2017ing}.
\subsection{Modular Objects}\label{App:Modular}
We define a weak Jacobi form of index $m\in\mathbb{Z}$ and weight $w\in\mathbb{N}$ as a holomorphic function $\phi:\,\mathbb{C}\times \mathbb{H}\rightarrow \mathbb{C}$ which transforms in the following manner
\begin{align}
&\phi\left(\frac{z}{c\rho+d},\frac{a\rho+b}{c\rho+d}\right)=(c\rho+d)^w\,e^{\frac{2\pi imcz^2}{c\rho+d}}\,\phi(z,\rho)\,,&&\forall\left(\begin{array}{cc}a & b \\ c & d\end{array}\right)\in SL(2,\mathbb{Z})\,,\nonumber\\
&\phi(z+k_1\rho+k_2,\rho)=e^{-2\pi im(k_1^2\rho+2k_1z)}\phi(z,\rho)\,,&&\forall k_{1,2}\in\mathbb{N}\,,
\end{align}
and which has the following Fourier expansion
\begin{align}
&\phi(z,\rho)=\sum_{n=0}^\infty\sum_{k\in\mathbb{Z}}c(n,k)\,Q_\rho^n\,e^{2\pi i zk}\,,&&\text{with} &&c(n,k)=(-1)^w\,c(n,-k)\,.
\end{align}
We define the following standard Jacobi forms of index $m=1$ and weights $w=-2$ and $w=0$
\begin{align}
&\fb(z;\rho)=\frac{\theta_1^2(z;\rho)}{\eta^6(\rho)}\,,&&\text{and} &&\fa(z;\rho)=8\sum_{i=2}^4\left(\frac{\theta_i(z;\rho)}{\theta_i(0;\rho)}\right)^2\,,\label{DefStandardJacobi}
\end{align}
where $\theta_{i=1,2,3,4}$ are the Jacobi theta functions, notably
\begin{align}
\theta_1(z;\rho)=-i\Qr^{1/8}e^{\pi iz}\prod_{n=1}^\infty\left(1-\Qr^n\right)\left(1-e^{2\pi iz}\,\Qr^n\right)\left(1-e^{-2\pi i z}\,\Qr^{n-1}\right)\,,\label{DefJacobiTheta1}
\end{align}
and $\eta$ the Dedekind eta function
\begin{align}
\eta(\rho)=\Qr^{1/24}\prod_{n=1}^\infty(1-\Qr^n)\,.\label{DefDedekindEta}
\end{align}
We note the followig shift symmetry of $\theta_1$
\begin{align}
\theta_1(z\pm\rho;\rho)=-\,Q_\rho^{-1/2}\,e^{\mp 2\pi i z}\,\theta_1(z;\rho)\,.\label{ThetaShift}
\end{align}

Furthermore, we define the Eisenstein series
\begin{align}
&E_{2n}(\rho)=1-\frac{4n}{B_{2n}}\sum_{k=1}^\infty\sigma_{2n-1}(k)\,Q_\rho^k\,,&&\text{with} &n\in\mathbb{N}&\,,
\end{align}
with $B_{2n}$ the Bernoulli numbers and $\sigma_{n}(k)$ the divisor sigma function. The $E_{2n}$ are holomorphic modular forms of weight $2n$, while $E_2$ can be completed into a quasi-modular form of weight $2$
\begin{align}
\widehat{E}_2(\rho,\bar{\rho})=E_2(\rho)-\frac{3}{\pi\,\text{Im}(\rho)}\,.
\end{align}
We also define \cite{Hohenegger:2019tii,Hohenegger:2020slq}
\begin{align}
\mathcal{I}_k(\widehat{a},\rho)=\sum_{n=1}^\infty\frac{n^{2k+1}}{1-Q_\rho^n}\left(\Qa^n+\frac{Q_\rho^n}{\Qa^n}\right)\,,\label{DefSeriesIk}
\end{align}
which can be computed iteratively
\begin{align}
&\mathcal{I}_k(\widehat{a},\rho)=D_{\widehat{a}}^{2k}\,\mathcal{I}_0(\widehat{a},\rho)\,,&&\text{with} &&D_{\widehat{a}}=\frac{1}{2\pi i}\frac{\partial}{\partial\widehat{a}}=\Qa\,\frac{\partial}{\partial \Qa}\,,\label{DefIkSeries}
\end{align}
where $\mathcal{I}_0$ can be written as
\begin{align}
\mathcal{I}_0(\widehat{a},\rho)=\frac{1}{(2\pi i)^2}\left[2\zeta(2)\,E_2(\rho)+\wp(\widehat{a};\rho)\right]=\frac{1}{(2\pi i)^2}\left[\mathbb{G}''(\widehat{a};\rho)+\frac{2\pi i}{\rho-\bar{\rho}}\right]\,.\label{DefI0E2}
\end{align}
Here $\wp$ is Weierstrass' elliptic function, which is defined as 
\begin{align}
\wp(z;\rho)=\frac{1}{z^2}+\sum_{k=1}^\infty 2(2k+1)\zeta(2k+2)\,E_{2k+2}(\rho)\,z^{2k}\,,\label{DefWeierstrassEllipticP}
\end{align}
and which can expressed as \cite{Gritsenko:1999nm,Israel:2016xfu}
\begin{align}
\wp(z;\rho)=\zeta(2)\,\frac{\fa(z;\rho)}{\fb(z;\rho)}\,.
\end{align}
$\mathbb{G}$ appearing in (\ref{DefI0E2}) is the two-point function of a free scalar field on the torus (and a prime denotes the derivative with respect to $z$)
\begin{align}
\mathbb{G}(z;\rho)=-\ln\left|\frac{\theta_1(z;\rho)}{\theta'_1(0;\rho)}\right|^2-\frac{\pi}{2\text{Im}(\rho)}\,(z-\bar{z})^2\,.\label{DefGpropFun}
\end{align} 
For use in Section~\ref{Sect:ExpandPartitionFunctionExamples}, we introduce the following derivatives of the Eisenstein series
\begin{align}
\mathfrak{d}E_{2n}(\rho)=Q_\rho\,\frac{d E_{2n}(\rho)}{d Q_\rho}\,,\label{DerEisenstein}
\end{align}
which are quasi-modular forms of weight $2n+2$ that can be written as homogeneous polynomials in $(E_2,E_4,E_6)$, \emph{e.g.}
\begin{align}
&\mathfrak{d}E_2=\frac{1}{12}(E_2^2-E_4)\,,\hspace{1cm}\mathfrak{d}E_4=\frac{1}{3}(E_2E_4-E_6)\,,\hspace{1cm} \mathfrak{d}E_6=\frac{1}{2}(E_2E_6-E_4^2)\,,\nonumber\\
&\mathfrak{d}E_8=\frac{2}{3}\,E_4\,(E_2E_4-E_6)\,,\hspace{1cm}\mathfrak{d}E_{10}=\frac{1}{6}(5E_2E_4E_6-3E_4^3-2E_6^2)\,.
\end{align}
These derivatives can be used to define meromorphic Jacobi forms of index $0$ and weight $2n$. Indeed, we have verified up to $n=6$ that in an expansion in powers of $\widehat{a}$ all $E_2$ cancel out in the following combinations
\begin{align}
&2n E_{2n}(\rho)\,\mathcal{I}_0(\widehat{a},\rho)+\mathfrak{d}E_{2n}(\rho)\,,&&\forall n>1\,.\label{HolomorphicCombinations}
\end{align}
Finally, we also define the \emph{Hecke operator}: let $J_{w,m}$ be the space of Jacobi forms of index $m$ and weight $w$, then
\begin{align}
\mathcal{H}_k:\,\,J_{w,m}(\Gamma)&\longrightarrow J_{w,km}(\Gamma)\nonumber\\
\phi(z,\rho)&\longmapsto \mathcal{H}_k(\phi(z,\rho)) =k^{w-1}\sum_{{d|k}\atop{b\text{ mod }d}}d^{-w}\,\phi\left(\frac{kz}{d},\frac{k\rho+bd}{d^2}\right)\,.\label{DefHeckeOperator}
\end{align}
\subsection{Elliptic Modular Graph Forms}\label{App:ModularGraphForms}
In this appendix we review important basic notation and definitions related to elliptic modular graph functions and forms (eMFG). Our conventions follow mainly \cite{DHoker:2020hlp,DHoker:2022dxx,DHoker:2018mys}.

\emph{Modular graph functions} \cite{DHoker:2015wxz,DHoker:2016mwo,DHoker:2017pvk,Zerbini:2017usf,Gerken:2018jrq,Gerken:2019cxz,Gerken:2020yii,Dorigoni:2021jfr,Dorigoni:2021ngn,Dorigoni:2022npe} are maps from graphs of $N$ vertices connected by (decorated and oriented) edges, to the space of modular invariant functions. They can be generalised to \emph{modular graph forms} which are maps from such graphs into the space of (non-holomorphic) modular forms. Such maps are important in the study of scattering amplitudes in string theory and (supersymmetric) gauge theories (see \emph{e.g.} \cite{Green:1999pv,Green:2008uj}) and have been systematically studied in recent years (see \emph{e.g.} \cite{DHoker:2015gmr,DHoker:2015sve,Zerbini:2015rss}). The concept of modular graph functions and forms can further be generalise to \emph{elliptic modular graph} (functions) \emph{forms} (eMFG) \cite{DHoker:2020hlp,DHoker:2022dxx} as maps from graphs to (non-holomorphic) elliptic functions on the torus that are (invariant) covariant under modular transformations. A simple example is the function $\mathbb{G}$ defined in (\ref{DefGpropFun}), which is the scalar Greens-function (\emph{i.e.} the 2-point function) of a free scalar on the torus. Let 
\begin{align}
\Lambda&=\{z\in\mathbb{C}|z=m\,\rho+n\text{ for }(m,n)\in\mathbb{Z}^2\}\,,\nonumber\\
\Lambda'&=\{z\in\mathbb{C}|z=m\,\rho+n\text{ for }(m,n)\in\mathbb{Z}^2\text{ and }(m,n)\neq (0,0)\}\,,
\end{align}
then $\mathbb{G}(z;\rho)$ can also be written in the following form
\begin{align}
&\mathbb{G}(z;\rho)=\frac{\rho-\bar{\rho}}{2\pi i}\sum_{p\in\Lambda'}\frac{\chi_p(z;\rho)}{|p|^2}\,.
\end{align}
Here we have introduced
\begin{align}
\chi_p(z;\rho)=e^{2\pi i(nr-ms)}&&\text{with} &&\begin{array}{l}z=r\,\rho+s\hspace{0.5cm}\text{ for }r,s\in[0,1]\\ p=m\,\rho+n\in\Lambda'\,,\end{array}
\end{align}
which satisfy the relations
\begin{align}
&\chi_{p_1+p_2}(z;\rho)=\chi_{p_1}(z;\rho)\,\chi_{p_2}(z;\rho)\,,&&\chi_{p}(z_1+z_2;\rho)=\chi_p(z_1;\rho)\,\chi_p(z_2;\rho)\,,&&\chi_{-p}(z;\rho)=\chi_p(-z,\rho)\,.\nonumber
\end{align}
A building block for more sophisticated eMGF is the \emph{Kronecker-Eisenstein series} \cite{Kronecker,Brown2011MultipleEP,Broedel:2015hia}
\begin{align}
&\Omega(u,v;\rho) =\sum_{p\in\Lambda}\frac{\chi_p(u;\rho)}{v-p}=\frac{1}{v}+\sum_{p\in\Lambda'}\frac{\chi_p(u;\rho)}{v-p}\,, &&\forall u,v\in\mathbb{C}\,.
\end{align}
As a function of $u$, $\Omega$ is invariant under $u\to u+\kappa$ for any $\kappa\in\Lambda$ and as a function of $v$, $\Omega$ is meromorphic with simple poles for $v\in\Lambda$. It can also be written in terms of Jacobi-theta functions (\ref{DefJacobiTheta1}), which make these properties more tangible
\begin{align}
&\Omega(u,v;\rho) = \text{exp}\left( 2\pi i v\frac{{\rm Im}(u)}{{\rm Im}(\rho)}\right) \frac{\theta_1(u+v;\rho) \theta_1^{\prime}(0;\rho)}{\theta_1(u;\rho)\theta_1(v;\rho)}\,,&&\forall u,v\in\mathbb{C}\,,\label{DefOmega}
\end{align}
where $\theta'_1(0;\rho)=2\pi \eta^3(\rho)$, with $\eta$ the Dedekind eta function defined in (\ref{DefDedekindEta}). 

We also remark the following important relation between a product of two $\Omega$ with related arguments and the difference of two Weierstrass elliptic functions introduced in (\ref{DefWeierstrassEllipticP})
\begin{align}
\Omega(u,v;\rho)\,\Omega(u,-v;\rho)=-4\pi^2 \eta^6(\rho)\,\frac{\theta_1(u+v;\rho)\theta_1(u-v;\rho)}{\theta_1^2(u;\rho)\,\theta_1^2(v;\rho)}=\wp(u) - \wp(v)\,,&&\forall u,v\in\mathbb{C}\,.\label{IdOmegaWeierstrass}
\end{align}
Using (see eq.~(\ref{DefI0E2})) $\wp(u)=\mathbb{G}''(u;\rho)-2\zeta(2)\,\widehat{E}_2(\rho)$, we can therefore write the combination (\ref{IdOmegaWeierstrass}) of Kronecker-Eisenstein series in terms of the difference of two scalar-Greens functions on the torus
\begin{align}
\Omega(u,v;\rho)\,\Omega(u,-v;\rho)=\mathbb{G}''(u;\rho)-\mathbb{G}''(v;\rho)\,.\label{KESpropagators}
\end{align}
\section{Propagator Building Blocks}\label{App:PropBuildingBlocks}
In this Appendix we review some of the objects introduced in \cite{Hohenegger:2020slq}. Our notation is slightly adapted to the current paper.
\subsection{External States}\label{App:ExternalStates}
In order to expand the free energy in a form that resembles Feynman diagrams, in \cite{Hohenegger:2020slq} certain 'external states' have been defined. In this work we shall use a more condensed notation for the leading contributions in $\epsilon_{1,2}$:
\begin{itemize}
\item free energy of the configuration $(1,1)$\\
The non-perturbative free energy associated with the web diagram $(1,1)$ was expanded in \cite{Hohenegger:2020slq} for $q=t=e^{2\pi i \epsilon}$ in the following form
\begin{align}
\mathcal{F}_{1,1}(\rho,S,\tau;\epsilon)=\ln \mathcal{Z}_{1,1}(\rho,S,\tau;\epsilon)=\sum_{n=0}^\infty\sum_{s=0}^\infty Q_\tau^n\,\epsilon^{2s-2}\,\buildH{n}{s}(S,\rho)\,.
\end{align}
In this work we shall use the following condensed notation for the coefficients with $s=0$, namely $\sth{r}(S,\rho):=\buildH{1}{s}(S,\rho)$ which are weak Jacobi forms of weight $-2$ and index $r$. They are related through a Hecke transformation (see \cite{Hohenegger:2020slq})
\begin{align}
\sth{r}(S,\rho)=\mathcal{H}_r\left(\sth{1}(S,\rho)\right)\,,
\end{align}
where the Hecke operator is defined in (\ref{DefHeckeOperator}). The first few instances are given by
\begin{align}
&\sth{1}=-\fb\,,\hspace{0.9cm}\sth{2}=-\frac{1}{16}\,\fa\,\fb\,,\hspace{0.9cm}\sth{3}=-\frac{\fb}{432}\,\left(\fa^2+12\,E_4\,\fb^2\right)\,.\label{ExtStatesH}
\end{align}

\item $W$ function \\
In \cite{Hohenegger:2015btj,Ahmed:2017hfr} the function $\buildW{1}{0}$ was found to be related to counting BPS states of a single M5-brane with M2-branes ending on both sides of it
\begin{align}
W(S,\rho;\epsilon)=\frac{\theta_1^2(S;\rho)-\theta_1(S+\epsilon;\rho)\,\theta_1(S-\epsilon;\rho)}{\theta_1^2(\epsilon;\rho)}=\sum_{s=0}^\infty\epsilon^{2s}\,\buildW{1}{s}(S,\rho)\,.
\end{align}
For the coefficients $s=0$ we shall use the shorthand notation $\stw{r}(S,\rho)=\buildW{r}{0}(S,\rho)$ in this article, where the $\stw{r}$ are quasi-Jacobi forms of weight $0$ and index $r$. Similar to the $\sth{r}$, they can (formally) also be related through Hecke transformations 
\begin{align}
\stw{r}(S,\rho)=\mathcal{H}_r\left(\stw{1}(S,\rho)\right)\,,
\end{align}
with $\mathcal{H}_r$ defined in (\ref{DefHeckeOperator}). The first few instances of $\stw{r}$ are given explicitly by
\begin{align}
&\stw{1}=\frac{1}{24}\,(\fa+2\,E_2\,\fb)\,,\hspace{1cm}\stw{2}=\frac{1}{384}\,\left(\fa^2+4\,E_2\,\fa\,\fb+4\,E_4\,\fb^2\right)\,,\label{FormW1}\\
&\stw{3}=\frac{1}{10368}\left(\fa^3+6 E_2\fa^2 \fb+12 E_4\fa\fb^2+8(9E_2E_4-8E_6)\right)\,.\label{StatesWExplicit}
\end{align}
\end{itemize}
\subsection{Two Points}\label{App:2PointProps}
In addition to the definition (\ref{DefLinesPropoldPaper}), we also introduce a graphical notation for the derivatives of the scalar two-point function (in the sense of (\ref{DefIkSeries})) by inserting $2k$-crosses 
\begin{align}
&\scalebox{1}{\parbox{2.5cm}{\begin{tikzpicture}[scale = 1.50]
\draw[fill=black] (-1.19,0) circle (0.05);
\draw[fill=black] (0.19,0) circle (0.05);
\node at (-1.15,-0.25) {\footnotesize $\widehat{b}_1$};
\node at (0.15,0.25) {\footnotesize $\widehat{b}_2$};
\draw[ultra thick,dashed] (-1.15,0) -- (0.15,0);
\draw[ultra thick] (-1,0.1) -- (-0.8,-0.1);
\draw[ultra thick] (-1,-0.1) -- (-0.8,0.1);
\draw[ultra thick] (-0,0.1) -- (-0.2,-0.1);
\draw[ultra thick] (-0,-0.1) -- (-0.2,0.1);
\node at (-0.475,0.1) {$\cdots$};
\node at (-0.5,-0.3) {$\underbrace{\hspace{1.6cm}}_{2k\text{-times}}$};
\end{tikzpicture}}}:=\mathcal{I}_k(\widehat{b}_2-\widehat{b}_1,\rho)=\frac{1}{(2\pi i)^2}\,\mathbb{G}^{(2k+2)}(\widehat{b}_2-\widehat{b}_1;\rho)\,,&&\forall k>1\,.\label{PropDer}
\end{align}
While the two-point function $\scalebox{1}{\parbox{1.05cm}{\begin{tikzpicture}[scale = 1.50]
\draw[fill=black] (0.19,0) circle (0.05);
\draw[fill=black] (-0.39,0) circle (0.05);
\draw[ultra thick,dashed] (-0.35,0) -- (0.15,0);
\end{tikzpicture}
}}$ in (\ref{DefLinesPropoldPaper}) (which is equal to (\ref{DefI0E2})) is a quasi-Jacobi form, the quantities with derivatives in eq.~(\ref{PropDer}) are meromorphic Jacobi forms since only the term $\mathcal{O}(\widehat{a}^0)$ in (\ref{DefI0E2}) contains $E_2$. 

For vanishing $\epsilon$, the two-point function (\ref{DefLinesPropoldPaper}) can furthermore be related to (\ref{DefLinesPropUnrefined}) in the following manner
\begin{align}
\scalebox{1}{\parbox{2.4cm}{\begin{tikzpicture}[scale = 1.50]
\draw (0.19,0) circle (0.05);
\draw (-0.69,0) circle (0.05);
\node at (-0.7,-0.2) {\footnotesize $(\widehat{b}_1,\emptyset)$};
\node at (0.15,0.25) {\footnotesize $(\widehat{b}_2,\Box)$};
\draw[ultra thick] (-0.65,0) -- (0.15,0);
\end{tikzpicture}}
}\,+\,\scalebox{1}{\parbox{2.4cm}{\begin{tikzpicture}[scale = 1.50]
\draw (0.19,0) circle (0.05);
\draw (-0.69,0) circle (0.05);
\node at (-0.7,-0.2) {\footnotesize $(\widehat{b}_1,\Box)$};
\node at (0.15,0.25) {\footnotesize $(\widehat{b}_2,\emptyset)$};
\draw[ultra thick] (-0.65,0) -- (0.15,0);
\end{tikzpicture}}
}\bigg|_{\epsilon=0}&=2\,\left(\wp(\widehat{b}_2-\widehat{b}_1;\rho)-\wp(S;\rho)\right)\nonumber\\
&=2\left[\mathbb{G}''(\widehat{b}_2-\widehat{b}_1;\rho)-\frac{\pi^2}{3}\,\widehat{E}_2(\rho)\right]-2\,\wp(S,\rho)\nonumber\\
&=2(2\pi i)^2\left(\scalebox{1}{\parbox{1.8cm}{\begin{tikzpicture}[scale = 1.50]
\draw[fill=black] (0.19,0) circle (0.05);
\draw[fill=black] (-0.69,0) circle (0.05);
\node at (-0.65,-0.25) {\footnotesize $\widehat{b}_1$};
\node at (0.2,0.25) {\footnotesize $\widehat{b}_2$};
\draw[ultra thick,dashed] (-0.65,0) -- (0.15,0);
\end{tikzpicture}
}}+\frac{E_2}{12}+\frac{\wp(S,\rho)}{24\zeta(2)}\right)\,.
\end{align}
We notice that this two-point function is manifestly modular and holomorphic, however, (compared to (\ref{DefLinesPropoldPaper})) it is also a function of $S$. 

\subsection{Three Points}\label{App:3PointProps}
In this appendix we describe combinations of objects resembling three-point functions, which are relevant to describe the contributions to the non-perturbative partition function for $N=3$. We start by defining the quantities that appear in the leading instanton contribution in (\ref{N3LeadingInstanton})
{\allowdisplaybreaks\begin{align}
&\scalebox{1}{\parbox{2.2cm}{\begin{tikzpicture}[scale = 1.50]
\draw[fill=black] (-0.89,0) circle (0.05);
\draw[fill=black] (-0.01,0) circle (0.05);
\draw[fill=black] (-0.48,0.69) circle (0.05);
\node at (-1,-0.25) {\footnotesize $\widehat{b}_1$};
\node at (0.1,-0.25) {\footnotesize $\widehat{b}_2$};
\node at (-0.48,0.95) {\footnotesize $\widehat{b}_3$};
\draw[ultra thick,dashed] (-0.85,0) -- (-0.05,0);
\end{tikzpicture}
}}\hspace{0.25cm}+\hspace{0.25cm}\scalebox{1}{\parbox{2.2cm}{\begin{tikzpicture}[scale = 1.50]
\draw[fill=black] (-0.89,0) circle (0.05);
\draw[fill=black] (-0.01,0) circle (0.05);
\draw[fill=black] (-0.48,0.69) circle (0.05);
\node at (-1,-0.25) {\footnotesize $\widehat{b}_1$};
\node at (0.1,-0.25) {\footnotesize $\widehat{b}_2$};
\node at (-0.48,0.95) {\footnotesize $\widehat{b}_3$};
\draw[ultra thick,dashed] (-0.01,0) -- (-0.48,0.69);
\end{tikzpicture}
}}\hspace{0.25cm}+\hspace{0.25cm}\scalebox{1}{\parbox{2.2cm}{\begin{tikzpicture}[scale = 1.50]
\draw[fill=black] (-0.89,0) circle (0.05);
\draw[fill=black] (-0.01,0) circle (0.05);
\draw[fill=black] (-0.48,0.69) circle (0.05);
\node at (-1,-0.25) {\footnotesize $\widehat{b}_1$};
\node at (0.1,-0.25) {\footnotesize $\widehat{b}_2$};
\node at (-0.48,0.95) {\footnotesize $\widehat{b}_3$};
\draw[ultra thick,dashed] (-0.89,0) -- (-0.48,0.69);
\end{tikzpicture}
}}=\frac{1}{(2\pi i)^2}\sum_{\ell=1}^3\sum_{j\neq \ell}\left(\mathbb{G}''(\widehat{b}_\ell-\widehat{b}_j;\rho)+\frac{2\pi i}{\rho-\bar{\rho}}\right)\,,\nonumber\\
&\scalebox{1}{\parbox{2.2cm}{\begin{tikzpicture}[scale = 1.50]
\draw[fill=black] (-0.89,0) circle (0.05);
\draw[fill=black] (-0.01,0) circle (0.05);
\draw[fill=black] (-0.48,0.69) circle (0.05);
\node at (-1,-0.25) {\footnotesize $\widehat{b}_1$};
\node at (0.1,-0.25) {\footnotesize $\widehat{b}_2$};
\node at (-0.48,0.95) {\footnotesize $\widehat{b}_3$};
\draw[ultra thick,dashed] (-0.85,0) -- (-0.05,0);
\draw[ultra thick,dashed] (-0.89,0) -- (-0.48,0.69);
\end{tikzpicture}
}}\hspace{0.25cm}+\hspace{0.25cm}\scalebox{1}{\parbox{2.2cm}{\begin{tikzpicture}[scale = 1.50]
\draw[fill=black] (-0.89,0) circle (0.05);
\draw[fill=black] (-0.01,0) circle (0.05);
\draw[fill=black] (-0.48,0.69) circle (0.05);
\node at (-1,-0.25) {\footnotesize $\widehat{b}_1$};
\node at (0.1,-0.25) {\footnotesize $\widehat{b}_2$};
\node at (-0.48,0.95) {\footnotesize $\widehat{b}_3$};
\draw[ultra thick,dashed] (-0.01,0) -- (-0.48,0.69);
\draw[ultra thick,dashed] (-0.85,0) -- (-0.05,0);
\end{tikzpicture}
}}\hspace{0.25cm}+\hspace{0.25cm}\scalebox{1}{\parbox{2.2cm}{\begin{tikzpicture}[scale = 1.50]
\draw[fill=black] (-0.89,0) circle (0.05);
\draw[fill=black] (-0.01,0) circle (0.05);
\draw[fill=black] (-0.48,0.69) circle (0.05);
\node at (-1,-0.25) {\footnotesize $\widehat{b}_1$};
\node at (0.1,-0.25) {\footnotesize $\widehat{b}_2$};
\node at (-0.48,0.95) {\footnotesize $\widehat{b}_3$};
\draw[ultra thick,dashed] (-0.89,0) -- (-0.48,0.69);
\draw[ultra thick,dashed] (-0.01,0) -- (-0.48,0.69);
\end{tikzpicture}
}}=\frac{1}{(2\pi i)^4}\sum_{\ell=1}^3\prod_{j\neq \ell}\left(\mathbb{G}''(\widehat{b}_\ell-\widehat{b}_j;\rho)+\frac{2\pi i}{\rho-\bar{\rho}}\right)\,.\label{Tier1Correlators2Prop}
\end{align}}
For higher instanton levels, we also introduce the shorthand notation
{\allowdisplaybreaks
\begin{align}
&\scalebox{1}{\parbox{2.8cm}{\begin{tikzpicture}[scale = 1.50]
\draw[fill=black] (-1.19,0) circle (0.05);
\draw[fill=black] (0.19,0) circle (0.05);
\draw[fill=black] (-0.48,1.19) circle (0.05);
\node at (-1.2,-0.25) {\footnotesize $\widehat{b}_1$};
\node at (0.3,-0.25) {\footnotesize $\widehat{b}_2$};
\node at (-0.48,1.45) {\footnotesize $\widehat{b}_3$};
\draw[thick,domain=0:-300,<-] plot ({0.15*cos(\x)-0.48},{0.15*sin(\x)+0.5});
\node at (-0.48,0.5) {\tiny $\sum$};
\draw[ultra thick,dashed] (-1.15,0) -- (0.15,0);
\draw[ultra thick,dashed] (-1.175,0.05) -- (-0.5,1.15);
\draw[ultra thick,dashed] (0.175,0.05) -- (-0.46,1.15);
%
%
\end{tikzpicture}
}}\hspace{0.25cm}=\hspace{0.25cm}\frac{1}{(2\pi)^6}\prod_{1\leq i<j\leq 3}\left(\mathbb{G}''(\widehat{b}_j-\widehat{b}_i;\rho)+\frac{2\pi i}{\rho-\bar{\rho}}\right)\,,\nonumber\\[24pt]
&\scalebox{1}{\parbox{2.8cm}{\begin{tikzpicture}[scale = 1.50]
\draw[fill=black] (-1.19,0) circle (0.05);
\draw[fill=black] (0.19,0) circle (0.05);
\draw[fill=black] (-0.48,1.19) circle (0.05);
\node at (-1.2,-0.25) {\footnotesize $\widehat{b}_1$};
\node at (0.3,-0.25) {\footnotesize $\widehat{b}_2$};
\node at (-0.48,1.45) {\footnotesize $\widehat{b}_3$};
\draw[thick,domain=0:-300,<-] plot ({0.15*cos(\x)-0.48},{0.15*sin(\x)+0.5});
\node at (-0.48,0.5) {\tiny $\sum$};
\draw[ultra thick,dashed] (-1.175,0.05) -- (-0.5,1.15);
\draw[ultra thick,dashed] (0.175,0.05) -- (-0.46,1.15);
%
%
\end{tikzpicture}
}}\hspace{0.25cm}=\hspace{0.25cm}\frac{1}{(2\pi)^4}\sum_{\ell=1}^3\prod_{j\neq \ell}\left(\mathbb{G}''(\widehat{b}_\ell-\widehat{b}_j;\rho)+\frac{2\pi i}{\rho-\bar{\rho}}\right)\,,\nonumber\\[24pt]
&\scalebox{1}{\parbox{2.8cm}{\begin{tikzpicture}[scale = 1.50]
\draw[fill=black] (-1.19,0) circle (0.05);
\draw[fill=black] (0.19,0) circle (0.05);
\draw[fill=black] (-0.48,1.19) circle (0.05);
\node at (-1.2,-0.25) {\footnotesize $\widehat{b}_1$};
\node at (0.3,-0.25) {\footnotesize $\widehat{b}_2$};
\node at (-0.48,1.45) {\footnotesize $\widehat{b}_3$};
\draw[thick,domain=0:-300,<-] plot ({0.15*cos(\x)-0.48},{0.15*sin(\x)+0.5});
\node at (-0.48,0.5) {\tiny $\sum$};
\draw[ultra thick,dashed] (-1.175,0.05) -- (-0.5,1.15);
\draw[ultra thick,dashed] (0.175,0.05) -- (-0.46,1.15);
\begin{scope}[rotate=60,xshift=1.1cm,yshift=1.01cm]
\draw[ultra thick] (-0.9,0.1) -- (-0.7,-0.1);
\draw[ultra thick] (-0.9,-0.1) -- (-0.7,0.1);
\end{scope}
\begin{scope}[rotate=60,xshift=0.7cm,yshift=1.01cm]
\draw[ultra thick] (-0.9,0.1) -- (-0.7,-0.1);
\draw[ultra thick] (-0.9,-0.1) -- (-0.7,0.1);
\end{scope}
\end{tikzpicture}
}}\hspace{0.25cm}=\hspace{0.25cm}\frac{1}{(2\pi)^4}\sum_{\ell=1}^3\sum_{\mathcal{S}\in\{1,2,3\}\setminus\{\ell\}\atop{|\mathcal{S}|=2}}\prod_{(j_1,j_2)\in \mathcal{S}}\mathbb{G}^{(4)}(\widehat{b}_\ell-\widehat{b}_{j_1};\rho)\,\left(\mathbb{G}^{''}(\widehat{b}_\ell-\widehat{b}_{j_2};\rho)+\frac{2\pi i}{\rho-\bar{\rho}}\right)\,,\nonumber\\[24pt]
&\scalebox{1}{\parbox{2.8cm}{\begin{tikzpicture}[scale = 1.50]
\draw[fill=black] (-1.19,0) circle (0.05);
\draw[fill=black] (0.19,0) circle (0.05);
\draw[fill=black] (-0.48,1.19) circle (0.05);
\node at (-1.2,-0.25) {\footnotesize $\widehat{b}_1$};
\node at (0.3,-0.25) {\footnotesize $\widehat{b}_2$};
\node at (-0.48,1.45) {\footnotesize $\widehat{b}_3$};
\draw[thick,domain=0:-300,<-] plot ({0.15*cos(\x)-0.48},{0.15*sin(\x)+0.5});
\node at (-0.48,0.5) {\tiny $\sum$};
\draw[ultra thick,dashed] (-1.175,0.05) -- (-0.5,1.15);
\draw[ultra thick,dashed] (0.175,0.05) -- (-0.46,1.15);
\begin{scope}[rotate=60,xshift=1.1cm,yshift=1.01cm]
\draw[ultra thick] (-0.9,0.1) -- (-0.7,-0.1);
\draw[ultra thick] (-0.9,-0.1) -- (-0.7,0.1);
\end{scope}
\begin{scope}[rotate=60,xshift=0.7cm,yshift=1.01cm]
\draw[ultra thick] (-0.9,0.1) -- (-0.7,-0.1);
\draw[ultra thick] (-0.9,-0.1) -- (-0.7,0.1);
\end{scope}
\begin{scope}[rotate=-60,xshift=0.cm,yshift=0.175cm]
\draw[ultra thick] (-0.9,0.1) -- (-0.7,-0.1);
\draw[ultra thick] (-0.9,-0.1) -- (-0.7,0.1);
\end{scope}
\begin{scope}[rotate=-60,xshift=0.4cm,yshift=0.175cm]
\draw[ultra thick] (-0.9,0.1) -- (-0.7,-0.1);
\draw[ultra thick] (-0.9,-0.1) -- (-0.7,0.1);
\end{scope}
%
\end{tikzpicture}
}}\hspace{0.25cm}=\hspace{0.25cm}\frac{1}{(2\pi)^4}\sum_{\ell=1}^3\prod_{j\neq \ell}\mathbb{G}^{(4)}(\widehat{b}_\ell-\widehat{b}_j;\rho)\,,\nonumber
\end{align}}
where the symbol $\scalebox{1}{\parbox{0.5cm}{\begin{tikzpicture}[scale = 1.50]\draw[thick,domain=0:-300,<-] plot ({0.15*cos(\x)-0.48},{0.15*sin(\x)+0.5});
\node at (-0.48,0.5) {\tiny $\sum$};\end{tikzpicture}
}}$ indicates that a cyclic sum over all points has been taken into account.

\section{Propagator Polynomials}\label{App:Polynomials}
In this appendix, we provide the leading examples of the polynomials $\mathcal{P}_k(k_1,k_2)$ appearing in eq.~(\ref{FusionPropagators}):
{\allowdisplaybreaks
\begin{align}
\mathcal{P}_0&=1\,,\nonumber\\
\mathcal{P}_1&=4(k_1^2-k_1k_2+k_2^2)-1\,,\nonumber\\
\mathcal{P}_2&=16(k_1^4-k_1^3k_2+k_2^2k_2^2-k_1k_2^3+k_2^4)-8(3k_1^3-k_1^2k_2-k_1k_2^2+3k_2^2)\nonumber\\
&\hspace{0.6cm}-4(k_1^2-3k_1k_2+k_2^2)+6(k_1+k_2)\,,\nonumber\\
\mathcal{P}_3&=64(k_1^6-k_1^5k_2+k_1^4k_2-k_1^3k_2^3+k_1^2k_2^4-k_1k_2^5+k_2^6)\nonumber\\
&\hspace{0.6cm}-64(5k_1^5-3k_1^4k_2+k_1^3 k_2^2+k_1^2k_2^3-3k_1 k_2^4+5k_2^5)+32(15k_1^4-3k_1^2k_2-k_1^2k_2^2-3k_1k_2^3+15 k_2^4)\nonumber\\
&\hspace{0.6cm}-32(5k_1^3+k_1^2k_2+k_1k_2^2+5k_2^3)-4(31k_1^2-15k_1k_2+31k_2^2)+60(k_1+k_2)\,.
\end{align}}
We remark that these can also be characterised in the following way (which lends itself to generalisation to higher $k$):  
\begin{align}
&f_k(x)=\frac{x^{2k+1}}{1-Q_\rho^x}\,,&&\text{for} &&k\in\mathbb{N}^*\,,
\end{align}
then one can decompose the convolution of two such functions in the following way
\begin{align}
(f_{k_1}\star f_{k_2})(x):&=\int_{\mathbb{R}}dy\,f_{k_1}(y)\,f_{k_2}(x-y)
=\frac{1}{1-Q_\rho^x}\sum_{k=0}^{k_1+k_2+1}c_{k_1k_2k}\,\frac{x^{2k+1}}{\rho^{2(k_1+k_2-k+1)}}\nonumber\\
&=-\sum_{k=0}^{k_1+k_2+1}\,\frac{B_{2(k_1+k_2-k+1)}\,\mathcal{P}_k(k_1,k_2)}{(2k+1)!\,\rho^{2(k_1+k_2-k+1)}}\,f_k(x)\,.
\end{align}
We have furthermore observed a few further properties of the polynomials $\mathcal{P}_k(k_1,k_2)$. Indeed, the highest powers in each case follow the pattern $2^{2k}\sum_{a=0}^{2k}(-1)^ak_1^a k_2^{2k-a}$, while subleading contributions are characterised by particular symmetries. We have notably observed
\begin{align}
\mathcal{P}_k(k-k_1-1,k-k_2-1)=\mathcal{P}_{k}(k_1,k_2)\,,
\end{align}
and we have verified up to $k=10$ that they factorise for certain values of their arguments
{\allowdisplaybreaks
\begin{align}
&\mathcal{P}_k(k_1,-\tfrac{1}{2})=\mathcal{P}_k(k_1,k_1+\tfrac{1}{2})=\prod_{j=0}^{2k-1}(2k_1+1-j)\,,\nonumber\\
&\mathcal{P}_k(k_1,k_1-\tfrac{1}{2})=\mathcal{P}_k(k_1,k-\tfrac{1}{2})=\prod_{j=1}^{2k}(2k_1+1-j)\,,\nonumber\\
&\mathcal{P}_k(k_1,k_1)=\mathcal{P}_k(k_1,\tfrac{k-1}{2})=\frac{\prod_{j=0}^{2k}(2k_1+1-j)}{2k_1+1-k}\,.
\end{align}}
Finally, up to $k=10$ the $\mathcal{P}_k$ vanish for specific values of their arguments
\begin{align}
&\mathcal{P}_k(0,j)=0\,,&&\forall j=1,\ldots,k-2\,,\nonumber\\
&\mathcal{P}_k(\tfrac{j-1}{2},\tfrac{\ell-1}{2})=0\,,&&\forall \left\{\begin{array}{l}\ell=1,\ldots,2k-2-j\,,\\j=1,\ldots,2k\,.\end{array}\right.
\end{align}
We have also verified that the properties mentioned in this appendix uniquely fix the polynomial $\mathcal{P}_k$ up to $k=10$.

\begingroup
\sloppy
\printbibliography

@article{Hohenegger:2019tii,
    author = "Hohenegger, Stefan",
    title = "{From Little String Free Energies Towards Modular Graph Functions}",
    eprint = "1911.08172",
    archivePrefix = "arXiv",
    primaryClass = "hep-th",
    doi = "10.1007/JHEP03(2020)077",
    journal = "JHEP",
    volume = "03",
    pages = "077",
    year = "2020"
}

@article{Hohenegger:2020slq,
    author = "Hohenegger, Stefan",
    title = "{Diagrammatic Expansion of Non-Perturbative Little String Free Energies}",
    eprint = "2011.06323",
    archivePrefix = "arXiv",
    primaryClass = "hep-th",
    reportNumber = "LYCEN 2020-08",
    doi = "10.1007/JHEP04(2021)275",
    journal = "JHEP",
    volume = "04",
    pages = "275",
    year = "2021"
}

@article{Bastian:2017ing,
    author = "Bastian, Brice and Hohenegger, Stefan and Iqbal, Amer and Rey, Soo-Jong",
    title = "{Dual little strings and their partition functions}",
    eprint = "1710.02455",
    archivePrefix = "arXiv",
    primaryClass = "hep-th",
    doi = "10.1103/PhysRevD.97.106004",
    journal = "Phys. Rev. D",
    volume = "97",
    number = "10",
    pages = "106004",
    year = "2018"
}

@article{Hohenegger:2015cba,
    author = "Hohenegger, Stefan and Iqbal, Amer and Rey, Soo-Jong",
    title = "{M-strings, monopole strings, and modular forms}",
    eprint = "1503.06983",
    archivePrefix = "arXiv",
    primaryClass = "hep-th",
    reportNumber = "SNUST-15-02",
    doi = "10.1103/PhysRevD.92.066005",
    journal = "Phys. Rev. D",
    volume = "92",
    number = "6",
    pages = "066005",
    year = "2015"
}

@article{Hohenegger:2015btj,
    author = "Hohenegger, Stefan and Iqbal, Amer and Rey, Soo-Jong",
    title = "{Instanton-monopole correspondence from M-branes on $\mathbb S^1$ and little string theory}",
    eprint = "1511.02787",
    archivePrefix = "arXiv",
    primaryClass = "hep-th",
    doi = "10.1103/PhysRevD.93.066016",
    journal = "Phys. Rev. D",
    volume = "93",
    number = "6",
    pages = "066016",
    year = "2016"
}

@article{Hohenegger:2016eqy,
    author = "Hohenegger, Stefan and Iqbal, Amer and Rey, Soo-Jong",
    title = "{Self-Duality and Self-Similarity of Little String Orbifolds}",
    eprint = "1605.02591",
    archivePrefix = "arXiv",
    primaryClass = "hep-th",
    doi = "10.1103/PhysRevD.94.046006",
    journal = "Phys. Rev. D",
    volume = "94",
    number = "4",
    pages = "046006",
    year = "2016"
}

@article{Bastian:2018dfu,
    author = "Bastian, Brice and Hohenegger, Stefan and Iqbal, Amer and Rey, Soo-Jong",
    title = "{Beyond Triality: Dual Quiver Gauge Theories and Little String Theories}",
    eprint = "1807.00186",
    archivePrefix = "arXiv",
    primaryClass = "hep-th",
    doi = "10.1007/JHEP11(2018)016",
    journal = "JHEP",
    volume = "11",
    pages = "016",
    year = "2018"
}

@article{Bastian:2017ary,
    author = "Bastian, Brice and Hohenegger, Stefan and Iqbal, Amer and Rey, Soo-Jong",
    title = "{Triality in Little String Theories}",
    eprint = "1711.07921",
    archivePrefix = "arXiv",
    primaryClass = "hep-th",
    doi = "10.1103/PhysRevD.97.046004",
    journal = "Phys. Rev. D",
    volume = "97",
    number = "4",
    pages = "046004",
    year = "2018"
}

@article{DHoker:2020hlp,
    author = "D'Hoker, Eric and Kleinschmidt, Axel and Schlotterer, Oliver",
    title = "{Elliptic modular graph forms. Part I. Identities and generating series}",
    eprint = "2012.09198",
    archivePrefix = "arXiv",
    primaryClass = "hep-th",
    doi = "10.1007/JHEP03(2021)151",
    journal = "JHEP",
    volume = "03",
    pages = "151",
    year = "2021"
}

@article{Haghighat:2013gba,
    author = "Haghighat, Babak and Iqbal, Amer and Koz\c{c}az, Can and Lockhart, Guglielmo and Vafa, Cumrun",
    title = "{M-Strings}",
    eprint = "1305.6322",
    archivePrefix = "arXiv",
    primaryClass = "hep-th",
    doi = "10.1007/s00220-014-2139-1",
    journal = "Commun. Math. Phys.",
    volume = "334",
    number = "2",
    pages = "779--842",
    year = "2015"
}

@article{Haghighat:2013tka,
    author = "Haghighat, Babak and Kozcaz, Can and Lockhart, Guglielmo and Vafa, Cumrun",
    title = "{Orbifolds of M-strings}",
    eprint = "1310.1185",
    archivePrefix = "arXiv",
    primaryClass = "hep-th",
    doi = "10.1103/PhysRevD.89.046003",
    journal = "Phys. Rev. D",
    volume = "89",
    number = "4",
    pages = "046003",
    year = "2014"
}

@article{Hohenegger:2013ala,
    author = "Hohenegger, Stefan and Iqbal, Amer",
    title = "{M-strings, elliptic genera and $\mathcal{N} = 4$ string amplitudes}",
    eprint = "1310.1325",
    archivePrefix = "arXiv",
    primaryClass = "hep-th",
    reportNumber = "CERN-PH-TH-2013-237",
    doi = "10.1002/prop.201300035",
    journal = "Fortsch. Phys.",
    volume = "62",
    pages = "155--206",
    year = "2014"
}

@inproceedings{Witten:1995zh,
    author = "Witten, Edward",
    title = "{Some comments on string dynamics}",
    booktitle = "{STRINGS 95: Future Perspectives in String Theory}",
    eprint = "hep-th/9507121",
    archivePrefix = "arXiv",
    reportNumber = "IASSNS-HEP-95-63",
    pages = "501--523",
    month = "7",
    year = "1995"
}

@article{Kanazawa:2016tnt,
    author = "Kanazawa, Atsushi and Lau, Siu-Cheong",
    title = "{Local Calabi\textendash{}Yau manifolds of type $\tilde A$ via SYZ mirror symmetry}",
    eprint = "1605.00342",
    archivePrefix = "arXiv",
    primaryClass = "math.AG",
    doi = "10.1016/j.geomphys.2018.12.015",
    journal = "J. Geom. Phys.",
    volume = "139",
    pages = "103--138",
    year = "2019"
}

@article{Bhardwaj:2015oru,
    author = "Bhardwaj, Lakshya and Del Zotto, Michele and Heckman, Jonathan J. and Morrison, David R. and Rudelius, Tom and Vafa, Cumrun",
    title = "{F-theory and the Classification of Little Strings}",
    eprint = "1511.05565",
    archivePrefix = "arXiv",
    primaryClass = "hep-th",
    doi = "10.1103/PhysRevD.93.086002",
    journal = "Phys. Rev. D",
    volume = "93",
    number = "8",
    pages = "086002",
    year = "2016",
    note = "[Erratum: Phys.Rev.D 100, 029901 (2019)]"
}

@article{Aspinwall:1997ye,
    author = "Aspinwall, Paul S. and Morrison, David R.",
    title = "{Point - like instantons on K3 orbifolds}",
    eprint = "hep-th/9705104",
    archivePrefix = "arXiv",
    reportNumber = "RU-97-29, IASSNS-HEP-97-46",
    doi = "10.1016/S0550-3213(97)00516-6",
    journal = "Nucl. Phys. B",
    volume = "503",
    pages = "533--564",
    year = "1997"
}

@article{Seiberg:1997zk,
    author = "Seiberg, Nathan",
    title = "{New theories in six-dimensions and matrix description of M theory on T**5 and T**5 / Z(2)}",
    eprint = "hep-th/9705221",
    archivePrefix = "arXiv",
    reportNumber = "RU-97-42",
    doi = "10.1016/S0370-2693(97)00805-8",
    journal = "Phys. Lett. B",
    volume = "408",
    pages = "98--104",
    year = "1997"
}

@article{Intriligator:1997dh,
    author = "Intriligator, Kenneth A.",
    title = "{New string theories in six-dimensions via branes at orbifold singularities}",
    eprint = "hep-th/9708117",
    archivePrefix = "arXiv",
    reportNumber = "IASSNS-HEP-97-95, UCSD-PTH-22",
    doi = "10.4310/ATMP.1997.v1.n2.a5",
    journal = "Adv. Theor. Math. Phys.",
    volume = "1",
    pages = "271--282",
    year = "1998"
}

@article{Hanany:1997gh,
    author = "Hanany, Amihay and Zaffaroni, Alberto",
    title = "{Branes and six-dimensional supersymmetric theories}",
    eprint = "hep-th/9712145",
    archivePrefix = "arXiv",
    reportNumber = "CERN-TH-97-366, IASSNS-HEP-97-135",
    doi = "10.1016/S0550-3213(98)00355-1",
    journal = "Nucl. Phys. B",
    volume = "529",
    pages = "180--206",
    year = "1998"
}

@article{Brunner:1997gf,
    author = "Brunner, Ilka and Karch, Andreas",
    title = "{Branes at orbifolds versus Hanany Witten in six-dimensions}",
    eprint = "hep-th/9712143",
    archivePrefix = "arXiv",
    reportNumber = "HUB-EP-97-92",
    doi = "10.1088/1126-6708/1998/03/003",
    journal = "JHEP",
    volume = "03",
    pages = "003",
    year = "1998"
}

@article{Aharony:1999ks,
    author = "Aharony, Ofer",
    editor = "Lechtenfeld, O. and Louis, J. and Lust, Dieter and Nicolai, H.",
    title = "{A Brief review of 'little string theories'}",
    eprint = "hep-th/9911147",
    archivePrefix = "arXiv",
    reportNumber = "RUNHETC-99-42",
    doi = "10.1088/0264-9381/17/5/302",
    journal = "Class. Quant. Grav.",
    volume = "17",
    pages = "929--938",
    year = "2000"
}

@article{Kutasov:2001uf,
    author = "Kutasov, D.",
    editor = "Bachas, C. and Maldacena, Juan Martin and Narain, K. S. and Randjbar-Daemi, S.",
    title = "{Introduction to little string theory}",
    journal = "ICTP Lect. Notes Ser.",
    volume = "7",
    pages = "165--209",
    year = "2002"
}

@article{Haghighat:2018gqf,
    author = "Haghighat, Babak and Sun, Rui",
    title = "{M5 branes and Theta Functions}",
    eprint = "1811.04938",
    archivePrefix = "arXiv",
    primaryClass = "hep-th",
    doi = "10.1007/JHEP10(2019)192",
    journal = "JHEP",
    volume = "10",
    pages = "192",
    year = "2019"
}

@article{Bastian:2018jlf,
    author = "Bastian, Brice and Hohenegger, Stefan",
    title = "{Dihedral Symmetries of Gauge Theories from Dual Calabi-Yau Threefolds}",
    eprint = "1811.03387",
    archivePrefix = "arXiv",
    primaryClass = "hep-th",
    doi = "10.1103/PhysRevD.99.066013",
    journal = "Phys. Rev. D",
    volume = "99",
    number = "6",
    pages = "066013",
    year = "2019"
}

@article{Ahmed:2017hfr,
    author = "Ahmed, Ambreen and Hohenegger, Stefan and Iqbal, Amer and Rey, Soo-Jong",
    title = "{Bound states of little strings and symmetric orbifold conformal field theories}",
    eprint = "1706.04425",
    archivePrefix = "arXiv",
    primaryClass = "hep-th",
    doi = "10.1103/PhysRevD.96.081901",
    journal = "Phys. Rev. D",
    volume = "96",
    number = "8",
    pages = "081901",
    year = "2017"
}

@article{Iqbal:2010awa,
    author = "Iqbal, Amer and Nazir, Shaheen and Raza, Zahid and Saleem, Zain",
    title = "{Generalizations of Nekrasov-Okounkov Identity}",
    eprint = "1011.3745",
    archivePrefix = "arXiv",
    primaryClass = "math.CO",
    month = "11",
    year = "2010"
}

@article{Iqbal:2003ix,
    author = "Iqbal, Amer and Kashani-Poor, Amir-Kian",
    title = "{Instanton counting and Chern-Simons theory}",
    eprint = "hep-th/0212279",
    archivePrefix = "arXiv",
    reportNumber = "SLAC-PUB-9623, HUTP-02-A064, SU-ITP-02-48",
    doi = "10.4310/ATMP.2003.v7.n3.a4",
    journal = "Adv. Theor. Math. Phys.",
    volume = "7",
    number = "3",
    pages = "457--497",
    year = "2003"
}

@article{Iqbal:2003zz,
    author = "Iqbal, Amer and Kashani-Poor, Amir-Kian",
    title = "{SU(N) geometries and topological string amplitudes}",
    eprint = "hep-th/0306032",
    archivePrefix = "arXiv",
    reportNumber = "SLAC-PUB-9924, HUTP-03-A038, SU-ITP-03-12",
    doi = "10.4310/ATMP.2006.v10.n1.a1",
    journal = "Adv. Theor. Math. Phys.",
    volume = "10",
    number = "1",
    pages = "1--32",
    year = "2006"
}

@article{Iqbal:2004ne,
    author = "Iqbal, Amer and Kashani-Poor, Amir-Kian",
    title = "{The Vertex on a strip}",
    eprint = "hep-th/0410174",
    archivePrefix = "arXiv",
    reportNumber = "SMS-0402, SLAC-PUB-10804, SU-ITP-04-39",
    doi = "10.4310/ATMP.2006.v10.n3.a2",
    journal = "Adv. Theor. Math. Phys.",
    volume = "10",
    number = "3",
    pages = "317--343",
    year = "2006"
}

@article{Aganagic:2002qg,
    author = "Aganagic, Mina and Marino, Marcos and Vafa, Cumrun",
    title = "{All loop topological string amplitudes from Chern-Simons theory}",
    eprint = "hep-th/0206164",
    archivePrefix = "arXiv",
    reportNumber = "HUTP-02-A024",
    doi = "10.1007/s00220-004-1067-x",
    journal = "Commun. Math. Phys.",
    volume = "247",
    pages = "467--512",
    year = "2004"
}

@inproceedings{Gritsenko:1999nm,
    author = "Gritsenko, V.",
    title = "{Complex vector bundles and Jacobi forms}",
    eprint = "math/9906191",
    archivePrefix = "arXiv",
    month = "1",
    year = "1999"
}

@article{Israel:2016xfu,
    author = "Israel, Dan and Sarkis, Matthieu",
    title = "{Dressed elliptic genus of heterotic compactifications with torsion and general bundles}",
    eprint = "1606.08982",
    archivePrefix = "arXiv",
    primaryClass = "hep-th",
    doi = "10.1007/JHEP08(2016)176",
    journal = "JHEP",
    volume = "08",
    pages = "176",
    year = "2016"
}

@article{DHoker:2022dxx,
    author = "D'Hoker, Eric and Kaidi, Justin",
    title = "{Lectures on modular forms and strings}",
    eprint = "2208.07242",
    archivePrefix = "arXiv",
    primaryClass = "hep-th",
    month = "8",
    year = "2022"
}

@article{DHoker:2015wxz,
    author = {D'Hoker, Eric and Green, Michael B. and G\"urdogan, \"Omer and Vanhove, Pierre},
    title = "{Modular Graph Functions}",
    eprint = "1512.06779",
    archivePrefix = "arXiv",
    primaryClass = "hep-th",
    reportNumber = "DAMTP-2015-86, IPHT-T15-202, IHES-P-15-29",
    doi = "10.4310/CNTP.2017.v11.n1.a4",
    journal = "Commun. Num. Theor. Phys.",
    volume = "11",
    pages = "165--218",
    year = "2017"
}

@article{DHoker:2016mwo,
    author = "D'Hoker, Eric and Green, Michael B.",
    title = "{Identities between Modular Graph Forms}",
    eprint = "1603.00839",
    archivePrefix = "arXiv",
    primaryClass = "hep-th",
    reportNumber = "DAMTP-2016-21",
    doi = "10.1016/j.jnt.2017.11.015",
    journal = "J. Number Theor.",
    volume = "189",
    pages = "25--80",
    year = "2018"
}

@article{DHoker:2017pvk,
    author = "D'Hoker, Eric and Green, Michael B. and Pioline, Boris",
    title = "{Higher genus modular graph functions, string invariants, and their exact asymptotics}",
    eprint = "1712.06135",
    archivePrefix = "arXiv",
    primaryClass = "hep-th",
    reportNumber = "QMUL-PH-17-27, DAMTP-2017-45",
    doi = "10.1007/s00220-018-3244-3",
    journal = "Commun. Math. Phys.",
    volume = "366",
    number = "3",
    pages = "927--979",
    year = "2019"
}

@phdthesis{Zerbini:2017usf,
    author = "Zerbini, Federico",
    title = "{Elliptic multiple zeta values, modular graph functions and genus 1 superstring scattering amplitudes}",
    eprint = "1804.07989",
    archivePrefix = "arXiv",
    primaryClass = "math-ph",
    school = "Bonn U.",
    year = "2017"
}

@inproceedings{Zerbini:2018hgs,
    author = "Zerbini, Federico",
    title = "{Modular and Holomorphic Graph Functions from Superstring Amplitudes}",
    booktitle = "{KMPB Conference}: {Elliptic Integrals, Elliptic Functions and Modular  Forms in Quantum Field Theory}",
    eprint = "1807.04506",
    archivePrefix = "arXiv",
    primaryClass = "math-ph",
    doi = "10.1007/978-3-030-04480-0_18",
    pages = "459--484",
    year = "2019"
}

@article{Gerken:2018jrq,
    author = "Gerken, Jan E. and Kleinschmidt, Axel and Schlotterer, Oliver",
    title = "{Heterotic-string amplitudes at one loop: modular graph forms and relations to open strings}",
    eprint = "1811.02548",
    archivePrefix = "arXiv",
    primaryClass = "hep-th",
    doi = "10.1007/JHEP01(2019)052",
    journal = "JHEP",
    volume = "01",
    pages = "052",
    year = "2019"
}

@article{Gerken:2019cxz,
    author = "Gerken, Jan E. and Kleinschmidt, Axel and Schlotterer, Oliver",
    title = "{All-order differential equations for one-loop closed-string integrals and modular graph forms}",
    eprint = "1911.03476",
    archivePrefix = "arXiv",
    primaryClass = "hep-th",
    doi = "10.1007/JHEP01(2020)064",
    journal = "JHEP",
    volume = "01",
    pages = "064",
    year = "2020"
}

@article{Gerken:2020yii,
    author = "Gerken, Jan E. and Kleinschmidt, Axel and Schlotterer, Oliver",
    title = "{Generating series of all modular graph forms from iterated Eisenstein integrals}",
    eprint = "2004.05156",
    archivePrefix = "arXiv",
    primaryClass = "hep-th",
    doi = "10.1007/JHEP07(2020)190",
    journal = "JHEP",
    volume = "07",
    number = "07",
    pages = "190",
    year = "2020"
}

@phdthesis{Gerken:2020xte,
    author = "Gerken, Jan Erik",
    title = "{Modular Graph Forms and Scattering Amplitudes in String Theory}",
    eprint = "2011.08647",
    archivePrefix = "arXiv",
    primaryClass = "hep-th",
    doi = "10.18452/21829",
    school = "Humboldt U., Berlin, Humboldt U., Berlin",
    year = "2020"
}

@article{Bastian:2019hpx,
    author = "Bastian, Brice and Hohenegger, Stefan",
    title = "{Symmetries in A-type little string theories. Part I. Reduced free energy and paramodular groups}",
    eprint = "1911.07276",
    archivePrefix = "arXiv",
    primaryClass = "hep-th",
    doi = "10.1007/JHEP03(2020)062",
    journal = "JHEP",
    volume = "03",
    pages = "062",
    year = "2020"
}

@article{Bastian:2019wpx,
    author = "Bastian, Brice and Hohenegger, Stefan",
    title = "{Symmetries in A-type little string theories. Part II. Eisenstein series and generating functions of multiple divisor sums}",
    eprint = "1911.07280",
    archivePrefix = "arXiv",
    primaryClass = "hep-th",
    doi = "10.1007/JHEP03(2020)016",
    journal = "JHEP",
    volume = "03",
    pages = "016",
    year = "2020"
}

@article{Aganagic:2000gs,
    author = "Aganagic, Mina and Vafa, Cumrun",
    title = "{Mirror symmetry, D-branes and counting holomorphic discs}",
    eprint = "hep-th/0012041",
    archivePrefix = "arXiv",
    reportNumber = "HUTP-00-A047",
    month = "12",
    year = "2000"
}

@article{Bastian:2017jje,
    author = "Bastian, Brice and Hohenegger, Stefan",
    title = "{Five-Brane Webs and Highest Weight Representations}",
    eprint = "1706.08750",
    archivePrefix = "arXiv",
    primaryClass = "hep-th",
    doi = "10.1007/JHEP12(2017)020",
    journal = "JHEP",
    volume = "12",
    pages = "020",
    year = "2017"
}

@article{Aharony:1997bh,
    author = "Aharony, Ofer and Hanany, Amihay and Kol, Barak",
    title = "{Webs of (p,q) five-branes, five-dimensional field theories and grid diagrams}",
    eprint = "hep-th/9710116",
    archivePrefix = "arXiv",
    reportNumber = "IASSNS-HEP-97-113, RU-97-81, SU-ITP-97-40",
    doi = "10.1088/1126-6708/1998/01/002",
    journal = "JHEP",
    volume = "01",
    pages = "002",
    year = "1998"
}

@article{Aganagic:2003db,
    author = "Aganagic, Mina and Klemm, Albrecht and Marino, Marcos and Vafa, Cumrun",
    title = "{The Topological vertex}",
    eprint = "hep-th/0305132",
    archivePrefix = "arXiv",
    reportNumber = "CALT-68-2439, HUTP-03-A032, HU-EP-03-24, CERN-TH-2003-111",
    doi = "10.1007/s00220-004-1162-z",
    journal = "Commun. Math. Phys.",
    volume = "254",
    pages = "425--478",
    year = "2005"
}

@article{Iqbal:2007ii,
    author = "Iqbal, Amer and Kozcaz, Can and Vafa, Cumrun",
    title = "{The Refined topological vertex}",
    eprint = "hep-th/0701156",
    archivePrefix = "arXiv",
    doi = "10.1088/1126-6708/2009/10/069",
    journal = "JHEP",
    volume = "10",
    pages = "069",
    year = "2009"
}

@inproceedings{Nekrasov:2009rc,
    author = "Nekrasov, Nikita A. and Shatashvili, Samson L.",
    title = "{Quantization of Integrable Systems and Four Dimensional Gauge Theories}",
    booktitle = "{16th International Congress on Mathematical Physics}",
    eprint = "0908.4052",
    archivePrefix = "arXiv",
    primaryClass = "hep-th",
    reportNumber = "TCD-MATH-09-19, HMI-09-09, IHES-P-09-38",
    doi = "10.1142/9789814304634_0015",
    pages = "265--289",
    month = "8",
    year = "2009"
}

@article{Mironov:2009uv,
    author = "Mironov, A. and Morozov, A.",
    title = "{Nekrasov Functions and Exact Bohr-Zommerfeld Integrals}",
    eprint = "0910.5670",
    archivePrefix = "arXiv",
    primaryClass = "hep-th",
    reportNumber = "FIAN-TD-21-09, ITEP-TH-51-09",
    doi = "10.1007/JHEP04(2010)040",
    journal = "JHEP",
    volume = "04",
    pages = "040",
    year = "2010"
}

@article{Green:1999pv,
    author = "Green, Michael B. and Vanhove, Pierre",
    title = "{The Low-energy expansion of the one loop type II superstring amplitude}",
    eprint = "hep-th/9910056",
    archivePrefix = "arXiv",
    reportNumber = "DAMTP-1999-124, CERN-TH-99-200, SACLAY-SPH-T-99-071",
    doi = "10.1103/PhysRevD.61.104011",
    journal = "Phys. Rev. D",
    volume = "61",
    pages = "104011",
    year = "2000"
}

@article{Green:2008uj,
    author = "Green, Michael B. and Russo, Jorge G. and Vanhove, Pierre",
    title = "{Low energy expansion of the four-particle genus-one amplitude in type II superstring theory}",
    eprint = "0801.0322",
    archivePrefix = "arXiv",
    primaryClass = "hep-th",
    reportNumber = "DAMTP-2007-96, SPHT-T-07-126, UB-ECM-PF-07-29",
    doi = "10.1088/1126-6708/2008/02/020",
    journal = "JHEP",
    volume = "02",
    pages = "020",
    year = "2008"
}

@article{DHoker:2015gmr,
    author = "D'Hoker, Eric and Green, Michael B. and Vanhove, Pierre",
    title = "{On the modular structure of the genus-one Type II superstring low energy expansion}",
    eprint = "1502.06698",
    archivePrefix = "arXiv",
    primaryClass = "hep-th",
    reportNumber = "DAMTP-08-02-2015, IPHT-T15-012, IHES-P-15-04",
    doi = "10.1007/JHEP08(2015)041",
    journal = "JHEP",
    volume = "08",
    pages = "041",
    year = "2015"
}

@article{DHoker:2015sve,
    author = "D'Hoker, Eric and Green, Michael B. and Vanhove, Pierre",
    title = "{Proof of a modular relation between 1-, 2- and 3-loop Feynman diagrams on a torus}",
    eprint = "1509.00363",
    archivePrefix = "arXiv",
    primaryClass = "hep-th",
    reportNumber = "DAMTP-2015-48, IPhT-t15/134, IHES/P/15/07, DAMTP-2015-48, IPHT-T15-134, IHES-P-15-07",
    doi = "10.1016/j.jnt.2017.07.022",
    journal = "J. Number Theor.",
    volume = "196",
    pages = "381--419",
    year = "2019"
}

@article{Zerbini:2015rss,
    author = "Zerbini, Federico",
    title = "{Single-valued multiple zeta values in genus 1 superstring amplitudes}",
    eprint = "1512.05689",
    archivePrefix = "arXiv",
    primaryClass = "hep-th",
    doi = "10.4310/CNTP.2016.v10.n4.a2",
    journal = "Commun. Num. Theor. Phys.",
    volume = "10",
    pages = "703--737",
    year = "2016"
}

@article{Nekrasov:2002qd,
    author = "Nekrasov, Nikita A.",
    title = "{Seiberg-Witten prepotential from instanton counting}",
    eprint = "hep-th/0206161",
    archivePrefix = "arXiv",
    reportNumber = "ITEP-TH-22-02, IHES-P-04-22",
    doi = "10.4310/ATMP.2003.v7.n5.a4",
    journal = "Adv. Theor. Math. Phys.",
    volume = "7",
    number = "5",
    pages = "831--864",
    year = "2003"
}

@article{Moore:1997dj,
    author = "Moore, Gregory W. and Nekrasov, Nikita and Shatashvili, Samson",
    title = "{Integrating over Higgs branches}",
    eprint = "hep-th/9712241",
    archivePrefix = "arXiv",
    reportNumber = "ITEP-TH-62-97, HUTP-97-A089, YCTP-P23-97",
    doi = "10.1007/PL00005525",
    journal = "Commun. Math. Phys.",
    volume = "209",
    pages = "97--121",
    year = "2000"
}

@article{Lossev:1997bz,
    author = "Lossev, A. and Nekrasov, N. and Shatashvili, Samson L.",
    editor = "Baulieu, L. and Kazakov, V. and Picco, M. and Windey, Paul and Di Francesco, P. and Douglas, Michael R.",
    title = "{Testing Seiberg-Witten solution}",
    eprint = "hep-th/9801061",
    archivePrefix = "arXiv",
    reportNumber = "HUTP-97-A102, ITEP-TH-74-97",
    journal = "NATO Sci. Ser. C",
    volume = "520",
    pages = "359--372",
    year = "1999"
}

@article{Antoniadis:2010iq,
    author = "Antoniadis, I. and Hohenegger, S. and Narain, K. S. and Taylor, T. R.",
    title = "{Deformed Topological Partition Function and Nekrasov Backgrounds}",
    eprint = "1003.2832",
    archivePrefix = "arXiv",
    primaryClass = "hep-th",
    reportNumber = "CERN-PH-TH-2010-061, LMU-ASC-09-10",
    doi = "10.1016/j.nuclphysb.2010.04.021",
    journal = "Nucl. Phys. B",
    volume = "838",
    pages = "253--265",
    year = "2010"
}

@article{Antoniadis:2013bja,
    author = "Antoniadis, I. and Florakis, Ioannis and Hohenegger, S. and Narain, K. S. and Zein Assi, Ahmad",
    title = "{Worldsheet Realization of the Refined Topological String}",
    eprint = "1302.6993",
    archivePrefix = "arXiv",
    primaryClass = "hep-th",
    reportNumber = "CERN-PH-TH-2013-001",
    doi = "10.1016/j.nuclphysb.2013.07.004",
    journal = "Nucl. Phys. B",
    volume = "875",
    pages = "101--133",
    year = "2013"
}

@article{Antoniadis:2013mna,
    author = "Antoniadis, Ignatios and Florakis, Ioannis and Hohenegger, Stefan and Narain, K. S. and Zein Assi, Ahmad",
    title = "{Non-Perturbative Nekrasov Partition Function from String Theory}",
    eprint = "1309.6688",
    archivePrefix = "arXiv",
    primaryClass = "hep-th",
    reportNumber = "CERN-PH-TH-2013-223, MPP-2013-271",
    doi = "10.1016/j.nuclphysb.2014.01.006",
    journal = "Nucl. Phys. B",
    volume = "880",
    pages = "87--108",
    year = "2014"
}

@article{Antoniadis:2015spa,
    author = "Antoniadis, Ignatios and Florakis, Ioannis and Hohenegger, Stefan and Narain, K. S. and Zein Assi, Ahmad",
    title = "{Probing the moduli dependence of refined topological amplitudes}",
    eprint = "1508.01477",
    archivePrefix = "arXiv",
    primaryClass = "hep-th",
    reportNumber = "CERN-PH-TH-2015-159",
    doi = "10.1016/j.nuclphysb.2015.10.016",
    journal = "Nucl. Phys. B",
    volume = "901",
    pages = "252--281",
    year = "2015"
}

@article{Samsonyan:2017xdi,
    author = "Samsonyan, Marine and Angelantonj, Carlo and Antoniadis, Ignatios",
    editor = "Checchia, Paolo and others",
    title = "{$N=2^*$ (non-)Abelian theory in the $\Omega$ background from string theory}",
    doi = "10.22323/1.314.0546",
    journal = "PoS",
    volume = "EPS-HEP2017",
    pages = "546",
    year = "2017"
}

@article{Angelantonj:2017qeh,
    author = "Angelantonj, Carlo and Antoniadis, Ignatios and Samsonyan, Marine",
    title = "{A string realisation of $\varOmega$-deformed Abelian $\mathcal{N}=2^*$ theory}",
    eprint = "1702.04998",
    archivePrefix = "arXiv",
    primaryClass = "hep-th",
    reportNumber = "CERN-TH-2017-038",
    doi = "10.1016/j.nuclphysb.2017.07.015",
    journal = "Nucl. Phys. B",
    volume = "923",
    pages = "32--53",
    year = "2017"
}

@article{Angelantonj:2019qfw,
    author = "Angelantonj, Carlo and Antoniadis, Ignatios",
    title = "{The String Geometry Behind Topological Amplitudes}",
    eprint = "1910.03347",
    archivePrefix = "arXiv",
    primaryClass = "hep-th",
    doi = "10.1007/JHEP01(2020)005",
    journal = "JHEP",
    volume = "01",
    pages = "005",
    year = "2020"
}

@article{Angelantonj:2022dsx,
    author = "Angelantonj, Carlo and Antoniadis, Ignatios and Florakis, Ioannis and Jiang, Hongliang",
    title = "{Refined topological amplitudes from the \ensuremath{\Omega}-background in string theory}",
    eprint = "2202.13205",
    archivePrefix = "arXiv",
    primaryClass = "hep-th",
    reportNumber = "QMUL-PH-22-07",
    doi = "10.1007/JHEP05(2022)143",
    journal = "JHEP",
    volume = "05",
    pages = "143",
    year = "2022"
}

@article{Gopakumar:1998ii,
    author = "Gopakumar, Rajesh and Vafa, Cumrun",
    title = "{M theory and topological strings. 1.}",
    eprint = "hep-th/9809187",
    archivePrefix = "arXiv",
    reportNumber = "HUTP-98-A069",
    month = "9",
    year = "1998"
}

@article{Gopakumar:1998jq,
    author = "Gopakumar, Rajesh and Vafa, Cumrun",
    title = "{M theory and topological strings. 2.}",
    eprint = "hep-th/9812127",
    archivePrefix = "arXiv",
    reportNumber = "HUTP-98-A070",
    month = "12",
    year = "1998"
}

@article{Hollowood:2003cv,
    author = "Hollowood, Timothy J. and Iqbal, Amer and Vafa, Cumrun",
    title = "{Matrix models, geometric engineering and elliptic genera}",
    eprint = "hep-th/0310272",
    archivePrefix = "arXiv",
    reportNumber = "HUTP-03-A074",
    doi = "10.1088/1126-6708/2008/03/069",
    journal = "JHEP",
    volume = "03",
    pages = "069",
    year = "2008"
}

@article{Alday:2009aq,
    author = "Alday, Luis F. and Gaiotto, Davide and Tachikawa, Yuji",
    title = "{Liouville Correlation Functions from Four-dimensional Gauge Theories}",
    eprint = "0906.3219",
    archivePrefix = "arXiv",
    primaryClass = "hep-th",
    doi = "10.1007/s11005-010-0369-5",
    journal = "Lett. Math. Phys.",
    volume = "91",
    pages = "167--197",
    year = "2010"
}

@article{Nekrasov:2003rj,
    author = "Nekrasov, Nikita and Okounkov, Andrei",
    title = "{Seiberg-Witten theory and random partitions}",
    eprint = "hep-th/0306238",
    archivePrefix = "arXiv",
    reportNumber = "ITEP-TH-36-03, PUDM-2003, IHES-P-03-43",
    doi = "10.1007/0-8176-4467-9_15",
    journal = "Prog. Math.",
    volume = "244",
    pages = "525--596",
    year = "2006"
}

@article{Iqbal:2015fvd,
    author = "Iqbal, Amer and Kozcaz, Can and Yau, Shing-Tung",
    title = "{Elliptic Virasoro Conformal Blocks}",
    eprint = "1511.00458",
    archivePrefix = "arXiv",
    primaryClass = "hep-th",
    month = "11",
    year = "2015"
}

@article{Nieri:2015dts,
    author = "Nieri, Fabrizio",
    title = "{An elliptic Virasoro symmetry in 6d}",
    eprint = "1511.00574",
    archivePrefix = "arXiv",
    primaryClass = "hep-th",
    doi = "10.1007/s11005-017-0986-3",
    journal = "Lett. Math. Phys.",
    volume = "107",
    number = "11",
    pages = "2147--2187",
    year = "2017"
}

@article{Hayling:2017cva,
    author = "Hayling, Joseph and Papageorgakis, Constantinos and Pomoni, Elli and Rodriguez-Gomez, Diego",
    title = "{Exact Deconstruction of the 6D (2,0) Theory}",
    eprint = "1704.02986",
    archivePrefix = "arXiv",
    primaryClass = "hep-th",
    reportNumber = "QMUL-PH-17-06, DESY-17-030",
    doi = "10.1007/JHEP06(2017)072",
    journal = "JHEP",
    volume = "06",
    pages = "072",
    year = "2017"
}

@article{Mironov:2009qt,
    author = "Mironov, A. and Morozov, A.",
    title = "{The Power of Nekrasov Functions}",
    eprint = "0908.2190",
    archivePrefix = "arXiv",
    primaryClass = "hep-th",
    reportNumber = "FIAN-TD-18-09, ITEP-TH-34-09",
    doi = "10.1016/j.physletb.2009.08.061",
    journal = "Phys. Lett. B",
    volume = "680",
    pages = "188--194",
    year = "2009"
}

@article{Gaiotto:2009we,
    author = "Gaiotto, Davide",
    title = "{N=2 dualities}",
    eprint = "0904.2715",
    archivePrefix = "arXiv",
    primaryClass = "hep-th",
    doi = "10.1007/JHEP08(2012)034",
    journal = "JHEP",
    volume = "08",
    pages = "034",
    year = "2012"
}

@article{Aganagic:2015cta,
    author = "Aganagic, Mina and Haouzi, Nathan",
    title = "{ADE Little String Theory on a Riemann Surface (and Triality)}",
    eprint = "1506.04183",
    archivePrefix = "arXiv",
    primaryClass = "hep-th",
    month = "6",
    year = "2015"
}

@article{Aganagic:2013tta,
    author = "Aganagic, Mina and Haouzi, Nathan and Kozcaz, Can and Shakirov, Shamil",
    title = "{Gauge/Liouville Triality}",
    eprint = "1309.1687",
    archivePrefix = "arXiv",
    primaryClass = "hep-th",
    reportNumber = "ITEP-TH-30-13, SISSA-42-2013-FISI",
    month = "9",
    year = "2013"
}

@article{Aganagic:2014oia,
    author = "Aganagic, Mina and Haouzi, Nathan and Shakirov, Shamil",
    title = "{$A_n$-Triality}",
    eprint = "1403.3657",
    archivePrefix = "arXiv",
    primaryClass = "hep-th",
    month = "3",
    year = "2014"
}

@article{Shiraishi:1995rp,
    author = "Shiraishi, Jun'ichi and Kubo, Harunobu and Awata, Hidetoshi and Odake, Satoru",
    title = "{A Quantum deformation of the Virasoro algebra and the Macdonald symmetric functions}",
    eprint = "q-alg/9507034",
    archivePrefix = "arXiv",
    reportNumber = "YITP-U-95-30, DPSU-95-5, UT-715",
    doi = "10.1007/BF00398297",
    journal = "Lett. Math. Phys.",
    volume = "38",
    pages = "33--51",
    year = "1996"
}

@article{Awata:1996xt,
    author = "Awata, Hidetoshi and Kubo, Harunobu and Morita, Yoshifumi and Odake, Satoru and Shiraishi, Junichi",
    title = "{Vertex operators of the q Virasoro algebra: Defining relations, adjoint actions and four point functions}",
    eprint = "q-alg/9604023",
    archivePrefix = "arXiv",
    reportNumber = "DPSU-96-7, UT-750, EFI-96-14",
    doi = "10.1023/A:1007321109584",
    journal = "Lett. Math. Phys.",
    volume = "41",
    pages = "65--78",
    year = "1997"
}

@article{Haouzi:2017vec,
    author = "Haouzi, Nathan and Koz\c{c}az, Can",
    title = "{The ABCDEFG of little strings}",
    eprint = "1711.11065",
    archivePrefix = "arXiv",
    primaryClass = "hep-th",
    doi = "10.1007/JHEP06(2021)092",
    journal = "JHEP",
    volume = "06",
    pages = "092",
    year = "2021"
}

@article{Hohenegger:2020gio,
    author = "Hohenegger, Stefan and Iqbal, Amer",
    title = "{Symmetric orbifold theories from little string residues}",
    eprint = "2009.00797",
    archivePrefix = "arXiv",
    primaryClass = "hep-th",
    reportNumber = "LYCEN 2020-07",
    doi = "10.1103/PhysRevD.103.066004",
    journal = "Phys. Rev. D",
    volume = "103",
    number = "6",
    pages = "066004",
    year = "2021"
}

@article{Bastian:2018fba,
    author = "Bastian, Brice and Hohenegger, Stefan and Iqbal, Amer and Rey, Soo-Jong",
    title = "{Five-Dimensional Gauge Theories from Shifted Web Diagrams}",
    eprint = "1810.05109",
    archivePrefix = "arXiv",
    primaryClass = "hep-th",
    doi = "10.1103/PhysRevD.99.046012",
    journal = "Phys. Rev. D",
    volume = "99",
    number = "4",
    pages = "046012",
    year = "2019"
}

@article{Hohenegger:2016yuv,
    author = "Hohenegger, Stefan and Iqbal, Amer and Rey, Soo-Jong",
    title = "{Dual Little Strings from F-Theory and Flop Transitions}",
    eprint = "1610.07916",
    archivePrefix = "arXiv",
    primaryClass = "hep-th",
    doi = "10.1007/JHEP07(2017)112",
    journal = "JHEP",
    volume = "07",
    pages = "112",
    year = "2017"
}

@article{Wyllard:2009hg,
    author = "Wyllard, Niclas",
    title = "{A(N-1) conformal Toda field theory correlation functions from conformal N = 2 SU(N) quiver gauge theories}",
    eprint = "0907.2189",
    archivePrefix = "arXiv",
    primaryClass = "hep-th",
    doi = "10.1088/1126-6708/2009/11/002",
    journal = "JHEP",
    volume = "11",
    pages = "002",
    year = "2009"
}

@article{Awata:2009ur,
    author = "Awata, Hidetoshi and Yamada, Yasuhiko",
    title = "{Five-dimensional AGT Conjecture and the Deformed Virasoro Algebra}",
    eprint = "0910.4431",
    archivePrefix = "arXiv",
    primaryClass = "hep-th",
    doi = "10.1007/JHEP01(2010)125",
    journal = "JHEP",
    volume = "01",
    pages = "125",
    year = "2010"
}

@article{Schiappa:2009cc,
    author = "Schiappa, Ricardo and Wyllard, Niclas",
    title = "{An A(r) threesome: Matrix models, 2d CFTs and 4d N=2 gauge theories}",
    eprint = "0911.5337",
    archivePrefix = "arXiv",
    primaryClass = "hep-th",
    doi = "10.1063/1.3449328",
    journal = "J. Math. Phys.",
    volume = "51",
    pages = "082304",
    year = "2010"
}

@article{Awata:2010yy,
    author = "Awata, Hidetoshi and Yamada, Yasuhiko",
    title = "{Five-dimensional AGT Relation and the Deformed beta-ensemble}",
    eprint = "1004.5122",
    archivePrefix = "arXiv",
    primaryClass = "hep-th",
    doi = "10.1143/PTP.124.227",
    journal = "Prog. Theor. Phys.",
    volume = "124",
    pages = "227--262",
    year = "2010"
}

@article{Kimura:2015rgi,
    author = "Kimura, Taro and Pestun, Vasily",
    title = "{Quiver W-algebras}",
    eprint = "1512.08533",
    archivePrefix = "arXiv",
    primaryClass = "hep-th",
    doi = "10.1007/s11005-018-1072-1",
    journal = "Lett. Math. Phys.",
    volume = "108",
    number = "6",
    pages = "1351--1381",
    year = "2018"
}

@article{Kimura:2017hez,
    author = "Kimura, Taro and Pestun, Vasily",
    title = "{Fractional quiver W-algebras}",
    eprint = "1705.04410",
    archivePrefix = "arXiv",
    primaryClass = "hep-th",
    doi = "10.1007/s11005-018-1087-7",
    journal = "Lett. Math. Phys.",
    volume = "108",
    number = "11",
    pages = "2425--2451",
    year = "2018"
}

@book{Kimura:2020jxl,
    author = "Kimura, Taro",
    title = "{Instanton Counting, Quantum Geometry and Algebra}",
    eprint = "2012.11711",
    archivePrefix = "arXiv",
    primaryClass = "hep-th",
    doi = "10.1007/978-3-030-76190-5",
    isbn = "978-3-030-76189-9, 978-3-030-76190-5",
    publisher = "Springer",
    month = "7",
    year = "2021"
}

@article{Kimura:2016ebq,
    author = "Kimura, Taro",
    title = "{Double quantization of Seiberg-Witten geometry and W-algebras}",
    eprint = "1612.07590",
    archivePrefix = "arXiv",
    primaryClass = "hep-th",
    doi = "10.1090/pspum/100/01762",
    journal = "Proc. Symp. Pure Math.",
    volume = "100",
    pages = "405--431",
    year = "2018"
}

@article{Broedel:2015hia,
    author = "Broedel, Johannes and Matthes, Nils and Schlotterer, Oliver",
    title = "{Relations between elliptic multiple zeta values and a special derivation algebra}",
    eprint = "1507.02254",
    archivePrefix = "arXiv",
    primaryClass = "hep-th",
    reportNumber = "MITP-15-048",
    doi = "10.1088/1751-8113/49/15/155203",
    journal = "J. Phys. A",
    volume = "49",
    number = "15",
    pages = "155203",
    year = "2016"
}

@article{Kronecker,
    author = "Kronecker, L.",
    title = "{Zur Theorie der elliptischen Funktionen}",
   
    journal = "Abhandlungen d. Kgl. Preußischen Akademie",
   
    year = "1887"
}

@eprint{Brown2011MultipleEP,
  title={Multiple Elliptic Polylogarithms},
  author={Francis Brown and Andrey Levin},
eprint = "1110.6917",
    archivePrefix = "arXiv",
    primaryClass = "math.NT",

}

@article{Dorigoni:2021jfr,
    author = "Dorigoni, Daniele and Kleinschmidt, Axel and Schlotterer, Oliver",
    title = "{Poincar\'e series for modular graph forms at depth two. Part I. Seeds and Laplace systems}",
    eprint = "2109.05017",
    archivePrefix = "arXiv",
    primaryClass = "hep-th",
    reportNumber = "UUITP-41/21, DCPT-21/13",
    doi = "10.1007/JHEP01(2022)133",
    journal = "JHEP",
    volume = "01",
    pages = "133",
    year = "2022"
}

@article{Dorigoni:2021ngn,
    author = "Dorigoni, Daniele and Kleinschmidt, Axel and Schlotterer, Oliver",
    title = "{Poincar\'e series for modular graph forms at depth two. Part II. Iterated integrals of cusp forms}",
    eprint = "2109.05018",
    archivePrefix = "arXiv",
    primaryClass = "hep-th",
    reportNumber = "UUITP-42/21, DCPT-21/15",
    doi = "10.1007/JHEP01(2022)134",
    journal = "JHEP",
    volume = "01",
    pages = "134",
    year = "2022"
}

@article{DHoker:2018mys,
    author = "D'Hoker, Eric and Green, Michael B. and Pioline, Boris",
    title = "{Asymptotics of the $D^8 R^4$ genus-two string invariant}",
    eprint = "1806.02691",
    archivePrefix = "arXiv",
    primaryClass = "hep-th",
    reportNumber = "QMUL-PH-18-10, DAMTP-2018-24",
    doi = "10.4310/CNTP.2019.v13.n2.a3",
    journal = "Commun. Num. Theor. Phys.",
    volume = "13",
    pages = "351--462",
    year = "2019"
}

@article{Dorigoni:2022npe,
    author = "Dorigoni, Daniele and Doroudiani, Mehregan and Drewitt, Joshua and Hidding, Martijn and Kleinschmidt, Axel and Matthes, Nils and Schlotterer, Oliver and Verbeek, Bram",
    title = "{Modular graph forms from equivariant iterated Eisenstein integrals}",
    eprint = "2209.06772",
    archivePrefix = "arXiv",
    primaryClass = "hep-th",
    reportNumber = "UUITP-37/22",
    month = "9",
    year = "2022"
}

@article{Kimura:2016dys,
    author = "Kimura, Taro and Pestun, Vasily",
    title = "{Quiver elliptic W-algebras}",
    eprint = "1608.04651",
    archivePrefix = "arXiv",
    primaryClass = "hep-th",
    doi = "10.1007/s11005-018-1073-0",
    journal = "Lett. Math. Phys.",
    volume = "108",
    number = "6",
    pages = "1383--1405",
    year = "2018"
}
\endgroup

\end{document}